\documentclass{icfi}

\usepackage{amsmath,amssymb}
\usepackage{latexsym}
\usepackage{graphicx}
\usepackage{bbm}
\newcommand{\sspe}{< P s,s>}
\newcommand{\sspb}{\overline{< P s^{\prime},s >}}
\newcommand{\ssp}{< P s,s^{\prime} >}
\newcommand{\const}{(\ref{uno}),(\ref{dos}), and (\ref{tres})}
\newcommand{\be}{\nopagebreak[3]\begin{equation}}
\newcommand{\ee}{\end{equation}}
\newcommand{\ba}{\nopagebreak[3]\begin{eqnarray}}
\newcommand{\ea}{\end{eqnarray}}

\newcommand{\C}{\mathbb{C}}
\newcommand{\va}{\scriptscriptstyle}
\newcommand{\vani}{\scriptstyle}
\newcommand{\R}{\mathbb{R}}

\newcommand{\Hp}{{\cal H}_{phys}}
\newcommand{\Hg}{{\cal H}^{\va \cal G}_{ kin}}
\newcommand{\Hd}{{\cal H}^{\va \cal D}_{ kin}}
\newcommand{\Hk}{{\cal H}_{ kin}}
\newcommand{\PP}{{ P}}
\newcommand{\So}{{\widehat {S}}}

\newcommand{\BC}{BC }

\begin{document}
\markboth{Alejandro Perez} 
{INTRODUCTION TO LOOP QUANTUM GRAVITY AND SPIN FOAMS} 

%
\catchline{}{}{}{}{}
%

\title{INTRODUCTION TO LOOP QUANTUM GRAVITY AND SPIN FOAMS}

\author{\footnotesize ALEJANDRO PEREZ\footnote{Penn State
  University, 104 Davey Lab,
University Park, PA 16802,
USA.}}

\address{Institute for Gravitational Physics and Geometry, Penn State
  University, 104 Davey Lab\\
University Park, PA 16802,
USA\\ and \\
Centre de Physique Th\'eorique
CNRS \\
Luminy Case 907
F-13288 Marseille cedex 9, France\footnote{State completely without abbreviations, the
affiliation and mailing address, including country.}}



\maketitle


\begin{abstract}
These notes are a didactic overview of the non perturbative and
background independent approach to a quantum theory of gravity known as loop
quantum gravity. The definition of real connection variables
for general relativity, used as a starting point in the program,
is described in a simple manner. The main ideas leading to the
definition of the quantum theory are naturally introduced and the
basic mathematics involved is described. The main predictions of
the theory such as the discovery of Planck scale discreteness of
geometry and the computation of black hole entropy are reviewed.
The quantization and solution of the constraints is explained by
drawing analogies with simpler systems. Difficulties associated with
the quantization of the scalar constraint are discussed.

In a second part of the notes, the basic ideas behind the spin
foam approach are presented in detail for the simple solvable case of
2+1 gravity. Some results and ideas for four dimensional spin foams are
reviewed. \keywords{Quantum Gravity; Loop variables;
Non-perturbative methods; Path Integrals; Yang-Mills theory.}
\end{abstract}

\section{Introduction: why non perturbative quantum gravity?}

The remarkable experimental success of the standard model in the
description of fundamental interactions is the greatest achievement of
(relativistic) quantum field theory. Standard quantum field theory
provides an accomplished unification between the principles of quantum
mechanics and special relativity. The standard model is a very useful
example of a quantum field theory on the fixed background geometry of
Minkowski spacetime. As such, the standard model can only be regarded
as an approximation of the description of fundamental interactions
valid when the gravitational field is negligible. The standard model
is indeed a very good approximation describing particle physics in the
lab and in a variety of astrophysical situations because of the
weakness of the gravitational force at the scales of interest (see for
instance the lecture by Rog\'erio Rosenfeld\cite{rogelio}). Using the
techniques of quantum field theory on curved
spacetimes\cite{Wald:1995yp} one might hope to extend the
applicability of the standard model to situations where a non trivial,
but weak, gravitational field is present. These situations are thought
to be those where the spacetime curvature is small in comparison with
the Planck scale, although a clear justification for its regime of
validity in strong gravitational fields seems only possible when a
full theory of quantum gravity is available. Having said this, there
is a number of important physical situations where we do not have any
tools to answer even the simplest questions. In particular classical
general relativity predicts the existence of singularities in
physically realistic situations such as those dealing with black hole
physics and cosmology. Near spacetime singularities the classical
description of the gravitational degrees of freedom simply breaks
down. Questions related to the fate of singularities in black holes or
in cosmological situations as well as those related with apparent
information paradoxes are some of the reasons why we need a theory of
quantum gravity. This new theoretical framework---yet to be put
forward---aims at a consistent description unifying or, perhaps more
appropriately, underlying the principles of general relativity and
quantum mechanics.

The gravitational interaction is fundamentally different from all
the other known forces. The main lesson of general relativity is
that the degrees of freedom of the gravitational field are encoded
in the geometry of spacetime. The spacetime geometry is fully
dynamical: in gravitational physics the notion of absolute space
on top of which `things happen' ceases to make sense. The
gravitational field defines the geometry on top of which its own
degrees of freedom and those of matter fields propagate. This is
clear from the perspective of the initial value formulation of
general relativity, where, given suitable initial conditions on a
3-dimensional manifold, Einstein's equations determine the
dynamics that ultimately allows for the reconstruction of the
spacetime geometry with all the matter fields propagating on it. A
spacetime notion can only be recovered {\em a posteriori} once the
complete dynamics of the coupled geometry-matter system is worked
out. Matter affects the dynamics of the gravitational field and is
affected by it through the non trivial geometry that the latter
defines \footnote{\label{IVF}See for instance Theorem 10.2.2 in
\cite{Wald:1984rg}.}.  General relativity is not a theory of
fields moving on a curved background geometry; general relativity
is a theory of fields moving on top of each other\cite{book}.

In classical physics general relativity is not just a successful
description of the nature of the gravitational interaction.  
As a result of implementing the principles of general covariance,
general relativity provides the basic framework to assessing 
the physical world by cutting all ties to concepts of absolute space.
It represents the result of the long-line of developments that go all the way back to 
the thought experiments of Galileo about the relativity of the motion, to the
arguments of Mach about the nature of space and time, and finally to  the
magnificent conceptual synthesis of Einstein's:
the world is relational. There is no well defined notion of
absolute space and it only makes sense to describe physical
entities in relation to other physical entities. This conceptual
viewpoint is fully represented by the way matter and geometry play
together in general relativity. The full consequences of this in
quantum physics are yet to be unveiled\footnote{For a fascinating
account of the conceptual subtleties of general relativity see
Rovelli's book\cite{book}.}.

When we analyze a physical situation in pre-general-relativistic
physics we separate what we call the system from the relations to
other objects that we call the reference frame. The spacetime geometry
that we describe with the aid of coordinates and a metric is a
mathematical idealization of what in practice we measure using rods
and clocks. Any meaningful statement about the physics of the system is
a statement about the relation of some degrees of freedom in the
system with those of what we call the frame. The key point is that
when the gravitational field is trivial there exist a preferred set
of physical systems whose dynamics is very simple. These systems are
{\em inertial observers}; for instance one can think of them as given
by a spacial grid of clocks synchronized by the exchange of light
signals. These physical objects provide the definition of inertial
coordinates and their mutual relations can be described by a Minkowski
metric. As a result in pre-general relativistic physics we tend to
forget about rods and clocks that define (inertial) frames and we talk
about time $t$ and position $x$ and distances measure using the flat
metric $\eta_{ab}$\footnote{In principle we first use rods and clocks
to realize that in the situation of interests (e.g. an experiment at
CERN) the gravitational field is trivial and then we just encode this
information in a fix background geometry: Minkowski
spacetime.}. However, what we are really doing is comparing the
degrees of freedom of our system with those of a space grid of
world-lines of physical systems called {\em inertial observers}. From
this perspective, statements in special relativity are in fact
diffeomorphism invariant. The physics from the point of view of an
experimentalist---dealing with the system itself, clocks, rods, and
their mutual relations---is completely independent of coordinates. In
general relativity this property of the world is confronted head on:
only relational (coordinate-independent, or diffeomorphism invariant)
statements are meaningful (see discussion about the hole argument in
\cite{book}). There are no simple family of observers to define physical 
coordinates as in the flat case, so we use arbitrary labels (coordinates) and 
require the physics to be independent of them.

In this sense, the principles of general relativity state some
basic truth about the nature of the classical world. The far
reaching consequences of this in the quantum realm are certainly
not yet well understood.\footnote{A reformulation of quantum mechanics from a relational perspective
has been introduced by Rovelli\cite{Rovelli:1995fv} and further investigated in the context of conceptual puzzles of quantum 
mechanics\cite{Grot:1996xu,Gambini:2000ht,Gambini:2001pq,Gambini:2001ta,Gambini:2002tq}.
It is also possible that a radical change in the paradigms of quantum mechanics is necessary in the conceptual 
unification of gravity and the quantum\cite{Penrose:1992gt,Penrose:1995nd,Penrose:2000ic,Penrose:1994hp}.} 
However, it is very difficult to imagine that a notion of absolute 
space would be saved in the next step
of development of our understanding of fundamental physics. Trying to build
a theory of quantum gravity based on a notion of background
geometry would be, from this perspective, reminiscent of the
efforts by contemporaries of Copernicus of describing planetary
motion in terms of the geocentric framework.

\subsection{Perturbative quantum gravity}

Let us make some observations about the problems of standard
perturbative quantum gravity. In doing so we will revisit the
general discussion above, in a special situation. In standard
perturbative approaches to quantum gravity one attempts to
describe the gravitational interaction using the same techniques
applied to the definition of the standard model. As these
techniques require a notion of non dynamical background one
(arbitrarily) separates the degrees of freedom of the
gravitational field in terms of a background geometry $\eta_{ab}$
for $a,b=1\cdots 4$---fixed once and for all---and dynamical
metric fluctuations $h_{ab}$. Explicitly, one writes the spacetime
metric as 
\be 
\label{one} 
g_{ab}=\eta_{ab}+h_{ab}.
\end{equation} 
Notice that
the previous separation of degrees of freedom has no intrinsic
meaning in general relativity. In other words, for a generic space
time metric $g_{ab}$ we can write 
\begin{equation} 
 g_{ab}=\eta_{ab}+h_{ab}
=\tilde \eta_{ab}+\tilde h_{ab}, 
\end{equation} 
where $\eta_{ab}$ and $\tilde
\eta_{ab}$ can lead to different background 
light-cone structures of the underlying spacetime $(M,g_{ab})$; equally
natural choices of flat background metrics lead to different Minkowski
metrics in this sense. This is quite dangerous if we want to give any
physical meaning to the background, e.g., the light cone structures of
the two `natural' backgrounds will be generally different providing
different notions of causality!  Equation (\ref{one}) is used in the
classical theory in very special situations when one considers
perturbations of a given background $\eta_{ab}$. In quantum gravity
one has to deal with arbitrary superpositions of spacetimes; the above
splitting can at best be meaningful for very special semi-classical
states `peaked', so to say, around the classical geometry $\eta_{ab}$
with small fluctuations. It is very difficult to imagine how such a
splitting can be useful in considering general states with arbitrary
quantum excitations at all scales. Specially because of the dual role of the
gravitational field that simultaneously describes the geometry and
its own dynamical degrees of freedom. More explicitly, in the standard background
dependent quantization the existence of a fixed background
geometry is fundamental in the definition of the theory. For
instance, one expects fields at space-like separated points to
commute alluding to standard causality considerations. Even when
this is certainly justified in the range of applicability of the
standard model, in a background dependent quantization of gravity
one would be using the causal structure provided by the unphysical
background $\eta_{ab}$. Yet we know that the notion of causality
the world really follows is that of the full $g_{ab}$ (see
Footnote \ref{IVF}). This difficulty has been raised several times
(see for instance\cite{Wald:1984rg}). Equation (\ref{one}) could be
meaningful in special situations dealing with semi-classical issues,
but it does not seem to be of much use if one wants to describe the
fundamental degrees of freedom of quantum gravity.
\begin{figure}[h]\!\!\!\!\!\!
\centerline{\hspace{0.5cm} \(
\begin{array}{c}
\includegraphics[height=5cm]{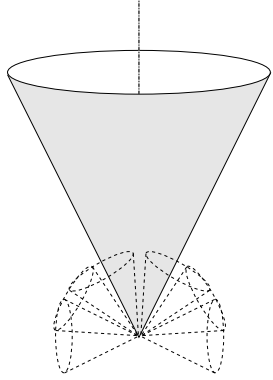}
\end{array}\) } \caption{The larger cone represents the light-cone at a point
  according to the ad hoc background $\eta_{ab}$. The smaller cones are a cartoon representation of
  the fluctuations of the true gravitational field represented by $g_{ab}$.}
\label{spinox}
\end{figure}

If we ignore all these issues and try to setup a naive
perturbative quantization of gravity we find that the theory is
non renormalizable. This can be expected from dimensional analysis
as the quantity playing the role of the coupling constant turns out to
be the Planck length $\ell_p$.  The non renormalizability of
perturbative gravity is often explained through an analogy with the (non-renormalizable) Fermi's four fermion effective description of the
weak interaction\cite{wein}. Fermi's four fermions theory is known
to be an effective description of the (renormalizable)
Weinberg-Salam theory.  The non renormalizable UV behavior of
Fermi's four fermion interaction is a consequence of neglecting
the degrees of freedom of the exchanged massive gauge bosons which are
otherwise disguised as the dimension-full coupling $\Lambda_{\va \rm Fermi}\approx 1/m_W^2$ at momentum transfer much
lower than the mass of the $W$ particle ($q^2<<m_W^2$).  A similar view is
applied to gravity to promote the search of a more fundamental
theory which is renormalizable or finite (in the perturbative
sense) and reduces to general relativity at low energies. From
this perspective it is argued that the quantization of
general relativity is a hopeless attempt to quantizing a
theory that does not contain the fundamental degrees of freedom.

These arguments, based on background dependent concepts, seem at
the very least questionable in the case of gravity.  Although one
should expect the notion of a background geometry to be useful in
certain semi-classical situations, the assumption that such
structure exists all the way down to the Planck scale is inconsistent
with what we know about gravity and quantum mechanics. General
considerations indicate that standard notions of space and time
are expected to fail near the Planck scale $\ell_p$ \footnote{ For
instance a typical example is to use a photon to measure distance.
The energy of the photon in our lab frame is given by
$E_{\gamma}=hc/\lambda$. We put the photon in a cavity and produce
a standing wave measuring the dimensions of the cavity in units of
the wavelength. The best possible precision is attained when the
Schwarzschild radius corresponding to energy of the photon is of the order of
its wavelength. Beyond that the photon can collapse to form a
black hole around some of the maxima of the standing wave. This
happens for a value $\lambda_c$ for which $\lambda_c\approx G
E_{\gamma}/c^4=h G/(\lambda_c c^3)$. The solution is
$\lambda_c\approx \sqrt{hG/c^3}$ which is Planck length.}. From
this viewpoint the non renormalizability of perturbative quantum
gravity is indicative of the inconsistency of the separation of
degrees of freedom in (\ref{one}).  The nature of spacetime is
expected to be very different from the classical notion in quantum
gravity.  The treatment that uses (\ref{one}) as the starting
point is assuming a well defined notion of background geometry at
all scales which directly contradicts these considerations.

In fact one could read the issues of divergences in an alternative
way.  The accepted jargon is that in a renormalizable theory the
microscopic physics does not affect low energy processes as its
effects can be recast in simple redefinition of a finite set of
parameters, i.e., renormalization. In the case of gravity this miracle
does not happen. All correlation functions blow up and one would need
to add an infinite number of corrections to the original Lagrangian
when integrating out microscopic physical degrees of freedom.  Now
imagine for the moment that a non perturbative quantization of general
relativity was available. If we now want to recover a `low energy
effective' description of a physical situation ---where (low energy)
physical considerations single out a preferred background---we would
expect the need of increasingly higher derivative correction terms to
compensate the fact the `high energy'\footnote{The notion of energy is
observer dependent in special relativity.  In the background
independent context is not even defined. We use the terminology `high
(low) energy' in this introduction to refer to the fundamental
(semi-classical) degrees of freedom in quantum gravity. However, the
reader should not take this too literally.  In fact if one tries to
identify the fundamental degrees of freedom of quantum gravity with a
preferred observer conflict with observation seem
inevitable\cite{Myers:2003fd,a17,Collins:2004bp}.} degrees of freedom know nothing
about that low energy background. In this sense the objections raised
above to the violent and un-natural splitting (\ref{one}) are related
to the issue of non renormalizability of general relativity.
The nature of the gravitational interaction is telling us that 
the standard paradigm of renormalization (based on the notion of 
a fixed background) is no longer applicable. A reformulation of
quantum field theory is necessary to cope with background independence. 

It is possible that new degrees of freedom would become important
at more fundamental scales. After all, that is the story of the
path that lead to the standard model where higher energy experiments
has often lead to the discovery of new interactions. 
It is also possible that including these degrees of freedom might be very important for the
consistency of the theory of quantum gravity. 
However, there is constraint that seem hardly avoidable: 
if we want to get a quantum theory that reproduces gravity
in the semi-classical limit we should have a background independent
formalism. In loop quantum gravity one stresses this viewpoint.
The hope is that learning how to define quantum field theory in
the absence of a background is a key ingredient in a recipe for quantum
gravity. Loop quantum gravity uses the quantization of general
relativity as a playground to achieve this goal.

I would like to finish this subsection with a quote of Weinberg's
1980 paper that would serve as introduction for the next section:
\vskip.5cm {\em ``It is possible that this problem}---the non
renormalizability of general relativity---{\em has arisen because
the usual flat-space formalism of quantum
  field theory simply cannot be applied to gravitation. After all, gravitation
  is a very special phenomenon, involving as it does the very topology of
  space and time''}
\vskip.5cm All these considerations make the case for a background
independent approach to quantum gravity. The challenge is to
define quantum field theory in the absence of any preestablished
notion of distance: quantum field theory without a metric.

\subsection{Loop quantum gravity}

Loop quantum gravity is an attempt to define a quantization of
gravity paying special attention to the conceptual lessons of
general relativity (the reader is encouraged to read  Rovelli's
book: {\em Quantum Gravity}\cite{book} as well as the  recent
review by Ashtekar and Lewandowski\cite{ash10}. For a detailed
account of the mathematical techniques involved in the
construction of the theory see Thiemann's book\cite{bookt}). The
theory is explicitly formulated in a background independent, and
therefore, non perturbative fashion. The theory is based on the
Hamiltonian (or canonical) quantization of general relativity in
terms of variables that are different from the standard metric variables. In terms of these variables general
relativity is cast into the form of a background independent
$SU(2)$ gauge theory partly analogous to $SU(2)$ Yang-Mills
theory.  The main prediction of loop quantum gravity (LQG) is the
discreteness\cite{lee1,c3} of the spectrum of geometrical operators
such as area and volume. The discreteness becomes important at the
Planck scale while the spectrum of geometric operators crowds very
rapidly at `low energy scales' (large geometries). This property of the
spectrum of geometric operators is consistent with the smooth
spacetime picture of classical general relativity.

Thus it is not surprising that perturbative approaches would lead
to inconsistencies. In splitting the gravitational field degrees
of freedom as in (\ref{one}) one is assuming the existence of a
background geometry which is smooth all the way down to the Planck
scale. As we consider contributions from `higher energies', this assumption is increasingly inconsistent
with the fundamental structure discovered in the non perturbative treatment.

The theory is formulated in four dimensions. Matter can be coupled
to the gravitational degrees of freedom and the inclusion of
matter has been studied in great detail. However, throughout these
lectures we will study the pure gravitational sector in order
simplify the presentation. On the super-symmetric extension of
canonical loop quantum gravity see
\cite{Jacobson:1987cj,Matschull:1993hy,Ling:2000su,Ling:2002ti,Ling:2000ss}.
A spin foam model for 3d super-gravity has been defined by Livine
and Oeckl\cite{Livine:2003hn}.

Despite the achievements of LQG there remain important issues to be addressed.
There are difficulties of a rather technical nature related to the
complete characterization of dynamics and the quantization of the so-called
scalar constraint. We will review these once we have introduced the basic formalism.
The most important question, however, remains open: whether the  semi-classical limit of
the LQG is consistent with general relativity and the Standard Model.

Before starting with the more technical material let us say a few
words regarding {\em unification}. In loop quantum gravity matter
is essentially added by coupling general relativity with it at the
classical level and then performing the (background independent)
canonical quantization. At this stage of development of the
approach it is not clear if there is some restriction in the kind
of interactions allowed or if degrees of freedom corresponding to
matter could arise in some natural way directly from those of
geometry. However, there is a unification which is addressed head
on by LQG: the need to find a common framework in which to
describe quantum field theories in the absence of an underlying
spacetime geometry.  The implications of this are far from being fully
developed.

\section{Hamiltonian formulation of general relativity: the new
variables}\label{hfnv}

In this section we introduce the variables that are used in the
definition of loop quantum gravity. We will present the $SU(2)$
Ashtekar-Barbero variables by starting with conventional ADM
variables, introducing triads at the canonical level, and finally
performing the well known canonical transformation that leads to
the new variables. This is, in my opinion, the simplest way to get
to the new variables. For a derivation that emphasizes the
covariant four dimensional character of these variables see
\cite{ash10}.

\subsection{Canonical analysis in ADM variables}

The action of general relativity in metric variables is given by
the Einstein-Hilbert action \be I[g_{\mu\nu}] =\frac{1}{2\kappa}
\int dx^4\ \sqrt{-g} R, \end{equation} where $\kappa=8\pi
G/c^3=8\pi\ell_p^2/\hbar$, $g$ is the determinant of the metric
$g_{ab}$ and $R$ is the Ricci scalar. The details of the
Hamiltonian formulation of this action in terms of ADM
variables can be found in\cite{Wald:1984rg}. One introduces a
foliation of spacetime in terms of space-like three dimensional
surfaces $\Sigma$. For simplicity we assume $\Sigma$ has no
boundaries. The ten components of the spacetime metric are
replaced by the six components of the induced Riemannian metric
$q_{ab}$ of $\Sigma$ plus the three components of the shift vector
$N_a$ and the lapse function $N$. In terms of these variables, after
performing the standard Legendre transformation, the action of
general relativity becomes \ba
\!\!\!\!\!\!\!\!\!\!\!\!\!\!\!\!\!\!\!\!\!\!\!\!\!\!\!\!&&
\nonumber I[q_{ab}, \pi^{ab}, N_a, N]= \frac{1}{2\kappa}\int dt
\int_{\Sigma} dx^3\ \ [ \pi^{ab} \dot q_{ab} \\ &&  + 2 N_b
{\nabla^{\va (3)}}_a(q^{-1/2} \pi^{ab})+N (q^{1/2}
[R^{\va(3)}-q^{-1}\pi_{cd}\pi^{cd}+\frac{1}{2}q^{-1}\pi^2])], \ea
where $\pi^{ab}$ are the momenta canonically conjugate to the
space metric $q_{ab}$, $\pi=\pi^{ab}q_{ab}$, ${\nabla^{\va
(3)}}_a$ is the covariant derivative compatible with the metric
$q_{ab}$, $q$ is the determinant of the space metric and
$R^{\va(3)}$ is the Ricci tensor of $q_{ab}$. The momenta
$\pi^{ab}$ are related to the extrinsic curvature
$K_{ab}$\footnote{The extrinsic curvature is given by
$K_{ab}=\frac{1}{2} {\cal L}_{n}q_{ab}$ where $n^a$ is the unit normal to $\Sigma$.} of $\Sigma$ by \be
\pi^{ab}=q^{-1/2} (K^{ab}-K q^{ab}) \end{equation} where $K=K_{ab}q^{ab}$ and
indices are raised with $q^{ab}$. Variations with respect to the
lapse and shift produce the four constraint equations:
\be\label{admc1} -V^b (q_{ab}, \pi^{ab})= 2 {\nabla^{\va
(3)}}_a(q^{-1/2} \pi^{ab})=0,\end{equation} and \be\label{admc2} -S(q_{ab},
\pi^{ab})=(q^{1/2}
[R^{\va(3)}-q^{-1}\pi_{cd}\pi^{cd}+{1}/{2}q^{-1}\pi^2])=0.\end{equation}
$V^b (q_{ab}, \pi^{ab})$ is the so-called vector constraint
and $S(q_{ab}, \pi^{ab})$ is the scalar constraint.  With this
notation the action can be written as \be \label{ADMA} I[q_{ab},
\pi^{ab},N_a, N]=\frac{1}{2\kappa} \int dt \int_{\Sigma} dx^3\
\left[ \pi^{ab} \dot q_{ab} - N_b V^b (q_{ab}, \pi^{ab})-N
S(q_{ab}, \pi^{ab})\right], \end{equation} where we identify the Hamiltonian
density ${\cal H}(q_{ab}, \pi^{ab},N_a,N)= N_b V^b (q_{ab},
\pi^{ab})+N S(q_{ab}, \pi^{ab})$. The Hamiltonian is a linear
combination of (first class) constraints, i.e., it vanishes
identically on solutions of the equations of motion. This is a
generic property of generally covariant systems. The symplectic
structure can be read off the previous equations, namely \be
\left\{\pi^{ab}(x),q_{cd}(y)\right\}=2 \kappa \
\delta^a_{(c}\delta^b_{d)} \delta(x,y), \ \ \ \
\left\{\pi^{ab}(x),\pi^{cd}(y)\right\}=\left\{q^{}_{ab}(x),q^{}_{cd}(y)\right\}=0
\end{equation} There are six configuration variables $q_{ab}$ and four
constraint equations (\ref{admc1}) and (\ref{admc2}) which implies
the two physical degrees of freedom of gravity \footnote{This
counting of physical degrees of freedom is correct because
the constraints are first class\cite{Henneaux:1992ig}.}.

\subsection{Toward the new variables: the triad formulation}

Now we do a very simple change of variables. The idea is to use a
triad (a set of three 1-forms defining a frame at each point in
$\Sigma$) in terms of which the metric $q_{ab}$ becomes \be
\label{triad}q_{ab}=e^i_a e^{j}_b \delta_{ij},\end{equation} where
$i,j=1,2,3$. Using these variables we introduce the densitized
triad  \be\label{E} E^{a}_i:=\frac{1}{2}\epsilon^{abc}
\epsilon_{ijk} e^j_b e^k_c. \end{equation} Using this definition, the
inverse metric $q^{ab}$ can be related to the densitized triad as
follows \be\label{dtriad} q q^{ab}=E^a_iE^b_j\delta^{ij}. \end{equation} We
also define \be\label{missing} K_{a}^{i}:=\frac{1}{\sqrt{{\rm det}(E)}}K_{ab} E^{b}_j
\delta^{ij}. \end{equation} A simple exercise shows that one can write the
canonical term in (\ref{ADMA}) as 
\be \pi^{ab} \dot
q_{ab}=  -\pi_{ab}\dot q^{ab}= 2 E^a_i\dot K_a^i 
\end{equation} 
and that the
constraints $V^a(q_{ab},\pi^{ab})$ and $S(q_{ab},\pi^{ab})$ can
respectively be written as $V^a(E^a_i,K_a^i)$ and
$S(E^a_i,K_a^i)$. Therefore we can rewrite (\ref{ADMA}) in terms
of the new variables. However, the new variables are certainly
redundant, in fact we are using the nine $E^a_i$ to describe the
six components of $q^{ab}$. The redundancy has a clear geometrical
interpretation: the extra three degrees of freedom in the triad
correspond to our ability to choose different local frames $e_a^i$
by local $SO(3)$ rotations acting in the internal indices
$i=1,2,3$. There must then be an additional constraint in terms of
the new variables that makes this redundancy manifest. The missing
constraint comes from (\ref{missing}): we overlooked the fact that
$K_{ab}=K_{ba}$ or simply that $K_{[ab]}=0$. By inverting the
definitions (\ref{E}) and (\ref{missing}) in order to write
$K_{ab}$ in terms of $E^a_i$ and $K_a^i$ one can show that the
condition $K_{[ab]}=0$ reduces to \be G_i(E^a_j,K_a^j):=
\epsilon_{ijk}E^{a j} K_a^k = 0. \end{equation} Therefore we must include
this additional constraint to (\ref{ADMA}) if we want to use the
new triad variables. With all this the action of general
relativity becomes \ba \label{TETA} &&\nonumber I[E^a_j,
K_a^j,N_a, N,N^j]=  \\ && \frac{1}{\kappa}\int dt \int_{\Sigma}
dx^3\ \left[ E^a_i\dot K_a^i - N_b V^b (E^a_j,K_a^j)- N
S(E^a_j,K_a^j)-N^i G_i(E^a_j,K_a^j)\right], \ea where the explicit
form of the constraints in terms of triad variables can be worked
out from the definitions above. The reader is encouraged to do
this exercise but it is not going to be essential to understanding
what follows (expressions for the constraints can be found in reference\cite{ash}). 
The symplectic structure now becomes \be
\label{PT} \left\{E^{a}_j(x),K_{b}^i(y)\right\}=\kappa \,
\delta^a_{b}\delta^i_{j} \delta(x,y), \ \ \ \
\left\{E^{a}_j(x),E^{b}_i(y)\right\}=\left\{K_{a}^j(x),K_{b}^i(y)\right\}=0
\end{equation} The counting of physical degrees of freedom can be done as
before.

\subsection{New variables: the Ashtekar-Barbero connection variables}\label{NV}

The densitized triad (\ref{E}) transforms in the vector
representation of $SO(3)$ under redefinition of the triad
(\ref{triad}). Consequently, so does its conjugate momentum
$K_a^i$ (see equation (\ref{missing})). There is a natural $so(3)$-connection that defines the notion of
covariant derivative compatible with the triad. This connection is
the so-called {\em spin connection}  $\Gamma^{i}_a$ and is characterized as
the solution of Cartan's structure equations  \be
\partial_{[a} e^i_{b]} + \epsilon^{i}_{\ jk}\Gamma^j_{[a}
e^k_{b]} = 0 \end{equation} The solution to the previous equation can be written explicitly in terms of the triad components \be\label{19} \Gamma^i_a=
-\frac{1}{2} \epsilon^{ij}_{\ \ k}
e^b_j\left(\partial_{[a}e_{b]}^k+ \delta^{kl} \delta_{ms} e^{c}_l
e_a^m\partial_b e_c^s\right), \end{equation} where $e^a_i$ is the inverse
triad ($e^a_ie_a^j=\delta_{i}^j$). We can obtain an explicit function of the
densitized triad---$\Gamma^i_a(E_j^b)$---inverting (\ref{E})
from where \be\label{twenty} e_a^i=\frac{1}{2}\frac{\epsilon_{abc}
\epsilon^{ijk} E_j^b E_k^c}{\sqrt{|{\rm
    det}(E)|}} \ \ \ {\rm and} \ \ \ e^a_i=\frac{{\rm sgn}({\rm
    det}(E))\ E^a_i }{\sqrt{|{\rm
    det}(E)}|}.
\end{equation} The spin connection is an $so(3)$ connection that transforms
in the standard inhomogeneous way under local $SO(3)$
transformations. The
Ashtekar-Barbero\cite{Barbero:1995ud,Barbero:1994ap,Barbero:1993aa}
variables are defined by the introduction of a new connection
$A_a^i$ given by \be\label{immi} A_a^i=\Gamma_a^i +\gamma K_a^i,
\end{equation} where $\gamma$ is any non vanishing real number called the Immirzi
parameter\cite{immi}. The new variable is also an $so(3)$
connection as adding a quantity that transforms as a vector to a
connection gives a new connection. The remarkable fact about this
new variable is that it is in fact conjugate to $E_a^i$. More
precisely the Poisson brackets of the new variables are \be
\left\{E^{a}_j(x),A_{b}^i(y)\right\}=\kappa \, \gamma
\delta^a_{b}\delta^i_{j} \delta(x,y), \ \ \ \
\left\{E^{a}_j(x),E^{b}_i(y)\right\}=\left\{A_{a}^j(x),A_{b}^i(y)\right\}=0.
\end{equation} All the previous equations follow trivially from (\ref{PT})
except for $\left\{A_{a}^j(x),A_{b}^i(y)\right\}=0$ which requires
more calculations. The reader is invited to check it.

Using the connection variables the action becomes \ba \label{NEW}
&&\nonumber I[E^a_j, A_a^j,N_a, N,N^j]=  \\ && \frac{1}{\kappa}
\int dt \int_{\Sigma} dx^3\left[  E^a_i \dot A^a_i - N^b V_b
(E^a_j,A_a^j)- N S(E^a_j,A_a^j)-N^i G_i(E^a_j,A_a^j)\right], \ea
where the constraints are explicitly given by: \be\label{uno}
 V_b (E^a_j,A_a^j)=E_j^a F_{ab}-(1+\gamma^2) K_a^i G_i
\end{equation}
\be\label{dos}
S(E^a_j,A_a^j)= \frac{E^{a}_iE^b_j}{\sqrt{{\rm
    det}(E)}} \left(\epsilon^{ij}_{\ \ k} F^k_{ab}-2(1+\gamma^2)
K^i_{[a}K^j_{b]} \right)
\end{equation}
\be\label{tres}
 G_i(E^a_j,A_a^j)=D_aE^a_i,
\end{equation} where $F_{ab}=\partial_a A_b^i-\partial_b A_a^i+\epsilon^i_{\
jk}A_a^jA^k_b$ is the curvature of the connection $A_a^i$ and
$D_aE^a_i=\partial_a E^a_i+\epsilon_{ij}^{\ \ k} A^j_aE^a_k$ is
the covariant divergence of the densitized triad. We have seven
(first class) constraints for the 18 phase space variables
$(A_a^i, E^b_j)$. In addition to imposing conditions among the
canonical variables, first class constraints are generating
functionals of (infinitesimal) gauge transformations. From the
18-dimensional phase space of general relativity we end up with 11
fields necessary to coordinatize the constraint surface on which
the above seven conditions hold. On that 11-dimensional constraint
surface, the above constraint generate a seven-parameter-family of
gauge transformations. The reduce phase space is four dimensional
and therefore the resulting number of physical degrees of freedom
is {\em two}, as expected.

The constraint (\ref{tres}) coincides with the standard Gauss law
of Yang-Mills theory (e.g. $\vec{\nabla}\cdot \vec E=0$ in
electromagnetism). In fact if we ignore  (\ref{uno}) and
(\ref{dos}) the phase space variables $(A_a^i, E^b_j)$ together with the
Gauss law (\ref{tres}) characterize the physical phase space of an
$SU(2)$\footnote{The constraint structure does not distinguish
$SO(3)$ from $SU(2)$ as both groups have the same Lie algebra.
From now on we choose to work with the more fundamental (universal
covering) group $SU(2)$. In fact this choice is physically
motivated as $SU(2)$ is the gauge group if we want to include
fermionic matter\cite{c10bis,c11,baez11}.} Yang-Mills (YM) theory. The
gauge field is given by the connection $A_a^i$ and its conjugate
momentum is the electric field $E_j^b$. Yang-Mills theory is
a theory defined on a background spacetime geometry. Dynamics in
such a theory is described by a non vanishing Hamiltonian---the
Hamiltonian density of YM theory being ${\cal
H}=E_a^iE^a_i+B_a^iB^a_i$. General relativity is a generally
covariant theory and coordinate time plays no physical role.  The
Hamiltonian is a linear combination of constraints.\footnote{In the physics of the standard model we are used to identifying the coordinate $t$ 
with the physical time of a suitable family of observers. In the general covariant context of 
gravitational physics the coordinate time $t$ plays the role of a label with no physical 
relevance. One can arbitrarily change the way we coordinatize spacetime without affecting the physics.
This redundancy in the description of the physics (gauge symmetry) induces the appearance of constraints 
in the canonical formulation. The constraints in turn are the generating functions of these
gauge symmetries. The Hamiltonian generates evolution in coordinate time $t$ but because redefinition of $t$ is pure gauge, the 
Hamiltonian is a constraint itself, i.e. ${\cal H}=0$ on shell\cite{dirac,Henneaux:1992ig}.
More on this in the next section.} Dynamics is
encoded in the constraint equations \const. In this sense we can
regard general relativity in the new variables as a background
independent relative of $SU(2)$ Yang-Mills theory. We will see in
the sequel that the close similarity between these theories will
allow for the implementation of techniques that are very natural
in the context of YM theory.

\subsubsection{Gauge transformations}

Now let us analyze the structure of the gauge transformations
generated by the constraints \const. From the previous paragraph
it should not be surprising that the Gauss law (\ref{tres})
generates local $SU(2)$ transformations as in the case of YM
theory. Explicitly, if we define the smeared version of
(\ref{tres}) as \be G({\mathbbm{\alpha}})=\int_{\Sigma} dx^3 \
\alpha^i G_i(A_a^i,E^a_i)=\int_{\Sigma} dx^3 \alpha^i D_aE^a_i,
\end{equation} a direct calculation implies\be \delta_{\va G}
A_a^i=\left\{A_a^i,G({\mathbbm{\alpha}}) \right\}= -D_a \alpha^i \
\ \ {\rm and}\ \ \ \delta_{\va G}
E^a_i=\left\{E^a_i,G({\mathbbm{\alpha}})
\right\}=\left[E,\alpha\right]_i. \end{equation} If we write $A_a=A_a^i
\tau_i\in su(2)$ and $E^a=E^a_i \tau^i \in su(2)$, where $\tau_i$
are generators of $SU(2)$, we can write the finite version of the
previous transformation \be \label{finita}A^{\prime}_a=g A_a
g^{-1}+ g \partial_a g^{-1}\ \ \ {\rm and}\ \ \ E^{a\prime}=gE^ag^{-1} ,\end{equation}
which is the standard way the connection and the electric field
transform under gauge transformations in YM theory.

The vector constraint (\ref{uno}) generates three dimensional
diffeomorphisms of $\Sigma$. This is clear from the action of the
smeared constraint \be V(N^a)=\int_{\Sigma} dx^3 \ N^a V_a
(A_a^i,E^a_i) \end{equation} on the canonical variables 
\be\delta_{\va V}
A_a^i=\left\{A_a^i,V(N^a) \right\}= {\cal L}_{N} A_a^i \ \ \ {\rm
and}\ \ \   \delta_{\va V} E^a_i=\left\{E^a_i,V(N^a) \right\}={\cal
L}_{N} E^a_i, 
\end{equation} 
where ${\cal L}_{N}$ denotes the Lie derivative
in the $N^a$ direction. The exponentiation of these infinitesimal
transformations leads to the action of finite diffeomorphisms on
$\Sigma$.

Finally, the scalar constraint (\ref{dos}) generates coordinate
time evolution (up to space diffeomorphisms and local $SU(2)$ transformations). The
total Hamiltonian $H[\alpha, N^a, N]$ of general relativity can be
written as \be H(\alpha, N^a, N)=G(\alpha)+V(N^a)+S(N), \end{equation} where
\be S(N)=\int_{\Sigma} dx^3 \ N S (A_a^i,E^a_i).  \end{equation} Hamilton's equations of motion
are therefore \be \dot A_a^i=\left\{A_a^i,H(\alpha, N^a, N) \right\}=
\left\{A_a^i,S(N) \right\}+\left\{A_a^i,G(\alpha)
\right\}+\left\{A_a^i,V(N^a) \right\},\end{equation} and \be \dot
E^a_i=\left\{E^a_i,H(\alpha, N^a, N) \right\}= \left\{E^a_i,S(N)
\right\}+\left\{E^a_i,G(\alpha) \right\}+\left\{E^a_i,V(N^a)
\right\}.\end{equation} The previous equations define the action of $S(N)$ up
to infinitesimal $SU(2)$ and  diffeomorphism transformations given by
the last two terms and the values of $\alpha$ and $N^a$
respectively. In general relativity coordinate time evolution does not
have any physical meaning. It is analogous to  a $U(1)$ gauge
transformation in QED. 


\subsubsection{Constraints algebra}

Here we simply present the structure of the constraint algebra of
general relativity in the new variables. \be
\left\{G(\alpha),G(\beta) \right\}=G([\alpha,\beta]), \end{equation} where
$\alpha=\alpha^{i}\tau_i\in su(2)$, $\beta=\beta^{i}\tau_i\in
su(2)$ and $[\alpha,\beta]$ is the commutator in $su(2)$. \be
\left\{G(\alpha),V(N^a) \right\}= -G({\cal L}_{N}\alpha). \end{equation} \be
\left\{G(\alpha),S(N) \right\}= 0. \end{equation} \be \left\{V(N^a),V(M^a)
\right\}= V([N,M]^a), \end{equation} where
$[N,M]^a=N^b\partial_bM^a-M^b\partial_bN^a$ is the vector field
commutator. \be \left\{S(N),V(N^a) \right\}= -S({\cal L}_{N}N).
\end{equation} Finally \be \label{prob} \left\{S(N),S(M) \right\}=V(S^a)+
{\rm terms \ proportional\ to \ the \ Gauss
  \ constraint} ,
\end{equation} where for simplicity we are ignoring the terms proportional to
the Gauss law (the complete expression can be found in
\cite{ash10}) and \be S^{a}= \frac{E^a_iE^b_j \delta^{ij}}{|{\rm
det} E|}(N\partial_b M-M\partial_b N).\end{equation} Notice 
that  instead
of structure constants, the r.h.s. of (\ref{prob}) is written in
terms of field dependent structure functions. For this reason it
is said that the constraint algebra does not close in the BRS
sense.

\subsubsection{Ashtekar variables}

The connection variables introduced in this section do not have a
simple relationship with four dimensional fields. In particular
the connection (\ref{immi}) cannot be obtained as the pullback to
$\Sigma$ of a spacetime connection\cite{Samuel:2000ue}. Another
observation is that the constraints (\ref{uno}) and (\ref{dos})
dramatically simplify when $\gamma^2=-1$. Explicitly, for
$\gamma=i$ we have \be\label{unob}
 V^{\va SD}_b=E_j^a F_{ab}
\end{equation} \be\label{dosb} S^{\va SD}= \frac{E^{a}_iE^b_j}{\sqrt{{\rm
    det}(E)}} \ \epsilon^{ij}_{\ \ k} F^k_{ab}
\end{equation} \be\label{tresb}
 G^{\va SD}_i=D_aE^a_i,
\end{equation} where $SD$ stands for self dual; a notation that will become
clear below. Notice that with $\gamma=i$ the connection
(\ref{immi}) is complex\cite{ash} (i.e. $A_a\in sl(2,\C)$). To
recover real general relativity these variables must be
supplemented with the so-called reality condition that follows
from (\ref{immi}), namely \be \label{real} A_a^i+\bar
A^i_a=\Gamma_a^i(E). \end{equation} In addition to the simplification of the
constraints, the connection obtained for this choice of the
Immirzi parameter is simply related to a spacetime connection.
More precisely, it can be shown that $A_a$ is the pullback of
$\omega_{\mu}^{{\va +}IJ}$ ($I,J=1,\cdots 4$) where \be
\omega_{\mu}^{{\va +}IJ}=\frac{1}{2}(\omega_{\mu}^{IJ}-\frac{i}{2}
\epsilon^{IJ}_{\ \
  KL} \omega^{KL}_{\mu})
\end{equation} is the self dual part of a Lorentz connection
$\omega_{\mu}^{IJ}$. The gauge group---generated by the (complexified) Gauss
constraint---is in this case $SL(2,\C)$.

Loop quantum gravity was initially formulated in terms of these
variables. However, there are technical difficulties in defining the quantum
theory when the connection is valued in the Lie algebra of a non compact group.
Progress has been achieved constructing the quantum theory in terms of the real
variables introduced in Section \ref{NV}.

\subsection{Geometric interpretation of the new variables}\label{geo}

The geometric interpretation of the connection $A_a^i$, defined in
(\ref{immi}), is standard. The connection provides a definition of
{\em parallel transport} of $SU(2)$ spinors on the space manifold
$\Sigma$. The natural object is the $SU(2)$ element defining 
parallel transport along a path $e\subset \Sigma$  also called {\em holonomy}
denoted  $h_{e}[A]$, or more explicitly  \be \label{hol}h_{e}[A]=P
\exp-\int \limits_{e} A,\end{equation} where $P$ denotes a
path-order-exponential (more details in the next section).

The densitized triad---or electric field---$E_i^a$ also has a
simple geometrical meaning.  $E_i^a$ encodes the full background
independent Riemannian geometry of $\Sigma$ as is clear from
(\ref{dtriad}). Therefore, any geometrical quantity in space can
be written as a functional of $E^a_i$. One of the simplest is the
area $A_{S}[E^a_i]$ of a surface $S\subset \Sigma$ whose
expression we derive in what follows. Given a two dimensional
surface in $S\subset \Sigma$---with normal \be n_a=\frac{\partial
x^b}{\partial \sigma^1}\frac{\partial x^c}{\partial \sigma^2}
\epsilon_{abc}\end{equation} where $\sigma^1$ and $\sigma^2$ are local
coordinates on $S$---its area is given by  \be A_S[q^{ab}]=\int_S
\sqrt{h} \ d\sigma^1d\sigma^2, \end{equation} where $h={\rm det}(h_{ab})$
is the determinant of the metric $h_{ab}=q_{ab}-n^{-2}n_an_b$ induced on $S$ by
$q^{ab}$. From equation (\ref{dtriad}) it follows that
${\rm det}(q^{ab})={\rm det}(E^a_i)$. Let us contract
(\ref{dtriad}) with $n_an_b$, namely \be\label{dete} q
q^{ab}n_an_b= E^a_iE^b_j\delta^{ij}n_an_b. \end{equation} Now observe that
$q^{nn}=q^{ab}n_an_b$ is the $nn$-matrix element of the inverse of
$q_{ab}$. Through the well known formula for components of the
inverse matrix we have that \be
q^{nn}=\frac{{\det}(q_{ab}-n^{-2}n_an_b)}{{\det}(q_{ab})}=\frac{h}{q}.\end{equation}
But $q_{ab}-n^{-2} n_an_b$ is precisely the induced metric
$h_{ab}$. Replacing $q^{nn}$ back into (\ref{dete}) we conclude
that \be h= E^a_iE^b_j\delta^{ij}n_an_b. \end{equation} Finally we can write
the area of $S$ as an explicit functional of $E^a_i$:
\be\label{areac} A_S[E^a_i]=\int_S
\sqrt{E^a_iE^b_j\delta^{ij}n_an_b}\ d\sigma^1d\sigma^2. \end{equation} This
simple expression for the area of a surface will be very important
in the quantum theory.

\section{The Dirac program: the non perturbative quantization of GR}

The Dirac program\cite{dirac,Henneaux:1992ig} applied to the
quantization of generally covariant systems consists of the
following steps\footnote{There is another way to canonically
quantize a theory with constraints that is also developed by
Dirac. In this other formulation one solves the constraints at the
classical level to identify the physical or reduced phase space to
finally quantize the theory by finding a representation of the
algebra of physical observables in the physical Hilbert space
$\Hp$. In the case of four dimensional gravity this alternative
seem intractable due to the difficulty in identifying the true
degrees of freedom  of general relativity.}:
\begin{romanlist}[(ii)]
\item
Find a representation of the phase space variables of the theory
as operators
in an auxiliary or kinematical Hilbert space $\Hk$ satisfying the
standard commutation relations, i.e., $\{\ ,\ \}\rightarrow
-i/\hbar [\ ,\ ] $.

\item Promote the constraints to (self-adjoint) operators in $\Hk$. In the
case of gravity we must quantize the seven constraints $G_i(A,E)$,
$V_a(A,E)$, and $S(A,E)$.

\item Characterize the space of solutions of the constraints and define
the corresponding inner product that defines a notion of physical probability.
This defines the so-called physical Hilbert space $\Hp$.

\item Find a (complete) set of gauge invariant observables, i.e., operators
commuting with the constraints. They represent the questions that
can be addressed in the generally covariant quantum theory.

\end{romanlist}

\subsection{A simple example: the reparametrized particle}

Before going into the details of the definition of LQG we will
present the general ideas behind the quantization of generally
covariant systems using the simplest possible example: a non
relativistic particle. A non relativistic particle can be treated
in a generally covariant manner by introducing a non physical
parameter $t$ and promoting the physical time $T$ to a canonical
variable. For a particle in one dimension the standard action \be
S(p_X,X)=\int dT \left[ p_X \frac{dX}{dT}- H(p_X,X)\right] \end{equation} can be replaced by
the reparametrization invariant action \be
S_{rep}(P_T,P_X,T,X,N)=\int dt \left[ p_T \dot T +p_X \dot X- N
(p_T+H(p_X,X))\right],\end{equation} where the dot denotes derivative with
respect to $t$. The previous action is invariant under the
redefinition $t\rightarrow t^{\prime }=f(t)$. The variable $N$ is
a Lagrange multiplier imposing the scalar constraint
\be\label{c1d} C=p_T+H(p_X,X)=0.
\end{equation} 
Notice the formal similarity
with the action of general relativity (\ref{ADMA}) and
(\ref{TETA}). As in general relativity the Hamiltonian of the
system $H_{rep}=N (p_T+H(p_X,X))$ is zero on shell. It is easy to
see that on the constraint surface defined by (\ref{c1d})
$S_{rep}$ reduces to the standard $S$, and thus the new action
leads to the same classical solutions. The constraint $C$ is a
generating function of infinitesimal $t$-reparametrizations
(analog to diffeomorphisms in GR). This system is the simplest
example of generally covariant system.

Let us proceed and analyze the quantization of this action
according to the rules above.
\begin{romanlist}[(ii)]

\item We first define an auxiliary or {\em kinematical} Hilbert
space $\Hk$. In this case we can simply take $\Hk={\cal
L}^2(\R^2)$. Explicitly we use (kinematic) wave functions of
$\psi(X,T)$ and define the inner product \be
\label{fleme}<\phi,\psi>=\int dXdT \ \overline{\phi(X,T)}
\psi(X,T).\end{equation}

We next promote the phase space variables to self adjoint
operators satisfying the appropriate commutation relations. In
this case the standard choice is that $\widehat X$ and $\widehat
T$ act simply by multiplications and $\widehat
p_X=-i\hbar\partial/\partial X$ and $\widehat
p_T=-i\hbar\partial/\partial T$

\item The constraint becomes---this step is highly non trivial
in a field theory due to regularization issues: \be \widehat
C=-i\hbar\frac{\partial}{\partial T}-\hbar^2\frac{\partial^2}{\partial
X^2}+ V(X). \end{equation} Notice that the constraint equation $\widehat
C|\psi>=0$ is nothing else than the familiar Schroedinger equation.

\item The solutions of the quantum constraint are in this case
the solutions of Schroedinger equation. As it is evident from the
general form of Schroedinger equation we can characterize the set
of solutions by specifying the initial wave form at some $T=T_0$,
$\psi(X)=\psi(X,T=T_0)$. The physical Hilbert space is therefore
the standard $\Hp={\cal L}^2(\R)$ with physical inner product \be
<\phi,\psi>_p=\int dX \ \overline{\phi(X)} \psi(X). \end{equation} The
solutions of Schroedinger equation are not normalizable in $\Hk$
(they are not square-integrable with respect to (\ref{fleme}) due
to the time dependence impose by the Schroedinger equation). This
is a generic property of the solutions of the constraint when the
constraint has continuous spectrum (think of the eigenstates of
$\widehat P$ for instance).

\item Observables in this setting are easy to find. We are looking
for phase space functions commuting with the constraint. For
simplicity assume for the moment that we are dealing with a free
particle, i.e., $C=p_T+p_X^2/(2m)$. We have in this case the
following two independent observables: \be \widehat O_1=\widehat
X-\frac{\widehat p_X}{m}(\widehat T-T_0)\ \ \ \ {\rm and} \ \ \ \
\widehat O_2=\widehat p_X, \end{equation} where $T_0$ is just a $c$-number.
These are just the values of $X$ and $P$ at $T=T_0$. In the
general case where $V(X)\not=0$ the explicit form of these
observables as functions of the phase space variables will depend
on the specific interaction. Notice that in $\Hp$, as defined
above, the observables reduce to position $O_1=X$ and momentum
$O_2=p_X$ as in standard quantum mechanics.
\end{romanlist}

We have just reproduced standard quantum mechanics by quantizing
the reparametrization invariant formulation. As advertised in the general 
framework of the Dirac program applied to generally covariant systems, the
full dynamics is contained in the quantum constraints---here the
Schroedinger equation.

\subsection{The program of loop quantum gravity}

A formal description of the implementation of Dirac's program in the
case of gravity is presented in what follows. In the next section we
will start a more detailed review of each
of these steps.
\begin{romanlist}[(ii)]

\item In order to define the kinematical Hilbert space of general relativity in terms
 of the new variables we will choose the polarization where the connection
is  regarded as the configuration variable. The kinematical Hilbert
space consists of a suitable  set of functionals of the connection
$\psi[A]$ which are square integrable with respect to a suitable
(gauge invariant and diffeomorphism invariant) measure $d\mu_{\va
AL}[A]$ (called Ashtekar-Lewandowski measure\cite{ash3}). The
kinematical inner product is given by \be\label{alm}
<\psi,\phi>=\mu_{AL}[\overline \psi \phi]=\int d\mu_{\va AL}[A]\
\overline \psi[A] \phi[A]. \end{equation} In the next section we give the
precise definition of $\Hk$.

\item Both the Gauss constraint and the diffeomorphism constraint have a natural (unitary)
action on states on $\Hk$. For that reason the quantization (and subsequent solution)
is rather straightforward. The simplicity of these six-out-of-seven constraints is
a special merit of the use of connection variables as will become transparent in the sequel.

The scalar constraint (\ref{dos}) does not have a simple geometric
interpretation. In addition it is highly non linear which
anticipates the standard UV problems that plague quantum field
theory in the definition of products of fields (operator valued
distributions) at a same point.  Nevertheless, well defined
versions of the scalar constraint have been constructed. The fact
that these rigorously defined (free of infinities) operators exist
is again intimately related to the kind of variables used for the
quantization and some other special miracles occurring due to the
background independent nature of the approach. We emphasize that
the theory is free of divergences.

\item
 {\em Quantum
Einstein's equations} can be formally expressed now as:
\ba \nonumber
&& \widehat G_i(A,E)|\Psi> := \widehat{D_a E^a_i}|\Psi>=0\\
&&\nonumber\widehat V_a(A,E)|\Psi> :=\widehat{E^a_i
F^i_{ab}(A)}|\Psi>=0,\\ &&\widehat S(A,E)|\Psi> :=
[\widehat{{\sqrt{{\rm det}E}}^{-1}{E_i^a E_j^b
F^{ij}_{ab}(A)}}+\cdots]|\Psi>=0.\label{QEE}\ea As mentioned
above, the space of solutions of the first six equations is well
understood. The space of solutions of quantum scalar constraint
remains an open issue in LQG. For some mathematically consistent
definitions of $\widehat S$ the characterization of the solutions
is well understood\cite{ash10}. The definition of the physical
inner product is still an open issue. We will introduce the spin
foam approach in Section \ref{SFM} as a device for extracting
solutions of the constraints producing at the same time a
definition of the physical inner product in LQG. The spin foam
approach also aims at the resolution of some difficulties
appearing in the quantization of the scalar constraint that will
be discussed in Section \ref{lee}. It is yet not clear, however,
whether these consistent theories reproduce general relativity in
the semi-classical limit.

\item Already in classical gravity the construction of gauge
independent quantities is a subtle issue. At the present stage of
the approach physical observables are explicitly  known only in some
special cases. Understanding the set of physical observables is
however intimately related with the problem of characterizing the
solutions of the scalar constraint described before. We will
illustrate this by discussing simple examples of quasi-local Dirac
observable in Section \ref{lee}. For a vast discussion about this
issue we refer the reader to Rovelli's book\cite{book}.

\end{romanlist}

\section{Loop quantum gravity}\label{sect4}

\subsection{Definition of the kinematical Hilbert space}

\subsubsection{The choice of variables: Motivation}

As mentioned in the first item of the program formally described
in the last section, we need to define the vector space of
functionals of the connection and a notion of inner product to
provide it with a Hilbert space structure of $\Hk$. As emphasized
in Section \ref{geo}, a natural quantity associated with a
connection consists of the holonomy along a path (\ref{hol}). We
now give a more precise definition of it: Given a one dimensional
oriented path $e:[0,1]\subset\R\rightarrow\Sigma$ sending the
parameter $s\in [0,1]\rightarrow x^{\mu}(s)$, the holonomy
$h_{e}[A]\in SU(2)$ is denoted \be \label{holy}h_{e}[A]=P
\exp-\int_{e} A \; , \end{equation} where $P$ denotes a
path-ordered-exponential. More precisely, given the unique
solution $h_{e}[A,s]$ of the ordinary differential equation\be
\frac{d}{ds} h_{e}[A,s]+\dot x^{\mu}(s) A_{\mu} h_{e}[A,s]=0\end{equation}
with the boundary condition $h_{e}[A,0]=\mathbbm{1}$, the holonomy
along the path $e$ is defined as \be h_{e}[A]=h_{e}[A,1].\end{equation} The
previous differential equation has the form of a time dependent
Schroedinger equation, thus its solution can be formally written
in terms of the familiar series expansion \be
h_{e}[A]=\sum\limits_{n=0}^{\infty} \int\limits_{0}^{1} ds_1
\int\limits_{0}^{s_1} ds_2\ \cdots \! \int\limits_{0}^{s_{n-1}}
ds_n\  \dot x^{\mu_1}(s_1)\cdots \dot x^{\mu_n}(s_n)\
A_{\mu_1}(s_1)\cdots A_{\mu_n}(s_n), \end{equation} which is what the path
ordered exponential denotes in (\ref{holy}). Let us list some
important properties of the holonomy:
\begin{romanlist}[(b)]
 \item The definition of $h_{e}[A]$ is independent of the
parametrization of the path $e$.

\item The holonomy is a representation of the groupoid of oriented
paths. Namely, the holonomy of a path given by a single point is
the identity, given two oriented paths $e_1$ and $e_2$ such that
the end point of $e_1$ coincides with the starting point of $e_2$
so that we can define $e=e_1e_2$ in the standard fashion, then we
have \be\label{groupoid} h_{e}[A]=h_{e_1}[A]h_{e_2}[A],\end{equation} where
the multiplication on the right is the $SU(2)$ multiplication. We
also have that \be h_{e^{-1}}[A]=h^{-1}_{e}[A]. \end{equation} \item The
holonomy has a very simple behavior under gauge transformations.
It is easy to check from (\ref{finita}) that under a gauge
transformation generated by the Gauss constraint, the holonomy
transforms as \be\label{gghol}
h^{\prime}_{e}[A]=g(x(0))\ h_{e}[A]\ g^{-1}(x(1)). \end{equation}

\item The holonomy transforms in a very simple way under the
action of diffeomorphisms (transformations generated by the vector
constraint (\ref{uno})). Given $\phi \in {\rm Diff}(\Sigma)$ we
have \be\label{ddhol} h_{e}[\phi^*A]=h_{\phi^{-1}(e)}[A], \end{equation}
where $\phi^*A$ denotes the action of $\phi$ on the connection. In
other words, transforming the connection with a diffeomorphism is
equivalent to simply `moving' the path with $\phi^{-1}$.

\end{romanlist}

Geometrically the holonomy $h_{e}[A]$ is a functional of the
connection that provides a rule for the parallel transport of
$SU(2)$ spinors along the path $e$. If we think of it as a
functional of the path $e$ it is clear that it captures all the
information of the field $A_a^i$. In addition it has very simple
behavior under the transformations generated by six of the
constraints \const. For these reasons the holonomy is a  
natural choice of basic functional of the connection.

\subsubsection{The algebra of basic (kinematic)
observables}\label{cyll}

Shifting the emphasis from connections to holonomies leads to the
concept of {\em generalized connections}. A generalized connection
is an assignment of $h_{e}\in SU(2)$ to any path $e\subset \Sigma$.
In other words the fundamental observable is taken to be the
holonomy itself and not its relationship (\ref{holy}) to a smooth
connection. The algebra of kinematical observables is defined to
be the algebra of the so-called {\em cylindrical functions} of
generalized connections denoted ${\rm Cyl}$. The latter algebra
can be written as the union of the set of functions of generalized
connections defined on graphs $\gamma \subset \Sigma$, namely \be
\label{cyl} {\rm Cyl}=\cup_{\gamma} {\rm Cyl}_{\gamma},\end{equation} where
${\rm Cyl}_{\gamma}$ is defined as follows.

A graph ${\gamma}$ is defined as a collection of paths $e \subset
\Sigma$ ($e$ stands for {\em edge}) meeting at most at their
endpoints. Given a graph $\gamma\subset\Sigma$ we denote by $N_e$
the number of paths or edges that it contains. An element
$\psi_{\gamma,f} \in {\rm Cyl}_{\gamma}$ is labelled by a graph
$\gamma$ and a smooth function $f: SU(2)^{N_e}\rightarrow \C$, and
it is given by a functional of the connection defined as \be
\label{72}\psi_{\gamma,f}[A]:=f(h_{e_1}[A],h_{e_2}[A],\cdots
h_{e_{N_e}}[A]),\end{equation} where $e_i$ for $i=1,\cdots N_e$ are the edges
of the corresponding graph $\gamma$. The symbol $\cup_{\gamma}$ in
(\ref{cyl}) denotes the union of ${\rm Cyl}_{\gamma}$ for all
graphs in $\Sigma$. This is the algebra of basic observables upon
which we will base the definition of the kinematical Hilbert space
$\Hk$.

\begin{figure}[h!!!!!]
 \centerline{\hspace{0.5cm}\(
\begin{array}{ccc}
\includegraphics[height=1.5cm]{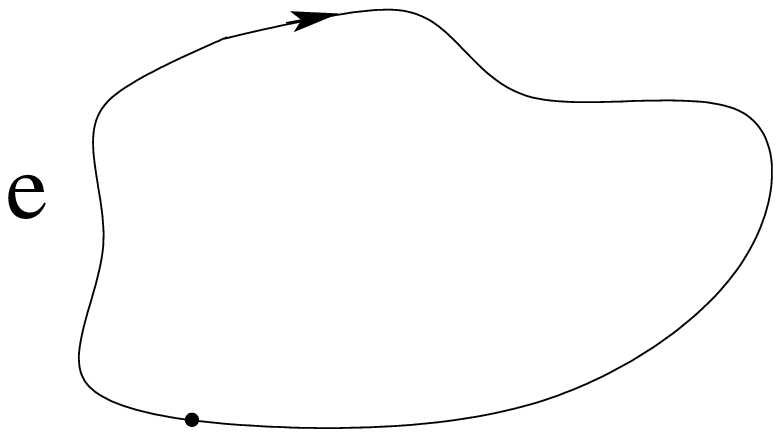}
\end{array}\ \ \ \ \ \ \ \ \
\begin{array}{ccc}
\includegraphics[height=1.5cm]{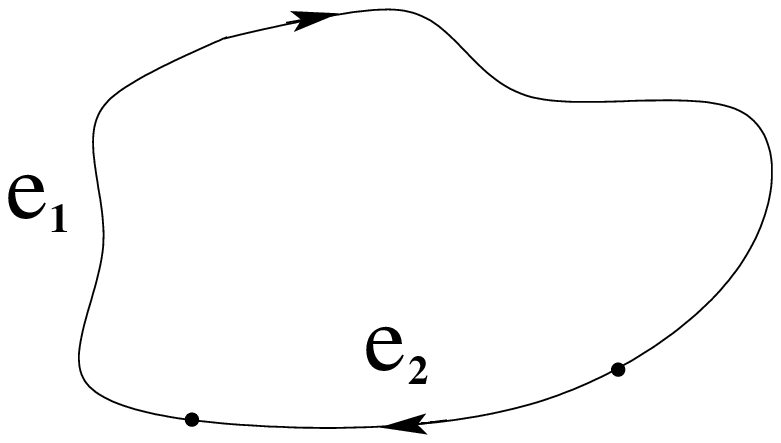}
\end{array} \ \ \ \ \ \ \ \ \
\begin{array}{ccc}
\includegraphics[height=2.3cm]{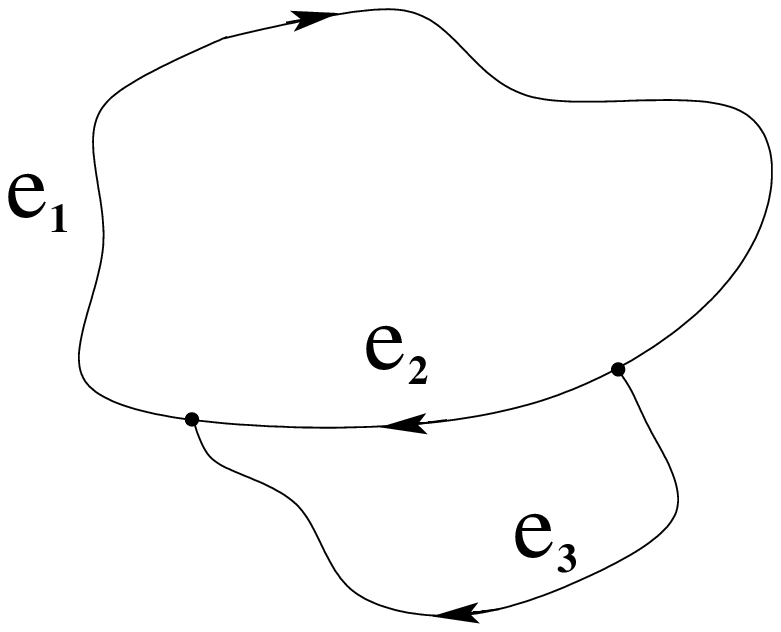}
\end{array}
\)} \caption{An example of three different graphs on which the
Wilson loop function (\ref{wl}) can be defined. The distinction
must be physically irrelevant.} \label{figloop}
\end{figure}

Before going into the construction of the representation of $Cyl$
that defines $\Hk$, it might be useful to give a few examples of
cylindrical functions. For obvious reason, $SU(2)$ gauge invariant
functions of the connection will be of particular interest in the
sequel. The simplest of such functions is the Wilson loop: given a
 closed loop $\gamma$ the Wilson loop is given by the trace of the
holonomy around the loop, namely \be\label{wl} W_{\gamma}[A]:={\rm
Tr}[h_{\gamma}[A]].\end{equation} Equation (\ref{gghol}) and the invariance
of the trace implies the $W_{\gamma}[A]$ is gauge invariant. The
Wilson loop $W_{\gamma}[A]$ is an element of ${\rm
Cyl}_{\gamma}\subset {\rm Cyl}$ according to the previous
definition. The graph consists of a single closed edge
($e=\gamma$) and an example is shown in Figure \ref{figloop}.
Notice, however, that we can also define $W_{\gamma}[A]$ as \be
W_{\gamma}[A]:= {\rm Tr}[h_{e_1}[A]h_{e_2}[A]], \end{equation} using
(\ref{groupoid}) in which case $W_{\gamma}[A] \in {\rm
Cyl}_{\gamma^{\prime}}$ ($\gamma^{\prime}$ is illustrated in the
center of Figure \ref{figloop}). Moreover, we can also think of
$W_{\gamma}[A] \in {\rm Cyl}_{\gamma^{\prime\prime}}$
($\gamma^{\prime\prime}$ is shown on the right of Figure
\ref{figloop}) as a function of $h_{e_1}[A]$ and $h_{e_2}[A]$ and
$h_{e_3}[A]$ (with trivial dependence on the third argument).
There are many ways to represent an element of ${\rm Cyl}$ as a
cylindrical function on a graph.  This flexibility in choosing the
graph will be important in the definition of $\Hk$ and its inner
product in the following section.

\begin{figure}[h!!!!!]
\centerline{\hspace{0.5cm} \(
\begin{array}{c}
\includegraphics[height=1.5cm]{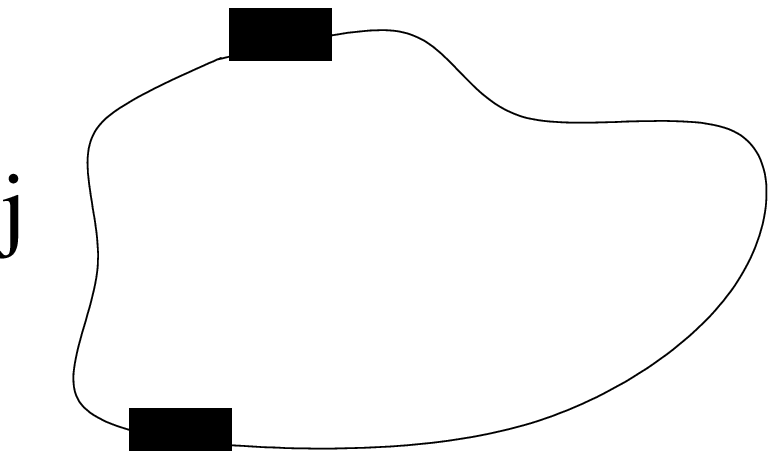}
\end{array}\ \ \ \ \ \ \ \ \ \begin{array}{c}
\includegraphics[height=2.2cm]{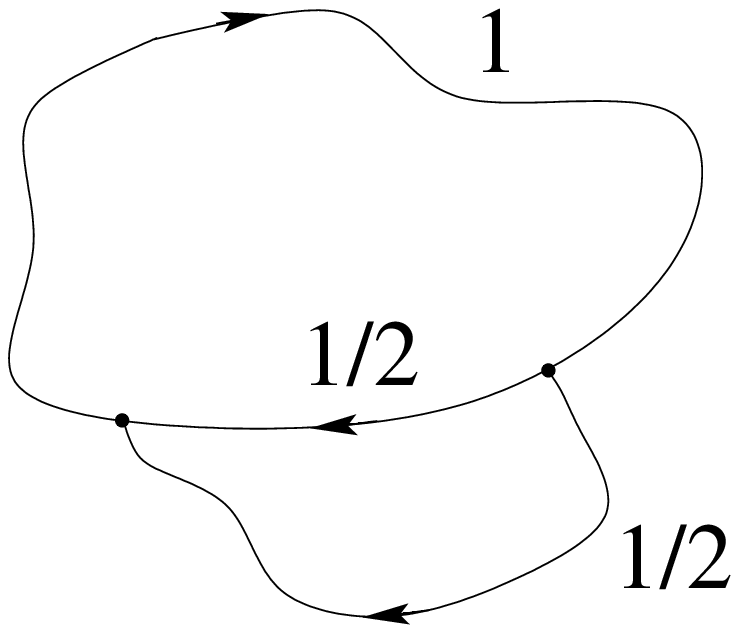}
\end{array}\ \ \ \ \ \ \ \ \ \begin{array}{c}
\includegraphics[height=2.2cm]{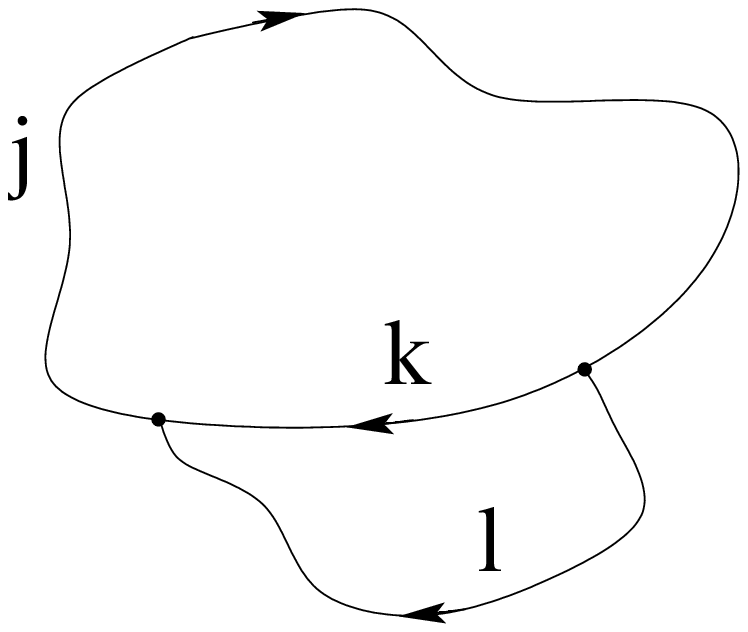}
\end{array}
\) } \caption{Examples of spin networks.}\label{spiny}
\end{figure}

Let us come back to our examples. There is a simple generalization
of the previous gauge invariant function. Given an arbitrary
representation matrix $M$ of $SU(2)$, then clearly
$W^{M}_{\gamma}[A]={\rm Tr}[M(h_{e}[A])]$ is a gauge invariant
cylindrical function. Unitary irreducible representation matrices
of spin $j$ will be denoted by $\stackrel{j}{\Pi}_{mm^{\prime}}$
for $-j \le m,m^{\prime}\le j$. The cylindrical function \be
W^{j}_{\gamma}[A]:= {\rm Tr}[\stackrel{j}{\Pi}(h_{e}[A])]\end{equation} is
the simplest example of {\em spin network} 
function\cite{pen,reis8,c4,baez10}. This function is represented as on the left
of Figure \ref{spiny}.

A more sophisticated spin network function can be associated to
the graph on the right of Figure \ref{figloop}. We take different
representation matrices of spins $1$, $1/2$ and $1/2$ evaluate
them on the holonomy along $e_1$, $e_2$, and $e_3$ respectively.
We define \be \Theta^{\va 1,1/2,1/2}_{\va e_1\cup e_2 \cup
e_3}[A]=\stackrel{1}{\Pi}(h_{e_1}[A])^{ij}
\stackrel{1/2}{\Pi}(h_{e_2}[A])_{AB}
\stackrel{1/2}{\Pi}(h_{e_3}[A])_{CD} \
\sigma_{i}^{AC}\sigma_{j}^{BD},\end{equation} where $i,j=1,2,3$ are vector
indices, $A,B,C,D=1,2$ are spinor indices, sum over repeated
indices is understood and $\sigma_{i}^{AC}$ are Pauli matrices. It
is easy to check that $\Theta^{\va 1,1/2,1/2}_{e_1\cup e_2 \cup
e_3}[A]$ is gauge invariant. This is because the Pauli matrices
are invariant tensors in the tensor product of representations
$1\otimes1/2\otimes1/2$ which is where gauge transformations act
on the nodes of the graph $e_1\cup e_2 \cup e_3$. Such spin
network function is illustrated on the  middle of Figure
\ref{spiny}. We can generalize this to arbitrary representations.
Given an invariant tensor $\iota\in j\otimes k\otimes
l$ the cylindrical function \ba \nonumber\!\!\!\!\!\Theta^{\va
j,k,l}_{\va e_1\cup e_2 \cup
e_3}[A]&=&\\
&& \!\!\!\!\!\!\!\!\!\stackrel{j}{\Pi}(h_{e_1}[A])_{m_1n_1}
\stackrel{k}{\Pi}(h_{e_2}[A])_{m_2n_2}
\stackrel{l}{\Pi}(h_{e_3}[A])_{m_3n_3} \
\iota^{m_1m_2m_3}\iota^{n_1n_2n_3}\ea is
gauge invariant by construction. The example is shown on
the right of Figure \ref{spiny}.

\begin{figure}[h!!!!!!!!!!!!!!!!!!!!!!!!!!!!!!!!!!!!!!!]
 \centerline{\hspace{0.5cm}\(
\begin{array}{ccc}
\includegraphics[height=3cm]{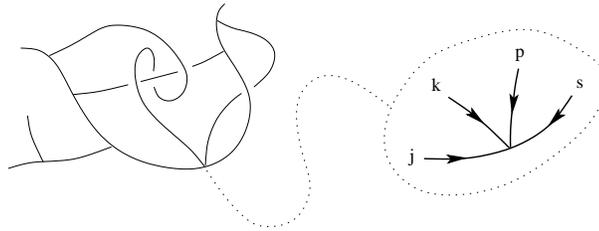}
\end{array}
\)} \caption{Schematic representation of the construction of a
spin network. To each node we associate an invariant vector in the
tensor product of irreducible representations converging at the
node. In this case we take $\iota^{n_1n_2n_3n_4}\in j \otimes k
\otimes p \otimes s$, and the relevant piece of spin network function
is $ \stackrel{\va j}{\Pi}(h_{e_1}[A])_{m_1n_1} \stackrel{\va
k}{\Pi}(h_{e_2}[A])_{\va m_2n_2} \stackrel{\va
p}{\Pi}(h_{e_3}[A])_{\va m_3n_3} \stackrel{\va
s}{\Pi}(h_{e_4}[A])_{\va m_4n_4}\iota^{\va n_1n_2n_3n_4}$.}
\label{inter}
\end{figure}
\begin{figure}[h!!!!!]
 \centerline{\hspace{0.5cm}\(
\begin{array}{ccc}
\includegraphics[height=3cm]{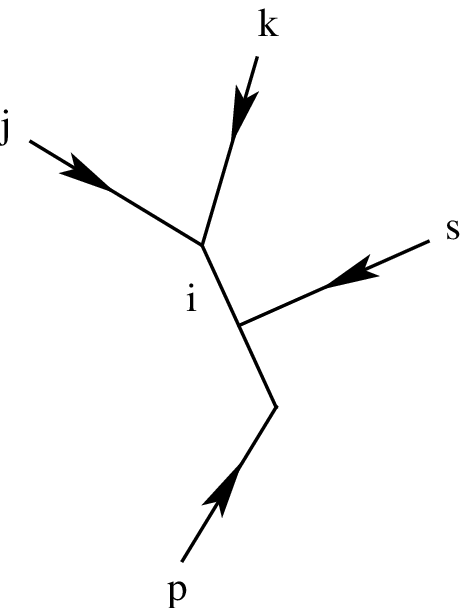}
\end{array}\ \ \ \ \ \ \ \ \
\begin{array}{ccc}
\includegraphics[height=3cm]{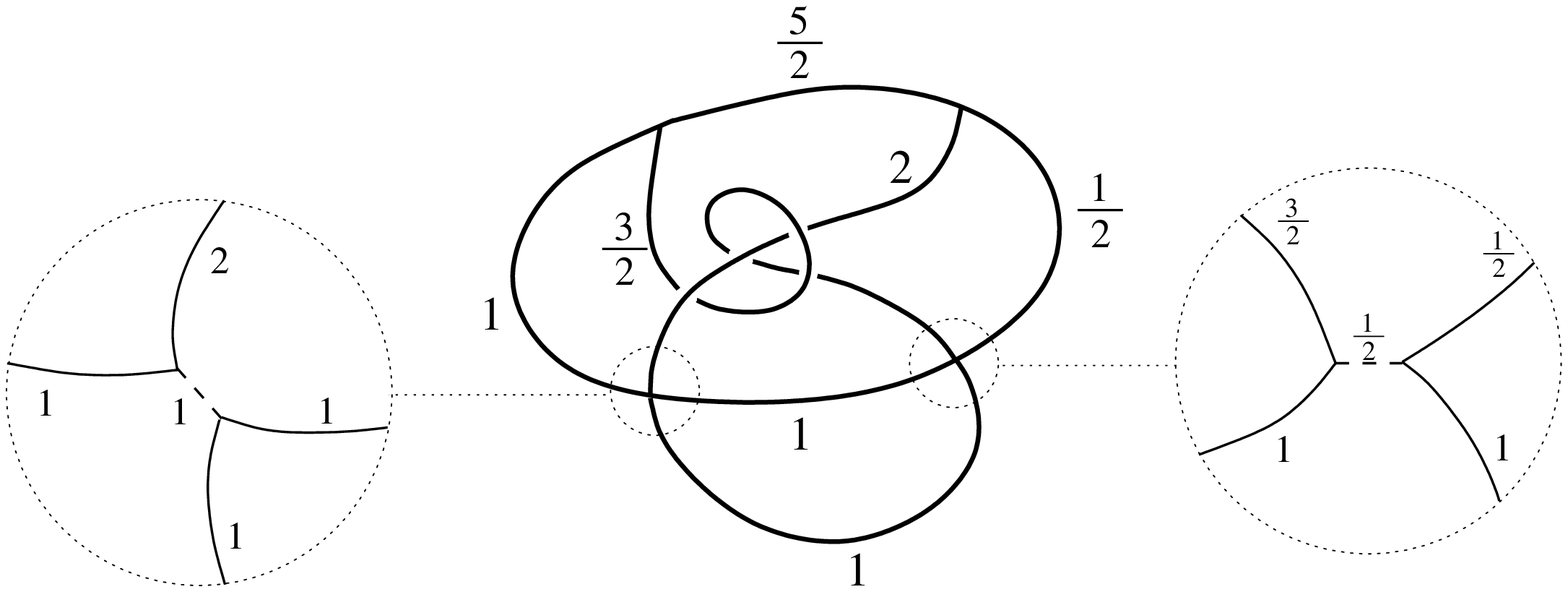}
\end{array}
\)} \caption{On the left: any invariant vector can be decomposed
in terms of the (unique up to normalization) three valent ones. At
each three node the standard rules of addition of angular momentum
must be satisfied for a non vanishing intertwiner to exist. On the
right: an example of spin network with the explicit decomposition
of intertwiners.} \label{snb}
\end{figure}

One can generalize the construction of these examples to the
definition of {\em spin networks} on arbitrary graphs $\gamma
\subset \Sigma$. The general construction is analogous to the one
in the previous examples. One labels the set of edges
$e\subset\gamma$ with spins $\{j_e\}$. To each node $n\subset
\gamma$ one assigns an invariant tensor, also called an {\em intertwiner},
$\iota_n$ in the tensor product of representations labelling the
edges converging at the corresponding node (see Figure
\ref{inter}). The spin network function is defined
\begin{equation}\label{sdef} s_{\va \gamma,
\{j_{e}\},\{\iota_{n}\}}[A]=\bigotimes_{n \subset \gamma} \
\iota_n\ \bigotimes_{e \subset \gamma}\
\stackrel{j_{e}}{\Pi}(h_{\va e}[A]) \;,
\end{equation}
where the indices of representation matrices and invariant tensors
is left implicit in order to simplify the notation. An example is
shown in Figures \ref{inter} and \ref{snb}.

Intertwiners in the tensor product of an arbitrary number of
irreducible representations can be expressed in terms of basic
intertwiners between three irreducible representations. In the
case of $SU(2)$ the latter are uniquely defined up to a
normalization; they are simply related to Clebsh-Gordon
coefficients---${\rm Inv}[j_1\otimes j_2 \otimes j_3]$ is either
trivial or one dimensional according to the standard rules for
addition of angular momentum. The construction is illustrated on
the left of Figure \ref{snb}, on the right we show an explicit
example of spin network with the nodes decomposed in terms of
three valent intertwiners..

So far we have introduced the algebra of functionals of
(generalized) connections $\rm Cyl$.  Spin networks where
presented here as special examples of elements of $\rm Cyl$, which
in addition are $SU(2)$ gauge invariant.  In the Subsection
\ref{GAUSS} we will show how spin network functions define a
complete basis of $\Hk$. But in order do that we must define
$\Hk$.

\subsubsection{The Ashtekar-Lewandowski representation of $\rm
Cyl$}\label{alr}

The cylindrical functions introduced in the previous section are
the candidates for states in $\Hk$. In this section we define
$\Hk$ and provide a representation of the algebra $\rm Cyl$.
Essentially what we need is the notion of a measure in the space of
generalized connections in order to give a meaning to the formal
expression (\ref{alm}) and thus obtain a definition of the kinematical
inner product. In order to do that we introduce a positive
normalized state (state in the algebraic QFT sense) $\mu_{AL}$ on
the ($C^{\star}$-algebra) $\rm Cyl$ as follows. Given a
cylindrical function $\psi_{\gamma,f}[A]\in {\rm Cyl}$ (as in
(\ref{72})) $\mu_{AL}(\psi_{\gamma,f})$ is defined as \be{
\mu_{AL}(\psi_{\gamma,f})=\int \prod\limits_{e\subset\gamma} dh_e\
f(h_{e_1},h_{e_2},\cdots h_{e_{N_e}}),} \end{equation} where $h_e\in SU(2)$
and $dh$ is the (normalized) Haar measure of $SU(2)$ \footnote{The
Haar measure of $SU(2)$ is defined by the following properties:
\[\int_{SU(2)} dg=1, \ \ \ {\rm and } \ \ \ dg=d(\alpha g)=d(g \alpha)=dg^{-1}\ \ \
\forall \alpha\in SU(2). \]}. The measure $\mu_{AL}$ is clearly
normalized as $\mu_{AL}(1)=1$ and positive \be
\label{+}\mu_{AL}(\overline{\psi_{\gamma,f}}\psi_{\gamma,f})=\int
\prod\limits_{e\subset\gamma} dh_e\
\overline{f(h_{e_1},h_{e_2},\cdots
h_{e_{N_e}})}f(h_{e_1},h_{e_2},\cdots h_{e_{N_e}})\ge 0 .\end{equation} Using
the properties of $\mu_{AL}$ we introduce the inner product
\ba\label{kip} \nonumber
<\psi_{\gamma,f},\psi_{\gamma^{\prime},g}>:&=&
\mu_{AL}(\overline{\psi_{\gamma,f}}\psi_{\gamma^{\prime},g})=\\
&=&\int \prod\limits_{e\subset \Gamma_{\gamma\gamma^{\prime}}}
dh_e\ \overline{f(h_{e_1},\cdots h_{e_{N_e}})} g(h_{e_1},\cdots
h_{e_{N_e}}),\ea where we use Dirac notation and the cylindrical
functions become wave functionals of the connection corresponding
to kinematical states
$\psi_{\gamma^{\prime},g}[A]=<A,\psi_{\gamma^{\prime},g}>=g(h_{e_1},\cdots
h_{e_{N_e}})$, and $\Gamma_{\gamma\gamma^{\prime}}$ is any graph
such that both $\gamma \subset \Gamma_{\gamma\gamma^{\prime}}$ and
$\gamma^{\prime} \subset \Gamma_{\gamma\gamma^{\prime}}$. The
state $\mu_{AL}$ is called the Ashtekar-Lewandowski
measure\cite{ash3}. The previous equation is the rigorous
definition of (\ref{alm}). The measure $\mu_{AL}$---through the
GNS construction\cite{Haag:1992hx}---gives a faithful
representation of the algebra of cylindrical functions (i.e.,
(\ref{+}) is zero if and only if $\psi_{\gamma,f}[A]=0$). The
kinematical Hilbert space $\Hk$ is the Cauchy completion of the
space of cylindrical functions $\rm Cyl$ in the
Ashtekar-Lewandowski measure. In other words, in addition to
cylindrical functions we add to $\Hk$ the limits of all the Cauchy
convergent sequences in the $\mu_{AL}$ norm. The operators
depending only on the connection act simply by multiplication in
the Ashtekar-Lewandowski representation. This completes the
definition of the kinematical Hilbert space $\Hk$.

\subsubsection{An orthonormal basis of $\Hk$.}

In this section we would like to introduce a very simple basis of
$\Hk$ as a preliminary step in the construction of a basis of the
Hilbert space of solutions of the Gauss constraint $\Hg$ At this
stage this is a simple consequence of the Peter-Weyl 
theorem\cite{Fuchs:1997jv}. The Peter-Weyl theorem can be viewed as a
generalization of Fourier theorem for functions on $S^1$. It
states that, given a function $f\in {\cal L}^2[SU(2)]$, it can be
 expressed as a sum over unitary irreducible representations of
$SU(2)$, namely
\footnote{\label{haary}The link with the $U(1)$ case is direct:
for $f\in {\cal L}^2[U(1)]$ we have $f(\theta)=\sum_n f_n \ {\rm
exp}(i n\theta)$, where ${\rm exp}(i n\theta)$ are unitary
irreducible representations of $U(1)$ and $f_n=(2\pi)^{-1}\int\
d\theta {\rm exp}(-i n\theta) f(\theta)$. The measure
$(2\pi)^{-1}d\theta$ is the Haar measure of $U(1)$.}
\be f(g)=\sum_{j} \sqrt{2j+1}\ f_j^{mm^{\prime}}
\stackrel{j}{\Pi}_{mm^{\prime}}(g), \end{equation} where \be\label{compo}
f_j^{mm^{\prime}}=\sqrt{2j+1}\int \limits_{SU(2)} \!\!dg\ \
{\stackrel{j}{\Pi}_{m^{\prime}m}}(g^{-1}) f(g), \end{equation} and $dg$ is
the Haar measure of $SU(2)$. This defines the harmonic analysis on
$SU(2)$. The completeness relation 
\be\label{deltadef}
\delta(g  h^{-1})=\sum_j (2j+1)
\stackrel{j}{\Pi}_{mm^{\prime}}(g)\stackrel{j}{\Pi}_{m^{\prime}m}(  h^{-1})=
\sum_j (2j+1) {\rm Tr}[\stackrel{j}{\Pi}(g  h^{-1})],
\end{equation} 
follows.
The previous equations imply the orthogonality relation for
unitary representations of $SU(2)$ \be \int \limits_{SU(2)}
\!\!dg\ \phi^j_{m^{\prime}m} \phi^{j^{\prime}}_{q^{\prime}q}=
\delta_{jj^{\prime}} \delta_{mq}\delta_{m^{\prime}q^{\prime}}, \end{equation}
where we have introduce the normalized representation matrices
${\phi}^{j}_{mn}:=\sqrt{2j+1}\stackrel{j}{\Pi}_{mn}$ for
convenience. Given an arbitrary cylindrical function
$\psi_{\gamma, f}[A]\in {\rm Cyl}$ we can use the Peter-Weyl
theorem and write \ba\label{expy} \nonumber
\psi_{\gamma,f}[A]&=&f(h_{e_1}[A],h_{e_2}[A],\cdots
\!h_{e_{N_e}}[A])=\\ &=& \!\!\!\!\sum_{j_1\cdots j_{N_e}}
f^{m_1\cdots m_{N_e}, n_1 \cdots n_{N_e}}_{j_1\cdots j_{N_e}}
{\phi}^{j_1}_{m_1n_1}\!(h_{e_1}[A])\cdots
{\phi}^{j_{N_e}}_{m_{N_e}n_{N_e}}\!(h_{e_{N_e}}[A]), \ea where
according to (\ref{compo}) $f^{m_1\cdots m_{N_e}, n_1 \cdots
n_{N_e}}_{j_1\cdots j_{N_e}}$ is just given by the kinematical
inner product of the cylindrical function with the tensor product
of irreducible representations, namely (\ref{kip}) \be
f^{m_1\cdots m_{N_e}, n_1 \cdots n_{N_e}}_{j_1\cdots
j_{N_e}}=<{\phi}^{j_1}_{m_1n_1}\cdots
{\phi}^{j_{N_e}}_{m_{N_e}n_{N_e}},\psi_{\gamma, f}>,\end{equation} where
$<,>$ is the kinematical inner product introduced in (\ref{kip}).
We have thus proved that the product of components of (normalized)
irreducible representations $\prod_{i=1}^{N_e}
{\phi}^{j_i}_{m_in_i}[h_{e_i}]$ associated with the $N_e$ edges
$e\subset \gamma$ (for all values of the spins $j$ and $-j\le
m,n\le j$ and for any graph $\gamma$) is a complete orthonormal
basis of $\Hk$!

\subsubsection{Solutions of the Gauss constraint: $\Hg$ and spin network
states}\label{GAUSS}

We are now interested in the solutions of the quantum Gauss
constraint; the first three of quantum Einstein's equations. These
solutions are characterized by the states in $\Hk$ that are
$SU(2)$ gauge invariant. These solutions define a new Hilbert
space that we call $\Hg$. We leave the subindex $kin$ to keep in
mind that there are still constraints to be solved on the way to 
$\Hp$. In previous sections we already introduced spin network
states as natural $SU(2)$ gauge invariant functionals of the
connection\cite{reis8,c4,baez10,lee2}. Now we will show how
these are in fact a complete set of orthogonal solutions of the
Gauss constraint, i.e., a basis of $\Hg$.

The action of the Gauss constraint is easily represented in $\Hk$.
At this stage it is simpler to represent finite $SU(2)$
transformations on elements of $\Hk$ (from which the infinitesimal
ones can be easily inferred) using (\ref{gghol}). Denoting ${\cal
U_{G}}[g]$ the operator generating a local $g(x)\in SU(2)$
transformation then its action can be defined directly on the
elements of the basis of $\Hk$ defined above, thus
\be\label{ggghol} {\cal
U_{G}}[g]{\phi}^{j}_{mn}[h_e]={\phi}^{j}_{mn}[g_s h_e g_t^{-1}],
\end{equation} where $g_s$ is the value of $g(x)$ at the source point of the
edge $e$ and $g_t$ the value of $g(x)$ at the target.
From the previous equation one can infer the action on an
arbitrary basis element, namely \be{\cal
U_{G}}[g]\prod_{i=1}^{N_e}
{\phi}^{j_i}_{m_in_i}[h_{e_i}]=\prod_{i=1}^{N_e}
{\phi}^{j_i}_{m_in_i}[g_{s_i} h_{e_i} g_{t_i}^{-1}].\end{equation} Now by
definition of the scalar product (\ref{kip}) and due to the
invariance of the Haar measure (see Footnote \ref{haary}) the
reader can easily prove that ${\cal U_{G}}[g]$ is a unitary
operator. From the definition it also follows that \be {\cal
U_{G}}[g_2]{\cal U_{G}}[g_1]={\cal U_{G}}[g_1g_2]. \end{equation} The
projection operator onto the set of states that are solutions of
the Gauss constraint can be obtained by group averaging
techniques. We can denote the projector $P_{\va \cal G}$ by \be
\label{pipin}P_{\va \cal G}=\int \limits D[g] \ {\cal
U_{G}}[g],\end{equation} where the previous expression denotes a formal
integration over all $SU(2)$ transformations. Its rigorous
definition is given by its action on elements of $\rm Cyl$. From
equation (\ref{ggghol}) the operator ${\cal U_{G}}[g]$ acts on
$\psi_{\gamma,g}\in {\rm Cyl}$ at the end points of the edges $e
\subset \gamma$, and therefore, so does $P_{\va \cal G}$. The
action of $P_{\va \cal G}$ on a given (cylindrical) state
$\psi_{\gamma,f}\in \Hk$ can therefore be factorized as follows:
\be \label{facy}P_{\va \cal G} \psi_{\gamma,f}= \prod_{n\subset
\gamma} P^{n}_{\va \cal G} \ \psi_{\gamma,f},\end{equation} where $P_{\cal
G}^n$ acts non trivially only at the node $n\subset \gamma$. In
this way we can define the action of $P_{\va \cal G}$ by focusing
our attention to a single node $n\subset \gamma$. For concreteness
let us concentrate on the action of $P_{\va \cal G}$ on an element
of $\psi_{\gamma,f}\in\Hk$ defined on the graph illustrated in
Figure \ref{inter}. The state $\psi_{\gamma,f}\in\Hk$ admits an
expansion in terms of the basis states as in (\ref{expy}). In
particular we concentrate on the action of $P_{\va \cal G}$ at the
four valent node thereto emphasized, let's call it $n_0\subset
\gamma$. In order to do that we can factor out of (\ref{expy}) the
(normalized) representation components $\phi^{j}_{mn}$
corresponding to that particular node and write \ba\label{expyn}
\nonumber \psi_{\gamma,f}[A]&=&\\ &=& \!\!\!\!\sum_{j_1\cdots
j_{4}} \left( {\phi}^{j_1}_{m_1n_1}\!(h_{e_1}[A])\cdots\!
{\phi}^{j_{4}}_{m_{4}n_{4}}\!(h_{e_{4}}[A])\right)\times
\left[{\rm REST}\right]_{j_1\cdots j_4}^{m_1\cdots m_{4}, n_1
\cdots n_{4}}[A], \ea where $\left[{\rm REST}\right]_{j_1\cdots
j_4}^{m_1\cdots m_{4}, n_1 \cdots n_{4}}[A]$ denotes what is left
in this factorization, and $e_1$ to $e_4$ are the four edges
converging at $n_0$ (see Figure \ref{inter}). We can define the
meaning of (\ref{pipin}) by giving the action of $P_{\cal
G}^{n_0}$ on ${\phi}^{j_1}_{m_1n_1}\!(h_{e_1}[A])\cdots\!
{\phi}^{j_{4}}_{m_{4}n_{4}}\!(h_{e_{4}}[A])$ as the action on a
general state can be naturally extended from there using
(\ref{facy}). Thus we define \be P^{n_0}_{\va \cal
G}{\phi}^{j_1}_{m_1n_1}\!(h_{e_1}[A])\cdots\!
{\phi}^{j_{4}}_{m_{4}n_{4}}\!(h_{e_{4}}[A])=\int dg
{\phi}^{j_1}_{m_1n_1}\!(g h_{e_1}[A])\cdots\!
{\phi}^{j_{4}}_{m_{4}n_{4}}\!(g h_{e_{4}}[A]), \end{equation} where $dg$ is
the Haar measure of $SU(2)$. Using the fact that
\be{\phi}^{j_1}_{mn}\!(g h[A])=\stackrel{j}{\Pi}_{mq}(g)\
{\phi}^{j}_{qn}\!(h[A])\end{equation} the action of $P_{\cal G}^{n_0}$ can be
written as \ba \label{di} \nonumber && P^{n_0}_{\va \cal
G}{\phi}^{j_1}_{m_1n_1}\!(h_{e_1}[A])\cdots\!
{\phi}^{j_{4}}_{m_{4}n_{4}}\!(h_{e_{4}}[A])= \\
&&=P^{n_0}_{m_1\cdots m_4, q_1\cdots q_4} \
{\phi}^{j_1}_{q_1n_1}\!(h_{e_1}[A])\cdots\!
{\phi}^{j_{4}}_{q_{4}n_{4}}\!(h_{e_{4}}[A]), \ea where \be
P^{n_0}_{m_1\cdots m_4, q_1\cdots q_4}=\int dg
\stackrel{j_1}{\Pi}_{m_1q_1}(g)\cdots
\stackrel{j_4}{\Pi}_{m_4q_4}(g). \end{equation} If we denote $V_{j_1\cdots
j_4}$ the vector space where the representation
$j_1\otimes\cdots\otimes j_4$ act, then previous equation defines a
map $P^{n_0}: V_{j_1\cdots j_4}\rightarrow V_{j_1\cdots j_4}$. Using
the properties of the Haar measure given in Footnote \ref{haary} one
can show that the map $P^{n_0}$ is indeed a projection (i.e.,
$P^{n_0}P^{n_0}=P^{n_0}$). Moreover, we also have \ba \nonumber &&
P^{n_0}_{m_1\cdots m_4, q_1\cdots q_4}
\stackrel{j_1}{\Pi}_{q_1n_1}(g)\cdots
\stackrel{j_4}{\Pi}_{q_4n_4}(g)=\\&&=
\stackrel{j_1}{\Pi}_{m_1q_1}(g)\cdots
\stackrel{j_4}{\Pi}_{m_4q_4}(g)P^{n_0}_{q_1\cdots q_4, n_1\cdots
n_4}=P^{n_0}_{m_1\cdots m_4, n_1\cdots n_4}, \ea i.e. $P^{n_0}$ is
right and left invariant. This implies that $P^{n_0}:V_{j_1\cdots
j_4}\rightarrow {\rm Inv}[V_{j_1\cdots j_4}]$, i.e., the
projection from $V_{j_1\cdots j_4}$ onto the ($SU(2)$) invariant
component of the finite dimensional vector space. We can choose an
orthogonal set of invariant vectors $\iota^{\alpha}_{m_1\cdots
m_4}$ (where $\alpha$ labels the elements), in other words an
orthonormal basis for ${\rm Inv}[V_{j_1\cdots j_4}]$ and write \be
P^{n_0}_{m_1\cdots m_4, n_1\cdots n_4} = \sum\limits_{\alpha}\
\iota^{\alpha}_{m_1\cdots m_4} \iota^{\alpha *}_{m_1\cdots
m_4},\end{equation} where $*$ denotes the dual basis element. In the same way
the action $P_{\cal G}$ on a node $n \subset \gamma$ of arbitrary
valence $\kappa$ is governed by the corresponding $P^n$ given
generally by \be P^{n}_{m_1\cdots m_{\kappa}, n_1\cdots
n_{\kappa}} = \sum\limits_{\alpha_{\kappa}}\
\iota^{\alpha_{\kappa}}_{m_1\cdots m_{\kappa}}
\iota^{\alpha_{\kappa} *}_{m_1\cdots m_{\kappa}}.\end{equation} According to
the tree decomposition of intertwiners in terms  of three valent
invariant vectors described around Figure \ref{snb},
$\alpha_{\kappa}$ is a $( \kappa-3)$-uple of spins.

Any solution of the Gauss constraint can be written as $P_{\cal G}
\psi$ for $\psi \in \Hk$. Equation (\ref{facy}) plus the obvious
generalization of (\ref{di}) for arbitrary nodes implies that the
result of the action of $P_{\cal G}$ on elements of $\Hk$ can be
written as a linear combination of products of representation
matrices $\phi^j_{mn}$ contracted with intertwiners, i.e. spin
network states as introduced as examples of elements of $\rm Cyl$
in (\ref{sdef}). Spin network states therefore form a complete
basis of the Hilbert space of solutions of the quantum Gauss law
$\Hg$!

\subsection{Geometric operators: quantization of the
triad}\label{goo}

We have introduced the set of basic configuration observables as
the algebra of cylindrical functions of the generalized
connections, and have defined the kinematical Hilbert space
through the Ashtekar-Lewandowski representation. By considering
finite gauge transformations we avoided quantizing the Gauss
constraint in the previous section, avoiding for simplicity, and
for the moment, the quantization of the triad field $E^a_i$
present in (\ref{tres}). In this section we will quantize the
triad field and will define a set of geometrical operators that
lead to the main physical prediction of LQG: discreteness of
geometry eigenvalues.

The triad $E^a_i$ naturally induces a two form with values in the
Lie algebra of $SU(2)$, namely, $E^a_i\epsilon_{abc}$. In the
quantum theory $E^a_i$ becomes an operator valued distribution. In
other words we expect integrals of the triad field with suitable
test functions to be well defined self adjoint operators in $\Hk$.
The two form naturally associated to $E^a_i$ suggests that the
smearing should be defined on two dimensional surfaces: \be
\label{flux} \widehat{E}[S,\alpha]=\int\limits_S
d\sigma^1d\sigma^2\frac{\partial x^{a}}{\partial \sigma^{1}}
\frac{\partial x^{b}}{\partial \sigma^{2}} \alpha^i
\widehat{E^a_i}\epsilon_{abc}=-i\hbar \kappa \gamma\int\limits_S
d\sigma^1d\sigma^2\frac{\partial x^{a}}{\partial \sigma^{1}}
\frac{\partial x^{b}}{\partial \sigma^{2}}\alpha^i
\frac{\delta}{\delta A_c^i} \epsilon_{abc},\end{equation} where $\alpha^i$ is
a smearing function with values on the Lie algebra of $SU(2)$. The
previous expression corresponds to the natural generalization of the notion
of electric flux operator in electromagnetism. In order to study
the action of (\ref{flux}) in $\Hk$ we notice that \ba \nonumber
&& \frac{\delta}{\delta A_c^i} h_{e}[A]= \frac{\delta}{\delta
A_c^i} \left( {\rm P}\ {\rm exp}\int ds \ \dot x^d(s) A_d^k \
\tau_k \right)=\\ && \ \ \ \ \ \ \ \ \ \ \ \ = \int ds \ \dot
x^c(s) \delta^{\va (3)}(x(s)-x) h_{e_1}[A]\tau_ih_{e_2}[A],\ea
where $h_{e_1}[A]$ and $h_{e_2}[A]$ are the holonomy along the two
new edges defined by the point at which the triad acts. Therefore
\ba &&\!\!\!\!\!\!\!\!\!\!\! \nonumber \widehat{E}[S,\alpha]h_e[A]=\\
&&\!\!\!\!\!\!\!\!\!\!\!= -i8\pi\ell_p^2\gamma\int d\sigma^1
d\sigma^2d\sigma^3 \frac{\partial x^a}{\partial \sigma^1}
\frac{\partial x^b}{\partial \sigma^2} \frac{\partial
x^c}{\partial s} \epsilon_{abc} \ \delta^{\va
(3)}(x(\sigma),x(s))\alpha^i h_{e_1}[A]\tau_i h_{e_2}[A].\ea
Finally using the definition of the delta distribution we obtain a
very simple expression for the action of the flux operator on the
holonomy integrating the previous expression. In the cases of
interest the result is: \be
\label{paniu1}\widehat{E}[S,\alpha]h_e[A]=-i8\pi\ell_p^2\gamma
\alpha^i h_{e_1}[A]\tau_i h_{e_2}[A]\ \ \ \ \ \ \ \ \ \ \ \
\begin{array}{c}
\includegraphics[width=3.5cm]{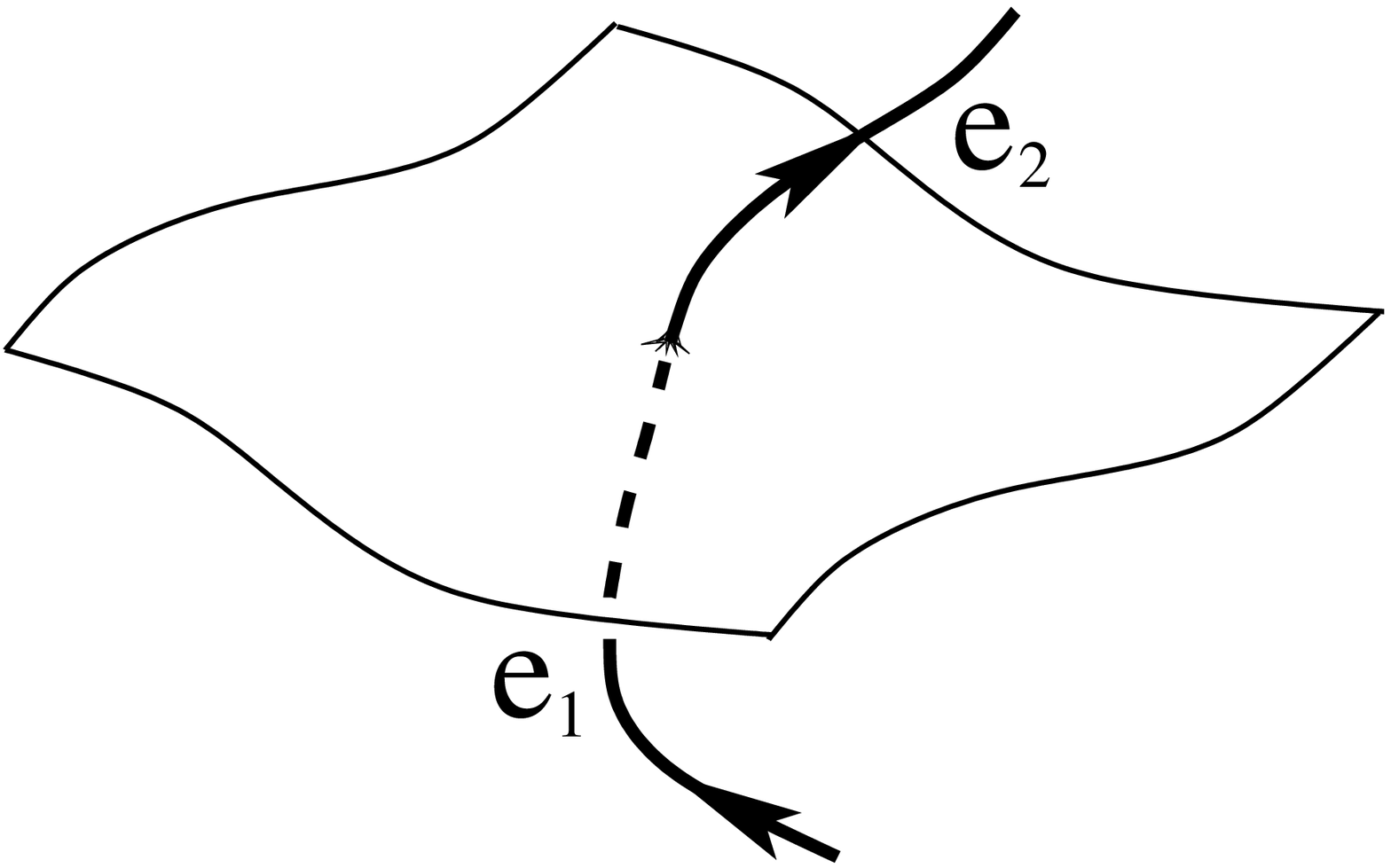}\end{array},\end{equation} and \be
\label{paniu2}\widehat{E}[S,\alpha]h_e[A]=i8\pi\ell_p^2\gamma
\alpha^i h_{e_1}[A]\tau_i h_{e_2}[A]\ \ \ \ \ \ \ \ \ \ \ \
\begin{array}{c}
\includegraphics[width=3.5cm]{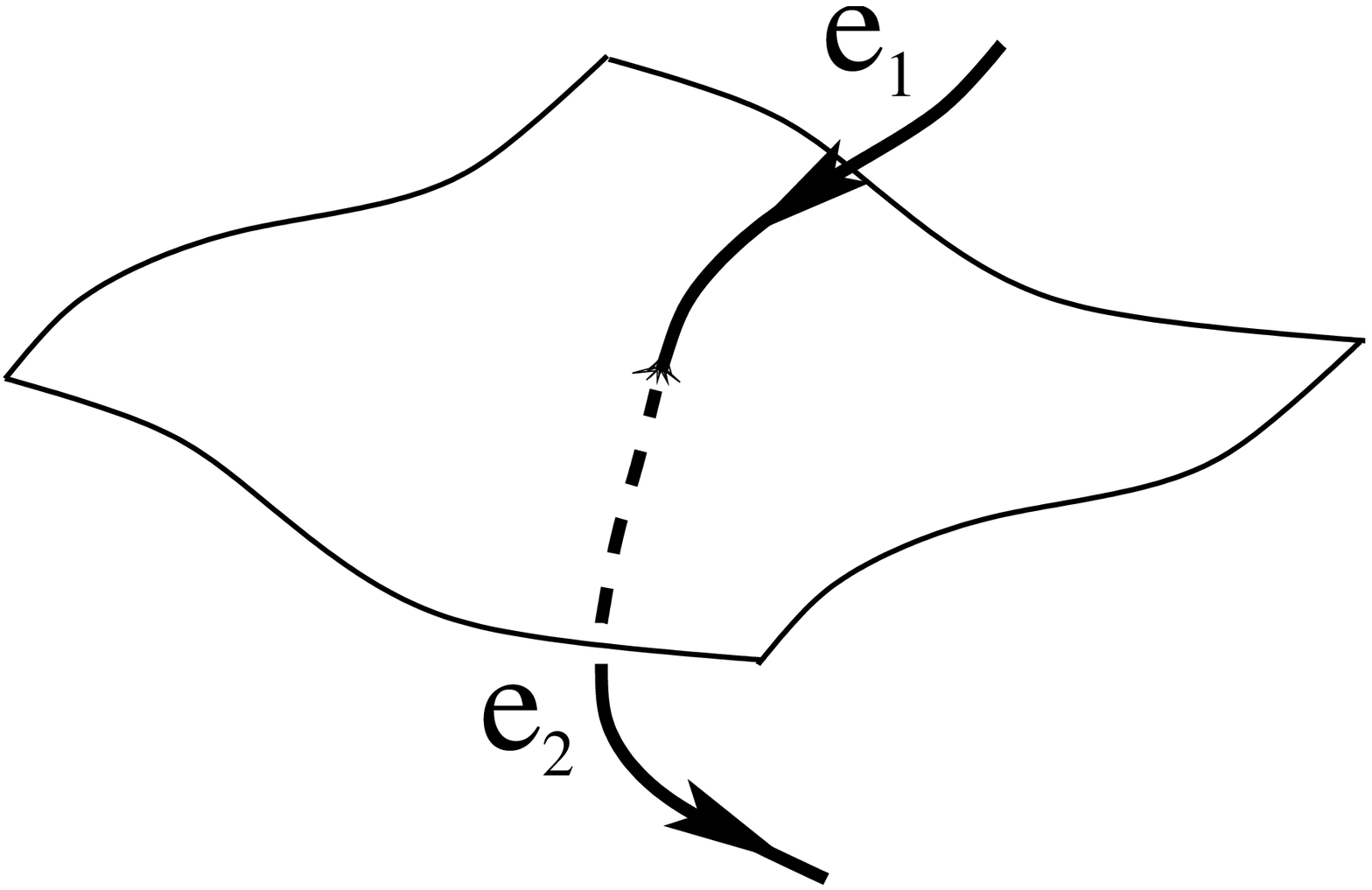}\end{array},\end{equation} and
$\widehat{E}[S,\alpha]h_{e_1}[A]=0$ when $e$ is tangential to $S$
or $e\cap S=0$.

From its action on the holonomy, and using basic $SU(2)$ representation theory, one can easily
obtain the action of $\widehat{E}[S,\alpha]$ to spin network
states and therefore to any state in $\Hk$. Using (\ref{kip}) one
can also verify that $\widehat{E}[S,\alpha]$ is self-adjoint. The
reader can also verify that it is $SU(2)$ gauge covariant. The
operators $\widehat{E}[S,\alpha]$ for all surfaces $S$ and all
smearing functions $\alpha$ contain all the information of the
quantum Riemannian geometry of $\Sigma$. In terms of the operators  
$\widehat{E}[S,\alpha]$ we can construct any geometric operator.

\subsubsection{Quantization of the Gauss constraint}

\begin{figure}[h!!!!!]
 \centerline{\hspace{0.5cm}\(
\begin{array}{ccc}
\includegraphics[height=4cm]{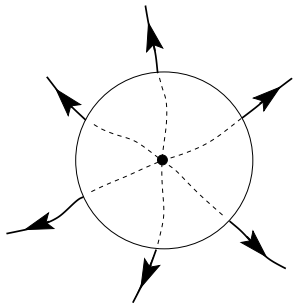}
\end{array}
\)}
\caption{The Gauss constraint imposes the net flux of non Abelian electric field be zero around nodes.} \label{fluxy}
\end{figure}

We have already imposed the Gauss constraint in Section
\ref{GAUSS} by direct construction of the Hilbert space of $SU(2)$
gauge invariant states $\Hg$ with a natural complete basis given
by the spin network states. Here we show that an important
identity for flux operators follows from gauge invariance. Given a
spin network node $n$ where $N_n$ edges converge, take a sphere
$S$ of radius $\epsilon$ (defined in some local coordinates)
centered at the corresponding node (see Figure \ref{fluxy}).  The
identity follows from the fact that for a gauge invariant node
\be\label{gaussy} \lim_{\epsilon\rightarrow 0}\widehat
E_S(\alpha)|\psi>=\sum \limits_{i=1}^{N_n} \widehat
E_{e_i}(\alpha)|\psi>=0, \end{equation} where $N_n$ is the number of edges at
the node, and $E_{e_i}(\alpha)$ is the flux operator through a
piece of the sphere that is punctured by only the edge $e_i$.
If we partition the sphere in pieces that are punctured by only
one edge, using (\ref{paniu1}) at each edge one notices that
(\ref{gaussy}) produces the first order term in an infinitesimal
gauge transformation $g_{\alpha}=\mathbbm{1}-\alpha^i\tau_i\in
SU(2)$ at the corresponding node: the operator acting in
(\ref{gaussy}) is indeed the quantum Gauss constraint action on
the given node!  Because the node is gauge invariant the action of
the such operator vanishes identically. The total quantum flux of
non-Abelian electric field must vanish according the Gauss
constraint.

\subsubsection{Quantization of the area}

The simplest of the geometric operators is the area of a
two-dimensional surface\cite{lee1,c3,ash2} $S\subset \Sigma$ which
classically depends on the triad $E^a_i$ as in (\ref{areac}). If
we introduce a decomposition of $S$ in two-cells, we can write the
integral defining the area as the limit of a Riemann sum, namely
\be A_S= \lim_{N\rightarrow \infty} A_S^N \end{equation} where the Riemann
sum can be expressed as \be A_S^N=\sum_{I=1}^N
\sqrt{E_i(S_I)E^i(S_I)} \end{equation} where $N$ is the number of cells, and
$E_i(S_I)$ corresponds to the flux of $E^a_i$  through the $I$-th
cell. The reader is invited to check that the previous limit does
in fact define the area of $S$ in classical geometry. The previous
expression for the area sets the path to the definition of the
corresponding quantum operator as it is written in terms of the
flux operators that we defined in the previous section. The
quantum area operator then simply becomes \be\label{are}
\widehat{A}_S= \lim_{N\rightarrow \infty} \widehat{A}_S^N,  \end{equation}
where we simply replace the classical $E_i(S_I)$ by $\widehat
E_i(S_I)$ according to (\ref{flux}). The important action to study
is that of $\widehat E_i(S_I)\widehat E^i(S_I)$ which on the
holonomy along a path that crosses $S_I$ only once is\ba \widehat
E_i(S_I) \widehat E^i(S_I) h_e[A]=i^2(8\pi\ell_p^2\gamma)^2
h_{e_1}[A]\tau_i\tau^i h_{e_2}[A]=(8\pi\ell_p^2\gamma)^2 (3/4)
h_{e}[A],\ea where we have used that $\tau_i=i\sigma_i/2$. The
action of the square of the flux through the cell $S_I$ is
diagonal on such holonomy! Using the definition of the unitary
irreducible representation of $SU(2)$ it follows that \ba \widehat
E_i(S_I) \widehat E^i(S_I)
\phi^j(h_e[A])_{mn}=(8\pi\ell_p^2\gamma)^2
(j(j+1))\phi^j(h_{e}[A])_{mn} ,\ea when the edge is that of an
arbitrary spin network state. The remaining important case is when
a spin network node is on $S_I$. A careful analysis shows that the
action is still diagonal in this case\cite{ash2}. Notice that the
cellular decomposition is chosen so that in the limit
$N\rightarrow\infty$ each $S_I$ is punctured at most at a single
point by either an edge (the case studied here) or a node.
\begin{figure}[h!!!!!]
 \centerline{\hspace{0.5cm}\(
\begin{array}{ccc}
\includegraphics[width=3.5cm]{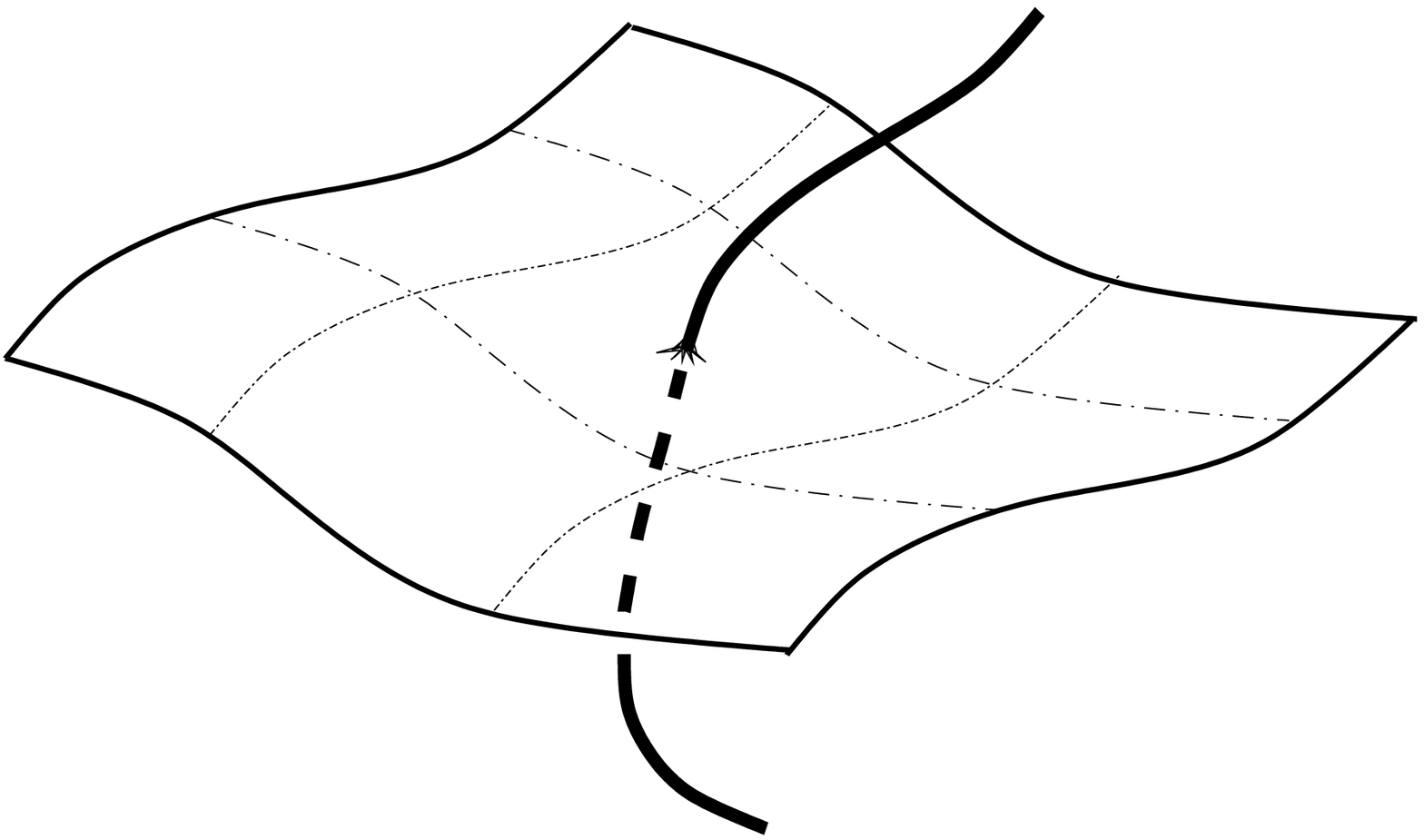}
\end{array}\ \ \ \ \ \ \ \ \
\begin{array}{ccc}
\includegraphics[width=3.5cm]{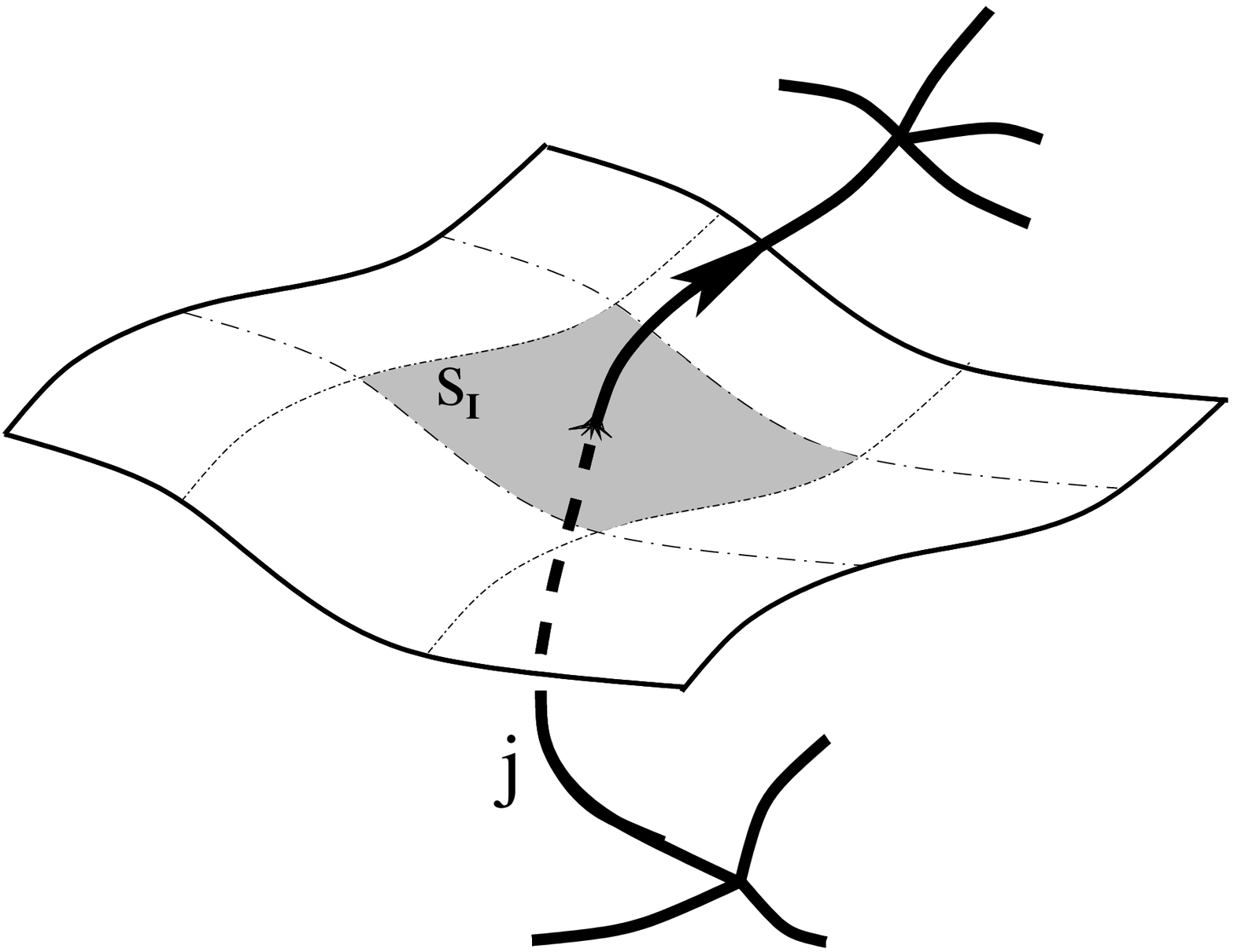}
\end{array}\ \ \ \ \ \ \ \ \
\begin{array}{ccc}
\includegraphics[width=4cm]{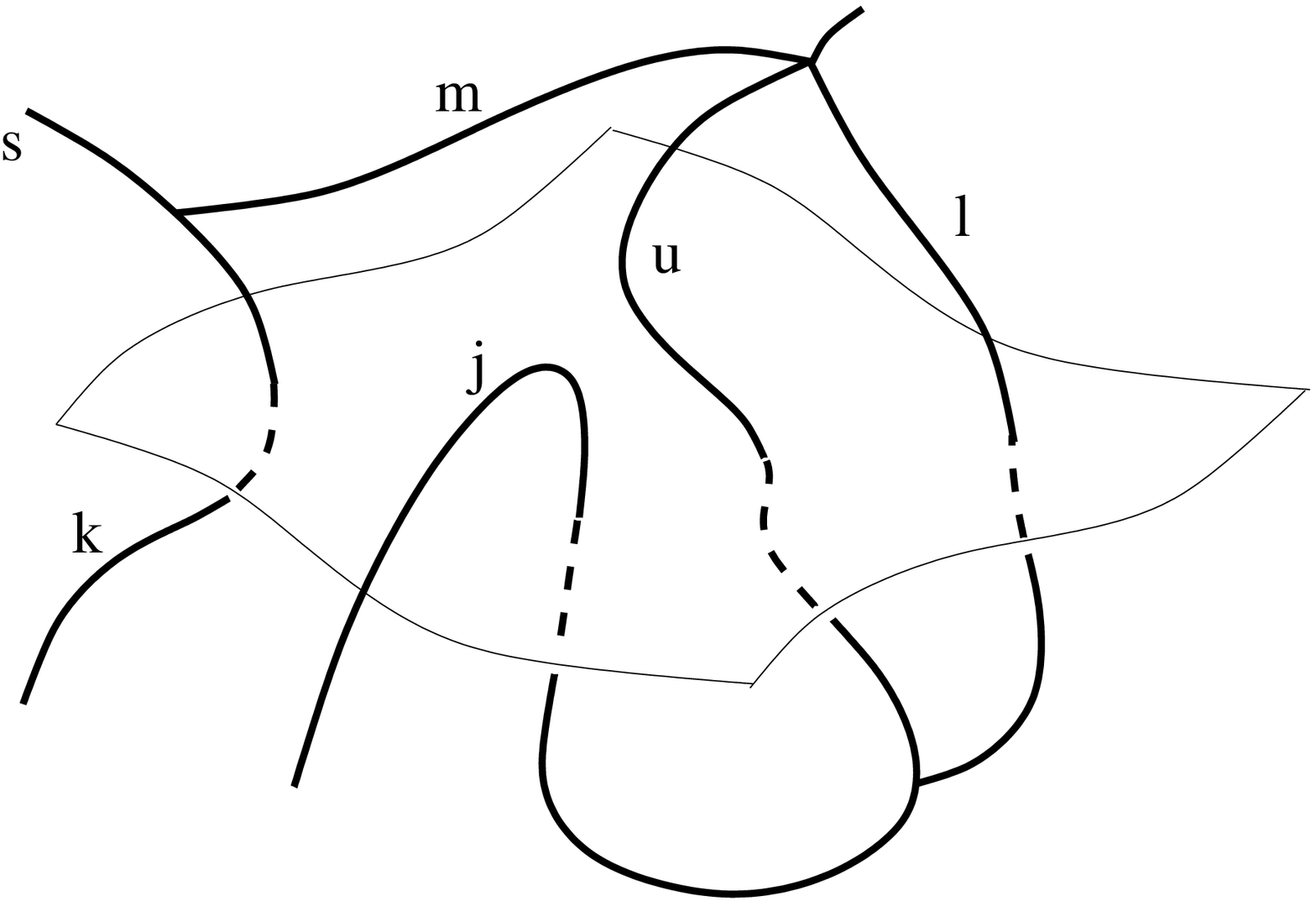}
\end{array}
\)} \caption{On the left: the regularization of (\ref{are}) is
defined so that in the limit in which the two cells are shrunk
they are  punctured by at most one edge. Center: The simplest
eigenstate of the area of $S$ is illustrated, a two cell at finite
regulator is emphasized. On the right: a generic eigenstate of the
area.} \label{supi}
\end{figure}

It is then a straightforward exercise to show that the action of
the area operator is diagonalized by the spin network states. Spin
network states are the eigenstates of the quantum area operator!
We have \be \widehat A_S|\psi>= 8\pi\ell_p^2 \gamma \sqrt{j(j+1)}
|\psi> ,\end{equation} for a single puncture and more generally \be \widehat
A_S|\psi>= 8\pi\ell_p^2\gamma \sum_p \sqrt{j_p(j_p+1)} |\psi> .\end{equation}
The eigenvalues when nodes lay on $S$ are also know in closed
form. We do not analyze that case here for lack of space; however,
the eigenvalues can be computed in a direct manner using the tools
that have been given here. The reader is encouraged to try
although the full answer can be found explicitly in the
literature\cite{ash2}. Notice that the spectrum of the area
operator depends on the value of the Immirzi parameter $\gamma$
(introduced in (\ref{immi})). This is a general property of
geometric operators. The spectrum of the area operator is 
clearly discrete.

\subsubsection{Quantization of the volume}

The volume of a three dimensional region $B\subset \Sigma$ is
classically given by \be V_B=\int_B \sqrt{q}\ d^3x,\end{equation} and it can
be
 expressed in terms of the triad operator. In fact using
(\ref{dtriad}) we conclude that \be q=|{\rm det
}(E)|=\left|\frac{1}{3!}\epsilon_{abc}E^a_iE^b_jE^c_j
\epsilon^{ijk}\right|.\end{equation} Therefore, \be V_B=\int_B
\sqrt{\left|\frac{1}{3!}\epsilon_{abc}E^a_iE^b_jE^c_j
\epsilon^{ijk}\right|}\ d^3x \end{equation} Following the similar
regularization techniques as in the case of the area operator we
write the previous integral as the limit of Riemann sums defined
in terms of a decomposition of $B$ in terms of three-cells (think
of a cubic lattice for concreteness), then we quantize the
regularized version\cite{lee1,c3,loll1,ash22} using the
fundamental flux operators of the previous section associated to
infinitesimal two cells. Explicitly, 
\be\label{pirulin} \widehat
V_B=\lim_{N \rightarrow \infty} \widehat V_B^{N} 
\end{equation} 
where 
\be\label{pirulin'}
\widehat V_B^{N}=\sum_{I}^N\sqrt{\
\left|\frac{1}{3!}\epsilon_{abc}\ \widehat E_i(S_I^a)\widehat
E_j(S_I^b)\widehat E_k(S_I^c)\ \epsilon^{ijk}\right|} 
\end{equation}
\begin{figure}[h!!!!!]
 \centerline{\hspace{0.5cm}\(
\begin{array}{c}
\includegraphics[width=4cm]{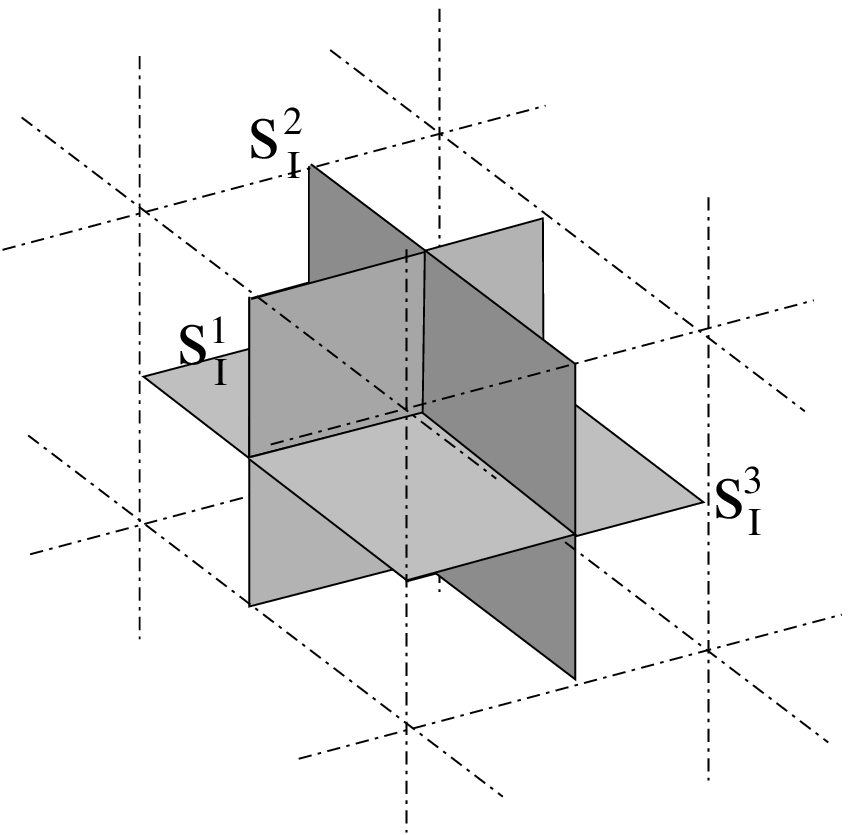}
\end{array}\ \ \ \ \ \ \ \ \ \ \ \ \ \ \ \ \ \ \begin{array}{c}
\includegraphics[width=4cm]{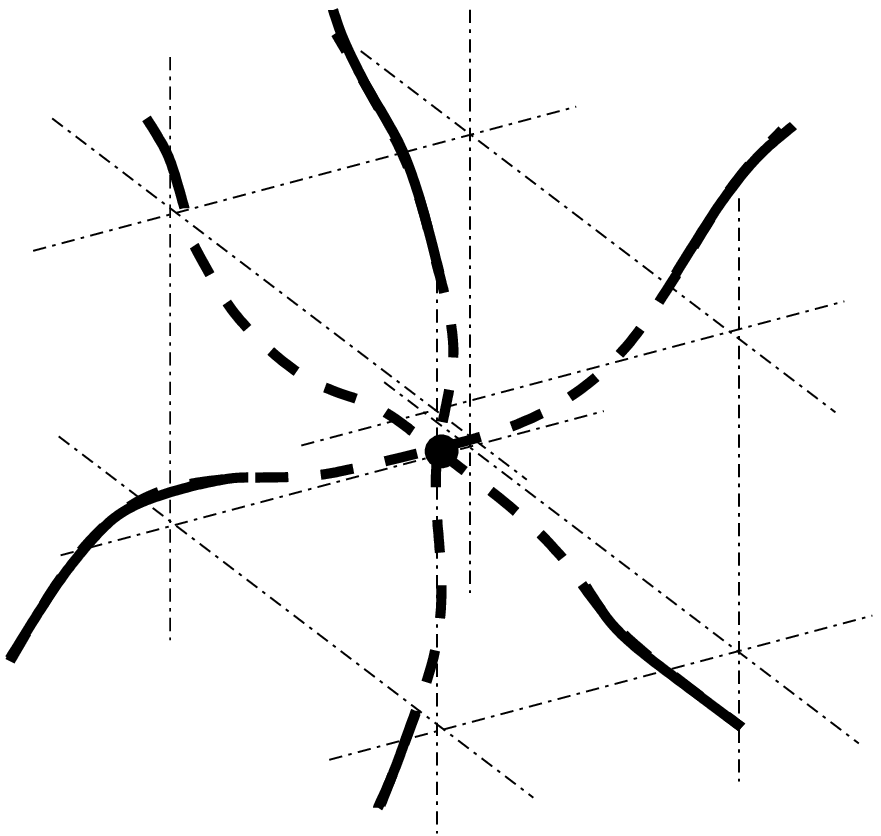}
\end{array}
\)} \caption{Regularization of the volume operator. The
infinitesimal two cells on which $E_i(S_I^a)$ are defined. The
cellular decomposition is shrunk in the limiting process so that
spin network nodes are contained in only one three cell.}
\label{voli}
\end{figure}
The way the surfaces $S_I^a$ are chosen for a given three-cell $I$ is
illustrated on the left of Figure \ref{voli}. The limit $N\rightarrow \infty$
is taken by keeping spin network nodes inside a single three-cell.

Let us finish this section by mentioning some general properties
of the volume operator. \begin{romanlist}[(ii)] \item There are at
least two consistent quantization of the volume operator. One is
sensitive to the differential structure at the node\cite{ash22}
while the other is fully combinatorial\cite{c3}.  The quantization
of the scalar constraint first introduced by Thiemann\cite{th2}
uses the version of volume operators that is sensitive to the
differential structure at the node.

\item Three valent nodes are annihilated by the volume operator.
This is a simple consequence of the Gauss constraint. The identity
(\ref{gaussy}) implies that for a three valent node one can write
one of the flux operators in (\ref{pirulin'}) as a linear
combination of the other two. The $\epsilon_{abc}$ in
(\ref{pirulin'}) makes the action of the operator equal to zero.

\item The action of the volume operator vanishes on nodes whose
edges lie on a plane, i.e., planar nodes.

\item The spectrum of the volume is discrete. The eigenvalue
problems is however more involved and an explicit closed formula
is known only in special
cases\cite{loll1,Loll:1995tp,Loll:1995wt,Thiemann:1996au}.
Recently new manipulations have lead to simplifications in the
spectral analysis of the volume operator\cite{Brunnemann:2004xi}.

\end{romanlist}

\subsubsection{Geometric interpretation of spin network states}

The properties of the area and volume operator provide a very
simple geometrical interpretation of spin network states. Edges of
spin networks carry quanta of area while volume quantum numbers
can be used to label nodes (plus some additional label to resolve
degeneracy when needed). This interpretation is fully background
independent and its combinatorial character implies that the
geometric information stored in a quantum state of geometry is
intrinsically diffeomorphism invariant. We can visualize a spin
network as a polymer like excitation of space geometry consisting
of volume excitations around spin network
nodes connected by spin network links representing 
area excitations of the (dual) surfaces separating nodes.  The picture is in complete agreement
with the general direction in which a background independent
formulation would be set up. The embedding or the coordinate
system that we choose to draw the spin network graph does not
carry any physical information. All the information about the
degrees of freedom of geometry (hence the gravitational field) is
contained in the combinatorial aspects of the graph (what is
connected to what) and in the discrete quantum numbers labelling
area quanta (spin labels of edges) and volume quanta (linear
combinations of intertwiners at nodes). The fact that spin
networks up to their embedding are physically relevant in LQG will
become more clear in the next section, when we solve the
diffeomorphism constraint.

\subsubsection{Continuum geometry}

After the discovery that geometric operators have discrete
spectra an obvious question is whether the discreteness is
compatible with the smooth geometry picture of the classical
theory. One can in fact check that the spectrum crowds very
rapidly when one gets to larger geometries as the spacing 
between eigenvalues decreases exponentially for large eigenvalues.

\subsection{Solutions of the diffeomorphism constraint: $\Hd$ and abstract spin networks}

In Section \ref{GAUSS} we solved the three quantum Einstein's
equations defining the Hilbert space of $SU(2)$ gauge invariant
states given by $\Hg\subset\Hk$. Now we turn our attention to the
next three quantum constraint equations, the vector constraint
(\ref{uno}). The technique that we apply to find the Hilbert space
of diffeomorphism invariant states $\Hd$ is analogous to the one
used to obtain $\Hg$. However, because the orbits of the
 diffeomorphisms are not compact, diffeomorphism invariant states
are not contained in the original $\Hk$. In relation to $\Hk$,
they have to be regarded as distributional states\cite{ash3}.

A simple example will allow us to introduce the basic ideas.
Consider an ordinary particle quantum system defined on a
cylinder. Assume the kinematical Hilbert space is given by square
integrable functions $\psi(\theta,z)$ or $\Hk={\cal L}^2(S^1\times
\R)$. In addition assume one has to solve the following constraint
equations: \be \widehat p_{\theta} \psi =0\ \ \ {\rm and}\ \ \
\widehat p_z\psi=0. \end{equation} The first constraint generates rotations
around the $z$-axis, i.e. it has compact orbits, and, in this
sense, is the analog of the Gauss constraint in LQG. The solutions
of the first constraint are wave functions invariant under
rotations around $z$, those that do not depend on $\theta$.
Therefore they are contained in the original Hilbert space $\Hk$
because, as the orbits of $p_{\theta}$ are compact, square
integrable functions $\psi(z)$ (i.e., independent of $\theta$) exist.
The second constraint has non compact orbits and in this sense is
the analog of the diffeomorphism constraint in LQG. The solution
of the second constraint are functions that do not depend on $z$.
They cannot be contained in $\Hk$, as functions $\psi(\theta)$
cannot be in $\Hk={\cal L}^2(S^1\times\R)$. However given a
suitable dense subset of $\Phi\subset\Hk$ of test functions, e.g.
functions of compact support, then any solution of the latter
constraint can be given a meaning as a distribution.  For instance a
solution of both constraints does not depend neither on $\theta$
nor on $z$. Its wave function corresponds to a constant
$(\psi_{0}|\theta,z>=c$ . The solution $(\psi_0|$ is clearly not
in $\Hk$. We use a rounded brackets in the notation to recall this
fact.  $(\psi_0|$ is in $\Phi^{\star}$, the topological dual of
$\Phi$, i.e., the set of linear functionals from $\Phi$ to $\C$.
Its action on any arbitrary function of compact support
$|\alpha>\in\Phi\subset\Hk$ is given by \be (\psi_0|\alpha>=\int
dz\ d\theta \ c \ \alpha(\theta, z), \end{equation} which is well defined.
The action of $(\psi_0|$ extracts the gauge invariant information
from the non gauge invariant state $|\alpha>$. As vector spaces we
have the relation $\Phi\subset\Hk\subset\Phi^{\star}$, usually
called the Gelfand triple.  In the case of LQG diffeomorphism
invariant states are in the dual of the cylindrical functions $\rm
Cyl$. The Gelfand triple of interest is ${\rm Cyl
}\subset\Hk\subset {\rm Cyl}^{\star}$. Diffeomorphism invariant
states have a well defined meaning as linear forms in $\rm
Cyl^{\star}$.

Let us now apply the same idea to define diffeomorphism invariant states.
Diffeomorphism transformations are easily represented in $\Hk$. We
denote ${\cal U_{D}}[\phi]$ the operator representing the action
of a diffeomorphism $\phi\in Diff(\Sigma)$. Its action can be
defined directly on the dense subset of cylindrical functions
${\rm Cyl}\subset\Hk$. Given $\psi_{\gamma,f}\in {\rm Cyl}$ as in
(\ref{72}) we have \be\label{dddhol} {\cal
U_{D}}[\phi]\psi_{\gamma,f}[A]=\psi_{\phi^{-1}\gamma,f}[A],\end{equation}
which naturally follows from (\ref{ddhol}). Diffeomorphisms act on
elements of $\rm Cyl$ (such as spin networks) by simply modifying
the underlying graph in the obvious manner. Notice that ${\cal
U_{D}}[\phi]$ is unitary according to the definition (\ref{kip}).

Notice that because  $(\ref{dddhol})$ is
not weakly continuous there is no well defined
notion of self-adjoint generator of infinitesimal diffeomorphisms in
$\Hk$\footnote{\label{ortho} This is because for any $\phi\in
{Diff}(\Sigma)$ the state ${\cal U_{D}}[\phi]\psi$ is orthogonal
to $\psi$ for a generic $\psi\in \Hk$.}. In other words the
unitary operator that implements a diffeomorphism transformation
is well defined but there is no corresponding self adjoint
operator whose exponentiation leads to $ {\cal U_{D}}[\phi]$.
Therefore, the  diffeomorphism constraint cannot be quantized in
the Ashtekar-Lewandowski representation. This is not really a problem as  ${\cal
U_{D}}[\phi]$ is all we need in the quantum theory to look
for diffeomorphism invariant states. In LQG one replaces
the second set of formal equations in (\ref{QEE}) by the well
defined equivalent requirement \be {\cal U_{D}}[\phi]\psi=\psi\end{equation} for
(distributional states) $\psi \in {\rm Cyl}^{\star}$. 

We are now
ready to explicitly write the solutions, namely \be\label{st}
([\psi_{\gamma,f}]|=\sum_{\phi \in Diff(\Sigma)}
<\psi_{\gamma,f}|{\cal U_{D}}[\phi]= \sum_{\phi \in Diff(\Sigma)}
<\psi_{\phi\gamma,f}|,\end{equation} were the sum is over all diffeomorphisms.
The brackets in $([\psi_{\gamma,f}]|$ denote that the
distributional state depends only on the equivalence class
$[\psi_{\gamma, f}]$ under diffeomorphisms. Clearly we have
$([\psi_{\gamma,f}]|{\cal U_{D}}[\alpha]=([\psi_{\gamma,f}]|$ for
any $\alpha\in Diff(\Sigma)$. Now we need to check that
$([\psi_{\gamma,f}]|\in{\rm Cyl}^{\star}$ so that the huge sum in
(\ref{st}) gives a finite result when applied to an element
$|\psi_{\gamma^{\prime},g}>\in {\rm Cyl}$, i.e., it is a well
defined linear form. That this is the case follows from
(\ref{kip}) (see the remark in Footnote \ref{ortho}) as in
\be([\psi_{\gamma,f}]|\psi_{\gamma^{\prime},g}> \end{equation} only a finite
number of terms from (\ref{st}) contribute. In fact for spin
networks with no discrete symmetries there is only one non trivial
contribution. 

The action of $([\psi_{\gamma,f}]|$ is
diffeomorphism invariant, namely
\be\label{pirupiru}([\psi_{\gamma,f}]|\psi_{\gamma^{\prime},g}>=([\psi_{\gamma,f}]|{\cal
U_{D}}[\phi]\psi_{\gamma^{\prime},g}>\end{equation} The inner product $<\ ,\
>_{diff}$ needed to promote the set of diffeomorphism invariant
states to the Hilbert space $\Hd$ is defined as \be
<[\psi_{\gamma,f}]|[\psi_{\gamma^{\prime},g}]>_{diff}=([\psi_{\gamma,f}]|\psi_{\gamma^{\prime},g}>\end{equation}
Due to (\ref{pirupiru}) the previous equation is well defined among
diffeomorphism equivalence classes of states under the action of diffeomorphisms and hence this is denoted by
the brackets on both sides.

\subsubsection{The quantization of the scalar constraint}

In this section we sketch the regularization and definition of the
last constraint to be quantized and solved: the scalar constraint.
The smeared version of the classical scalar constraint is
\ba\nonumber && S(N)=\int\limits_{\Sigma} dx^3 \ N
\frac{E_i^aE_j^b}{\sqrt{{\rm det}(E)}}\left(\epsilon^{ij}_{\ \ k}
F_{ab}^k-2(1+\gamma^2)K^i_{[a}K^j_{b]}\right)=\\  && =
S^{E}(N)-2(1+\gamma^2) \ T(N),\ea where
$K_a^i=\gamma^{-1}(A_a^i-\Gamma_a^i)$, and in the second line we
have introduced a convenient separation of the constraint into
what is called the Euclidean contribution $S^E(N)$ given by \be
S^{E}(N)=\int\limits_{\Sigma} dx^3 \ N
\frac{E_i^aE_j^b}{\sqrt{{\rm det}(E)}}\epsilon^{ij}_{\ \ k}
F_{ab}^k, \end{equation} and the extra piece \be T(N)=\int\limits_{\Sigma}
dx^3 \ N \frac{E_i^aE_j^b}{\sqrt{{\rm det}(E)}}K^i_{[a}K^j_{b]}.
\end{equation} The terms in the constraint look in fact very complicated. On
the one hand they are highly non linear which anticipates
difficulties in the quantization related to regularization issues
and potential UV divergences, factor ordering ambiguities, etc.
For instance the factor $1/{\rm det}(E)$ looks quite complicated
at first sight as does the spin connection $\Gamma_a^i$ in the
expression of $K_a^i$ (recall its definition (\ref{19}) and
(\ref{twenty}) in terms of the basic triad variables).

The crucial simplification of the apparently intractable problem
came from the ideas of Thiemann\cite{th1}. He observed that if one
introduces the phase space functional \be \bar K:=\int
\limits_{\Sigma} K_a^iE^a_i, \end{equation} then the following series of
identities hold: \be
K_a^i=\gamma^{-1}(A_a^i-\Gamma_a^i)=\frac{1}{\kappa
\gamma}\left\{A_a^i,\bar K\right\}, \end{equation} \be \bar
K=\frac{1}{\gamma^{3/2}} \left\{S^E(1),V\right\}, \end{equation} where
$V=\int \sqrt{{\rm det}(E)}$ is the volume of $\Sigma$, and
finally \be\label{ctriad} \frac{E_i^bE_j^c}{\sqrt{{\rm
det}(E)}}\epsilon^{ijk}\epsilon_{abc}=\frac{4}{\kappa\gamma}\left\{A^k_a,V\right\}.\end{equation}
With all this we can write the terms in the scalar constraint by
means of Poisson brackets among quantities that are simple enough
to consider their quantization. The Euclidean
constraint can be written as\be\label{seuc}
S^{E}(N)=\int\limits_{\Sigma} dx^3 \ N \ \epsilon^{abc}
\delta_{ij} F^i_{ab}\left\{ A^j_c,V \right\}, \end{equation} while the term
$T(N)$ becomes \be T(N)=\int\limits_{\Sigma} dx^3 \
\frac{N}{\kappa^2\gamma^3}\ \epsilon^{abc}\epsilon_{ijk}
\left\{A^i_a,\left\{S^E(1),V\right\}\right\}
\{A^j_b,\left\{S^E(1),V\right\}\} \left\{A^k_c,V\right\}. \end{equation} The
new form suggests that we can quantize the constraint by promoting the argument of the
Poisson brackets to operators and the Poisson brackets them self to commutators in
the standard way. One needs the volume operator $V$, whose
quantization was already discussed, and the quantization of the
connection and curvature. We present here the basic idea behind
the quantization of these. For a precise treatment the reader is
encouraged to read Thiemann's original
work\cite{th2,Thiemann:1996aw} and book\cite{bookt}. Given an
infinitesimal loop $\alpha_{ab}$ on the $ab$-plane with coordinate
area $\epsilon^2$, the curvature tensor can be regularized
observing that \be
h_{\alpha_{ab}}[A]-h^{-1}_{\alpha_{ab}}[A]=\epsilon^2
F_{ab}^i\tau_i+{\cal O}(\epsilon^4).\end{equation}  Similarly the Poisson
bracket $\left\{A^i_a,V\right\}$ is regularized as \be
h_{e_a}^{-1}[A]\left\{h_{e_a}[A],V\right\}=\epsilon
\left\{A^i_a,V\right\}+{\cal O}(\epsilon^2),\end{equation} where $e_a$ is a
path along the $a$-coordinate of coordinate length $\epsilon$.
With this we can write \ba \nonumber &&S^{E}(N)=\\&& \nonumber
=\lim_{\epsilon\rightarrow 0} \sum_I \ N_I \epsilon^3 \
\epsilon^{abc}{\rm
Tr}\left[F_{ab}(A)\left\{A_c,V\right\}\right]=\\&&=\lim_{\epsilon\rightarrow
0} \sum_I \ N_I \ \epsilon^{abc}{\rm Tr}\left[(
h_{\alpha^I_{ab}}[A]-h^{-1}_{\alpha^I_{ab}}[A])
h_{e^I_c}^{-1}[A]\left\{h_{e^I_c}[A],V\right\}\right], \ea where
in the first equality we have replaced the integral (\ref{seuc})
by a Riemann sum over cells of coordinate volume $\epsilon^3$ and
in the second line we have written it in terms of holonomies.
Notice that the explicit dependence on the cell size $\epsilon$
has disappeared in the last line.
\begin{figure}[h!!!!!]
 \centerline{\hspace{0.5cm}\(
\begin{array}{c}
\includegraphics[width=4cm]{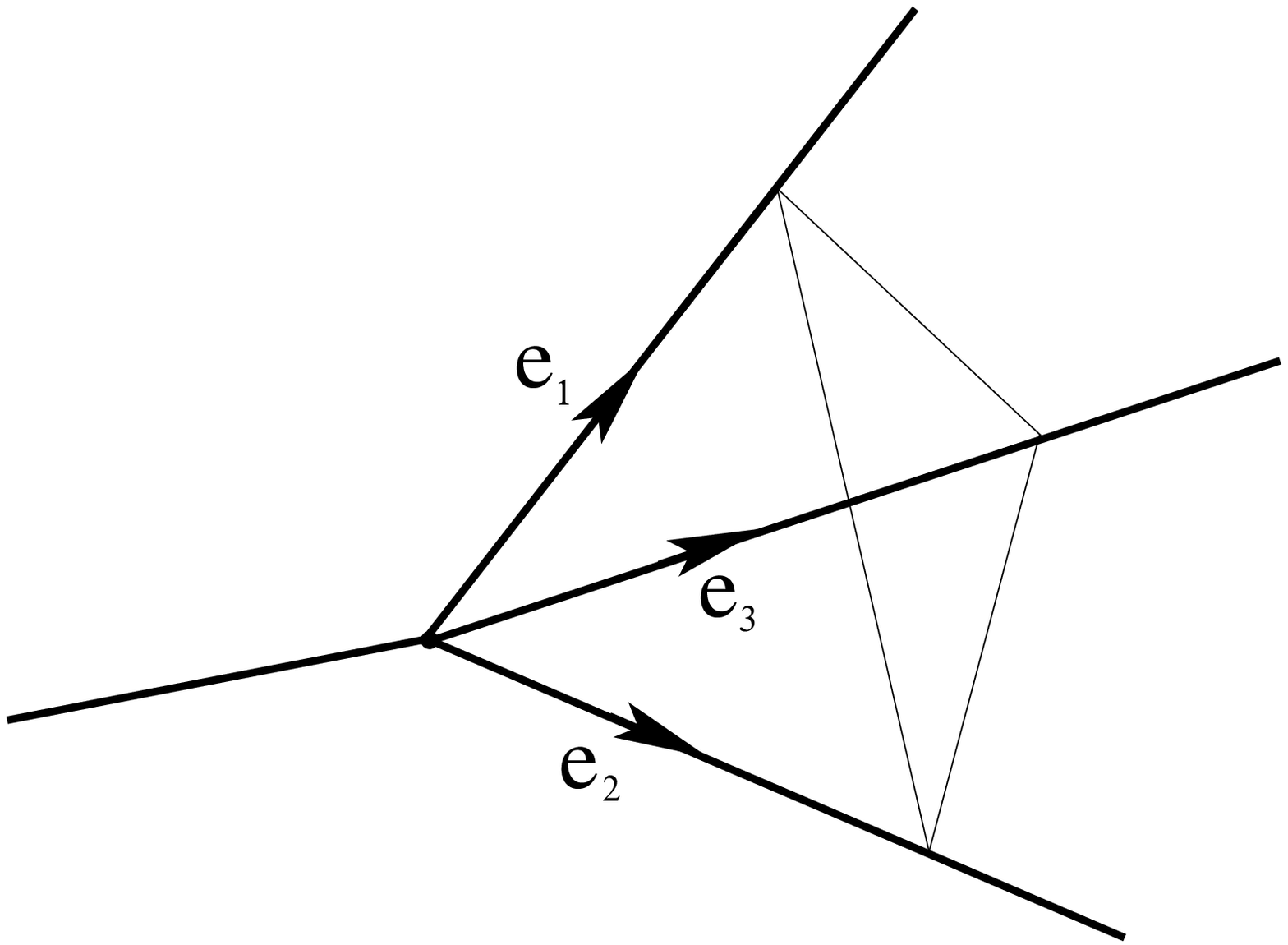}
\end{array}\ \ \ \ \ \ \ \ \ \ \ \ \begin{array}{c}
\includegraphics[width=2cm]{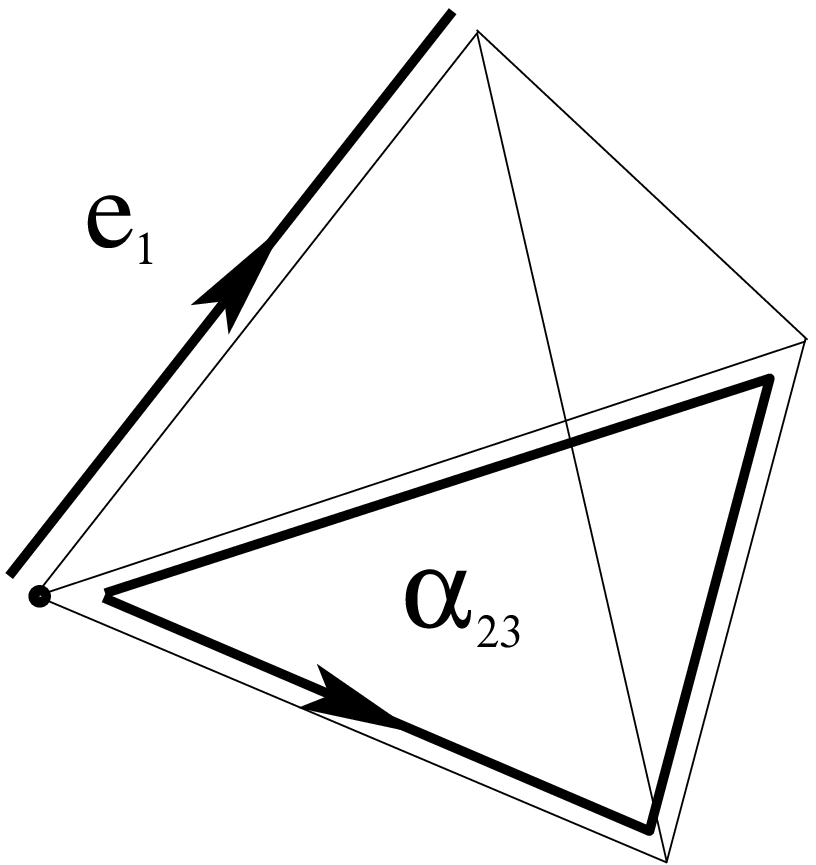}
\end{array}\ \ \ \begin{array}{c}
\includegraphics[width=2cm]{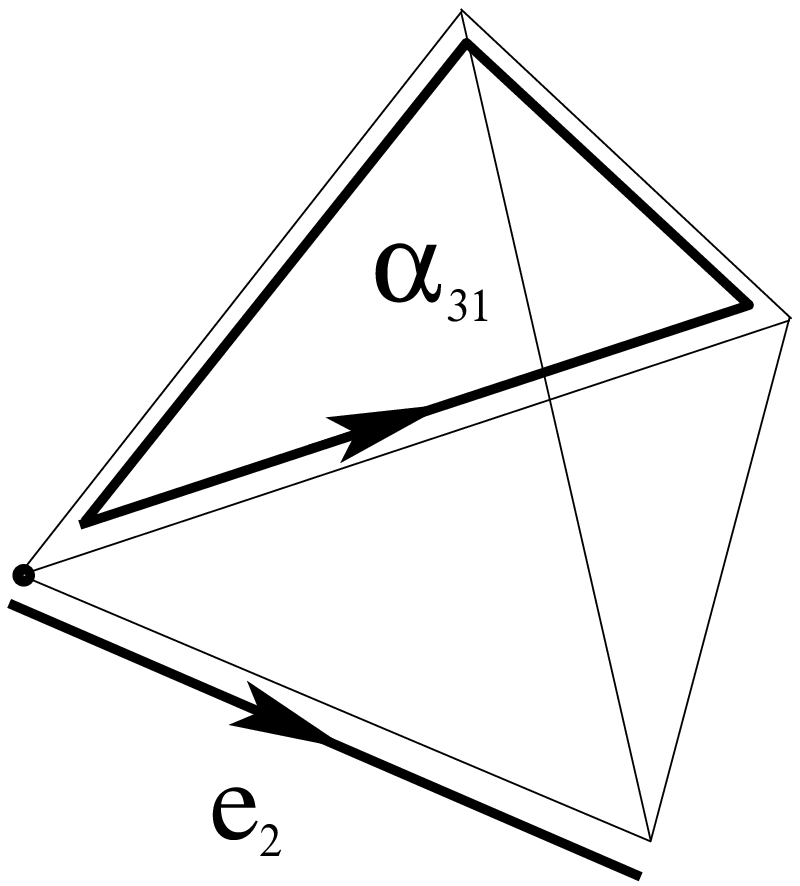}
\end{array}\ \ \ \begin{array}{c}
\includegraphics[width=2cm]{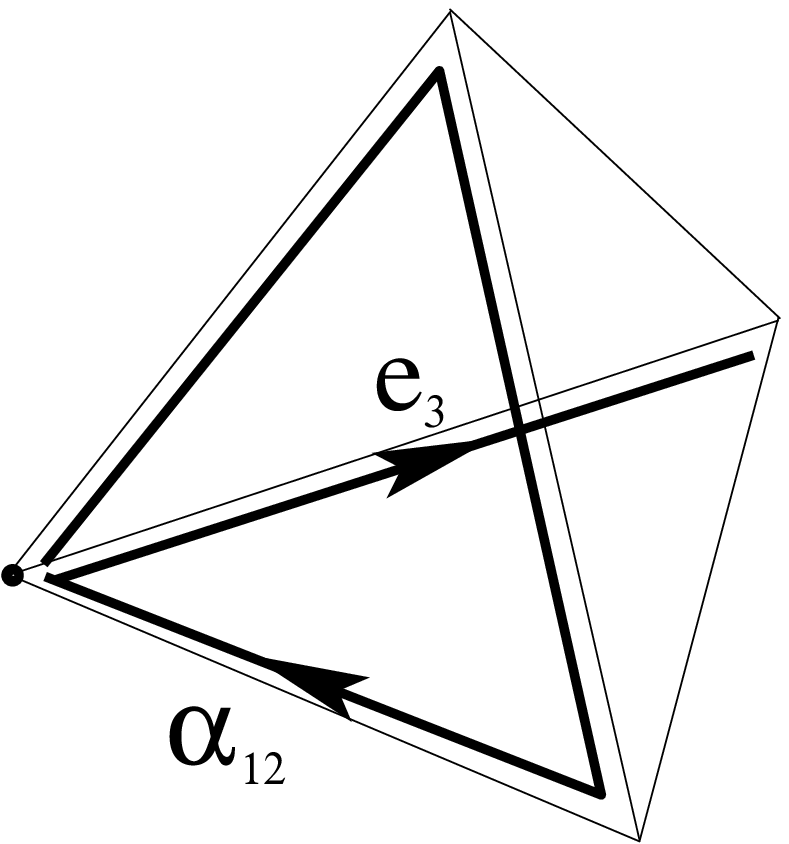}
\end{array}
\)} \caption{A tetrahedral infinitesimal cell adapted to a four
valent spin network node. On the right: the three non trivial
contributions of the cell to the regularized expression
(\ref{hami}). } \label{hamil}
\end{figure}
The cells are labelled with the index $I$ in analogy to the
regularization used for the area and volume in previous sections.
The loop $\alpha^I_{ab}$ is an infinitesimal closed loop of
coordinate area $\epsilon^2$ in the $ab$-plane associated to the
$I$-th cell, while the edge $e^{I}_a$ is the corresponding edge of
coordinate length $\epsilon$ dual to the $ab$-plane (see Figure
\ref{hamil} for a cartoon of the regularization first introduced by Thiemann). 
The idea now is to promote this regulated expression
to an operator by quantizing the ingredients of the formula:
notice that we already know how to do that as the expression
involves the holonomies and the volume, both well defined
operators in $\Hk$. The quantum constraint can formally be written
as \be \label{hami}\widehat S^{E}(N)=\lim_{\epsilon\rightarrow 0}
\sum_I \ N_I \ \epsilon^{abc}{\rm Tr}\left[( \widehat
h_{\alpha^I_{ab}}[A]-\widehat h^{-1}_{\alpha^I_{ab}}[A]) \widehat
h_{e^I_c}^{-1}[A]\left\{\widehat h_{e^I_c}[A],\widehat
V\right\}\right]. \end{equation} Now in order to have a rigorous definition
of $\widehat S^E(N)$ one needs to show that the previous limit
exists in the appropriate Hilbert space.

It is useful to describe some of the qualitative features of the
argument of the limit in (\ref{hami}) which we refer to as the
{\em regulated quantum scalar constraint} and we denote $\widehat
S_{\epsilon}(N)$. It is easy to see that the regulated quantum
scalar constraint acts only on spin network nodes, namely \be
\widehat S_{\epsilon}(N)\psi_{\gamma,f}=\sum_{n \gamma} N_n
\widehat S^n_{\epsilon}\ \psi_{\gamma,f} ,\end{equation} where $\widehat
S^{n}_{\epsilon}$ acts only on the node $n\subset \gamma$ and
$N_n$ is the value of the lapse $N(x)$ at the node.
 This is a simple consequence of the very same property of
the volume operator (\ref{pirulin}). Due to the action of the infinitesimal loop
operators representing the regularized curvature, the scalar
constraint modifies spin networks by creating
new links around nodes whose amplitudes depend on the details of
the action of the volume operator, the local spin labels and other
local features at the nodes. If we concentrate on the Euclidean
constraint, for simplicity, its action on four valent nodes can be
written as \ba\label{titi}
\!\!\!\!\!\!\!\!\!\!\!\!\!\!\!\!\!\!\!\!\!\!\!\!&& \nonumber \centerline{\hspace{-4cm}\(
\widehat S^{n}_{\epsilon} \begin{array}{c}
\includegraphics[width=2.5cm]{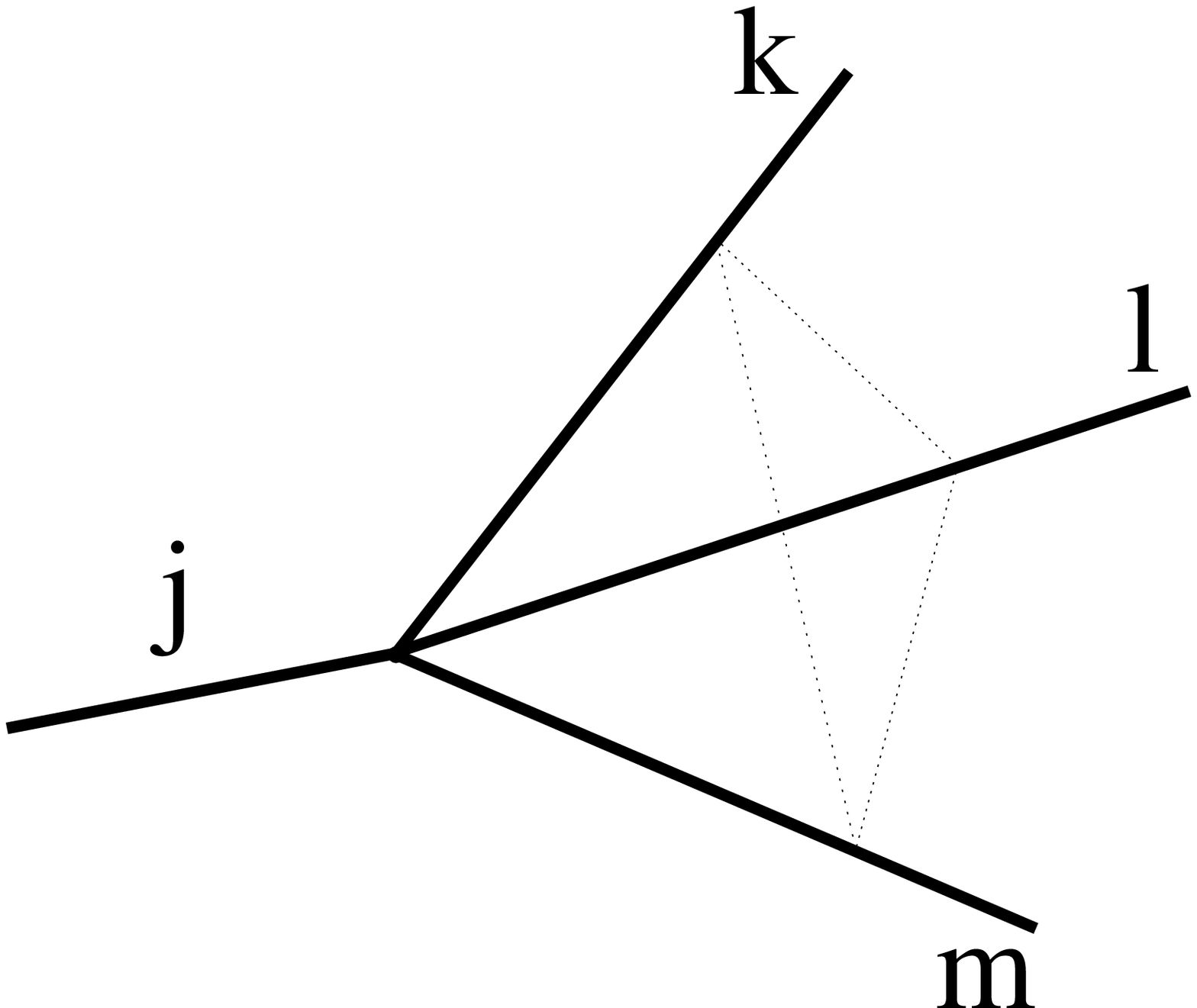}
\end{array} = \sum_{op} S_{jklm,opq} \begin{array}{c}
\includegraphics[width=2.5cm]{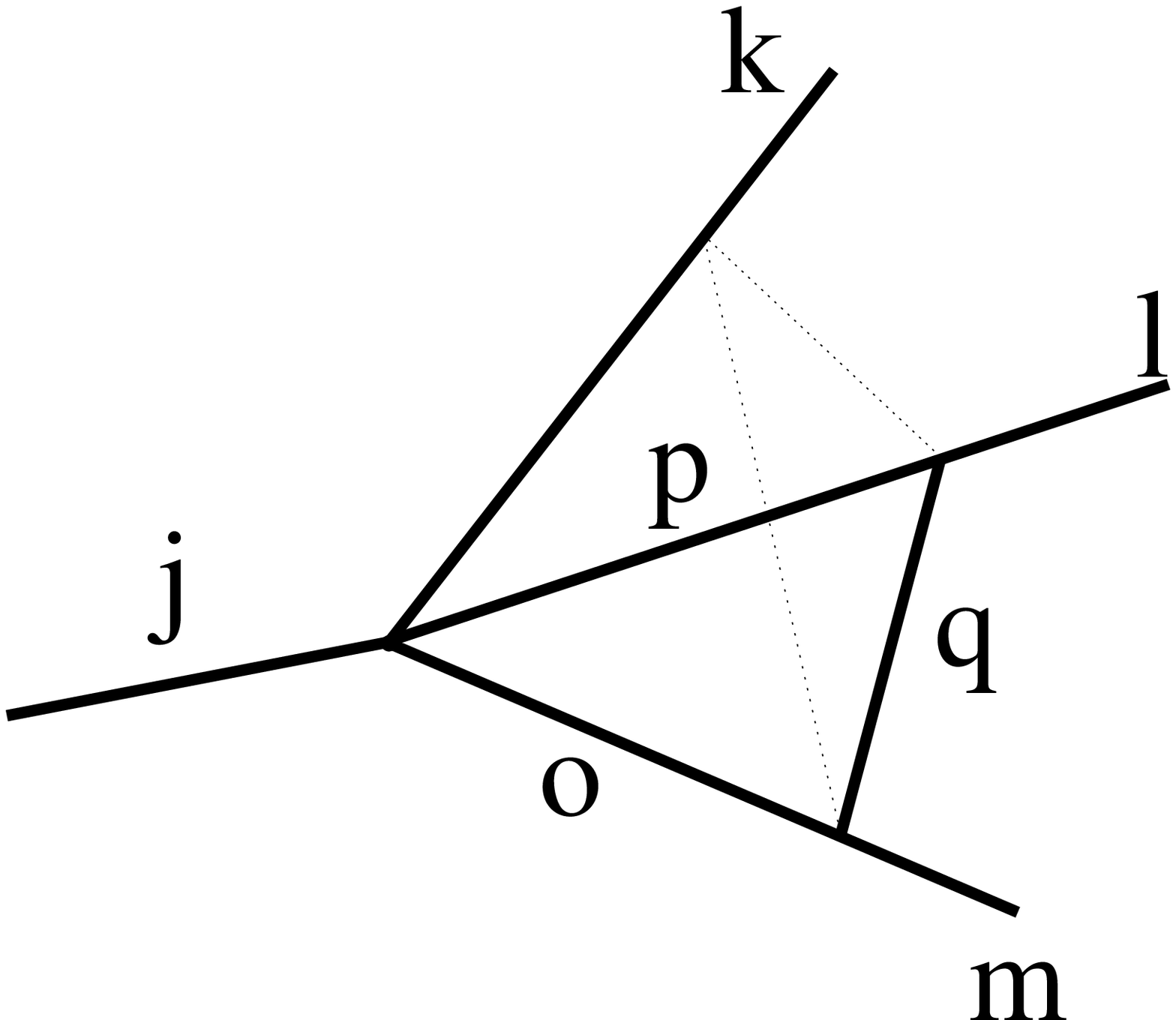}
\end{array}+\)} \\ &&  \centerline{\hspace{0.5cm}\(+ \sum_{op} S_{jlmk,opq} \begin{array}{c}
\includegraphics[width=2.5cm]{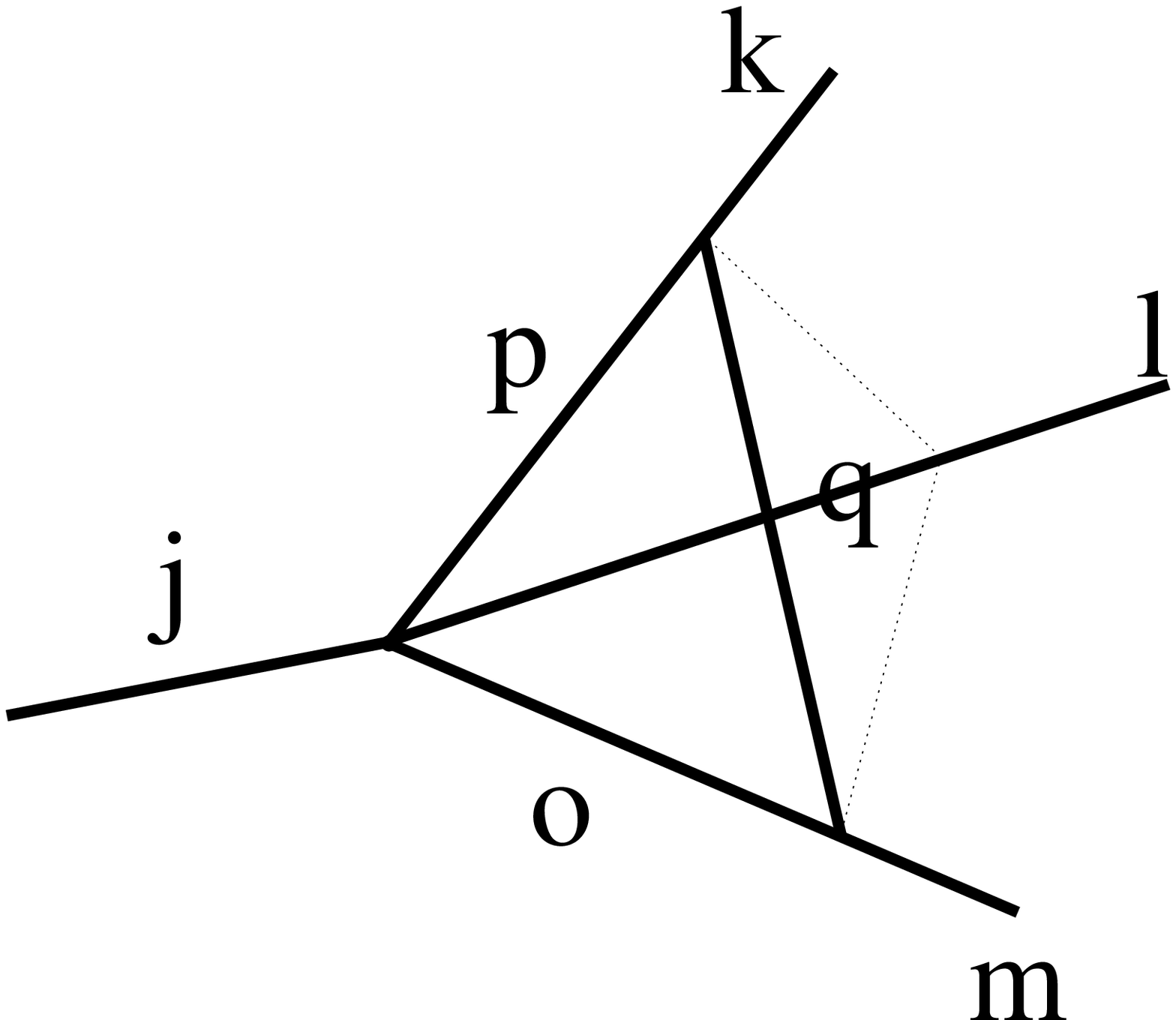}
\end{array} + \sum_{op} S_{jmkl,opq} \begin{array}{c}
\includegraphics[width=2.5cm]{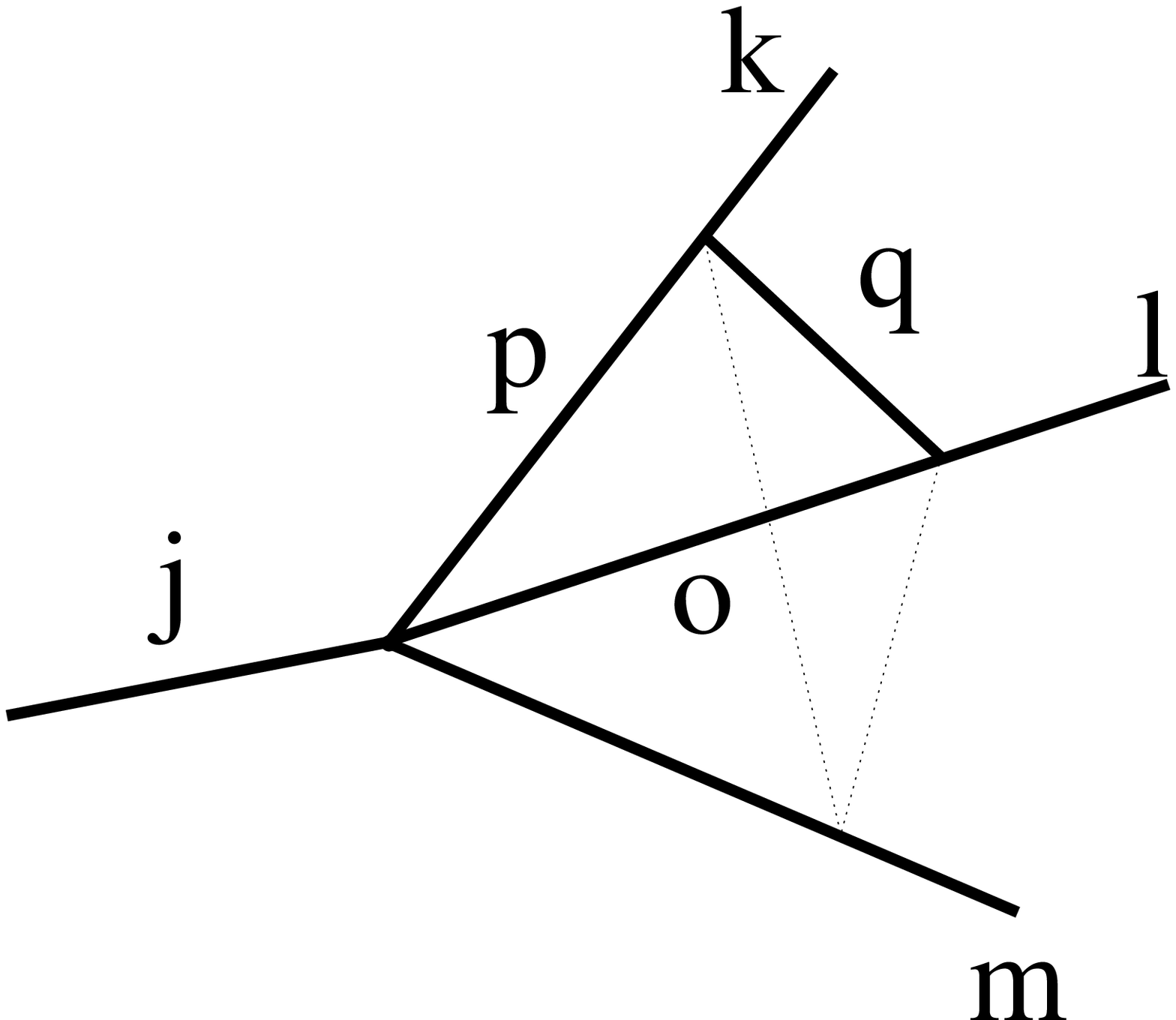}
\end{array}
\)}
,\ea
where $q=1/2$ in the case of (\ref{hami}) (we will see in the sequel that $q$ can be any arbitrary spin: this is one of the quantization ambiguities in the theory),
and $S_{jklm,opq}$ are coefficients that are independent of $\epsilon$ and can be computed explicitly from
(\ref{hami}).

Now we analyze the removal of the regulator. Since the only
dependence of $\epsilon$ is in the position of the extra link in
the resulting spin network states, the limit $\epsilon\rightarrow
0$ can be defined in the Hilbert space of diffeomorphism invariant
states $\Hd$. The key property is that in the diffeomorphism invariant context the
position of the new link in (\ref{titi}) is irrelevant.
Therefore, given a diffeomorphism invariant state $([\phi]|\in
\Hd\in {\rm Cyl}^{\star}$, as defined as in
(\ref{st}), the quantity $(\phi|\widehat S_{\epsilon}(N)|\psi>$ is well defined and independent of $\epsilon$. 
In other words the limit \be
(\phi|\widehat S(N)|\psi>=\lim_{\epsilon\rightarrow 0}\
(\phi|\widehat S_{\epsilon}(N)|\psi> \end{equation} exists trivially for any
given $\psi\in \Hk$.  A careful analysis on how this limit is
defined can be found in\cite{bookt}.

An important property of the definition of the quantum scalar
constraint is that the new edges added (or annihilated) are of a
very special character. For instance, not only do the new nodes in
(\ref{titi}) carry zero volume but also they are invisible to the
action of the quantum scalar constraint. The reason for that is
that the new three-valent nodes are planar and therefore the
action of the commutator of the holonomy with the volume operator
in (\ref{hami}) vanishes identically (recall the properties of the
volume operator at the end of Section \ref{goo}). For this reason
it is useful to refer to these edges as {\em exceptional edges}.

This property of Thiemann's constraint is indeed very important
for the consistency of the quantization. The non trivial
consistency condition on the quantization of the scalar constraint
corresponds to the quantum version of (\ref{prob}). The correct
commutator algebra is satisfied in the sense that for
diffeomorphism invariant states $(\phi|\in \Hd$ (defined as in
(\ref{st}))\be\label{anom} (\phi|[\widehat S(N),\widehat
S(M)]\psi>=0, \end{equation} for any $\psi \in \Hg$. The l.h.s. vanishes due
to the special property of exceptional edges as can be checked
by direct calculation. Notice  that r.h.s. of (\ref{prob}) is
expected to annihilate elements of $\Hd$---at least for the
appropriate factor ordering--- so that the previous equation is
in agreement with (\ref{prob}) and the quantization is said to be {\em anomaly-free}\cite{Thiemann:1997rv}.
All known\cite{c00,pul1,pul4} consistent
quantizations satisfy the Abelian property (\ref{anom}) in $\Hd$.
We will come back to this issue in the next Subsection.

Notice that (\ref{ctriad}) is the co-triad $e_a^i$ according to
(\ref{twenty}); this opens the way for the quantization of the
metric $q_{ab}=e^i_a e^j_b \delta_{ij}$ that is necessary for the
inclusion of matter. A well defined quantization of the scalar
constraint including Yang-Mills fields, scalar fields and fermions
has been put forward by Thiemann\cite{th5}.

\subsubsection{Solutions of the scalar constraint, physical observables, difficulties} \label{lee}

Here we briefly explore some of the generic consequences of the
theory constructed so far. The successful definition of the
quantum scalar constraint operator including the cases with
realistic matter couplings is a remarkable achievement of loop
quantum gravity. There are however some issues that we would like
to emphasize here.

There is a large degree of ambiguity on the definition of the
quantum scalar constraint. The nature of solutions or the dynamics
seems to depend critically on these ambiguities.  For instance it
is possible to arrive at a completely consistent quantization by
essentially replacing the holonomies in (\ref{hami})---defined in
the fundamental representation of $SU(2)$---by the corresponding
quantities evaluated on an arbitrary representation\cite{c00}. In
the applications of the theory to simple systems such as in loop
quantum cosmology this is known to have an important physical
effect\cite{Bojowald:2003mc}. Ambiguities are also present in the
way in which the paths defining the holonomies that regularize the
connection $A_a^i$ and the curvature $F^i_{ab}(A)$ in (\ref{hami})
are chosen. See for instance Section C in\cite{ash10} for an
alternative to Thiemann's prescription and a discussion of the
degree of ambiguity involved. There are factor ordering
ambiguities as well, which is evident from (\ref{hami}).
Therefore instead of a single theory we have infinitely many
theories which are mathematically consistent. A yet unresolved
issue is whether any of these theories is rich enough to reproduce
general relativity in the classical continuum limit.
\begin{figure}[h!!!!!]
 \centerline{\hspace{0.5cm}\(
\begin{array}{c}
\includegraphics[width=6cm]{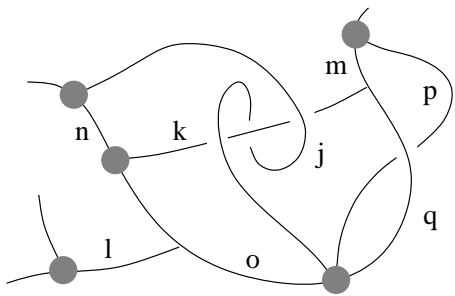}
\end{array}
\)} \caption{Solutions of the scalar constraint as dressed (diff-invariant) spin networks.
}
\label{dress}
\end{figure}

All the known consistent quantizations of the scalar constraint
satisfy property (\ref{anom}). Quantizations that satisfy this
property seem to share a property that is often referred to as
ultra-locality\cite{Smolin:1996fz}. This property can be
illustrated best in terms of the kind of solutions of the scalar
constraint.  We will keep the discussion here as general as
possible; therefore, we will study generic features of formal solutions. We
should point out that exact solutions are known for specific
quantizations of the scalar constraint. For instance, an algorithm
for constructing the general solution of the quantum scalar
constraint is described in detail by Ashtekar and Lewandowski\cite{ash10} in terms of the
quantum operator $\widehat S(N)$ introduced therein (see also
\cite{Thiemann:1996av}). These solutions satisfy the property
described below for generic formal solutions.

Quantization of the scalar constraint satisfying (\ref{anom}) act
on spin network nodes by adding (and/or annihilating\footnote{The version of quantum scalar constraint whose action is depicted in 
(\ref{titi}) is not self adjoint. One can introduce self adjoint definitions which contain a term that creates 
exceptional edges and another one  that  destroys them.}) {\em
exceptional} edges. As explained above these exceptional edges are
characterized by being invisible to subsequent actions of the
constraint. These exceptional edges are added (or destroyed) in
the local vicinity of nodes. For that reason, solutions of the
scalar constraint can be labelled by graphs with {\em `dressed'}
nodes as the one illustrated in Figure \ref{dress}. The shadowed
spheres denote certain (generally infinite) linear combinations of
spin networks with exceptional edges, diagrammatically
\ba\label{pirulo} \label{undress}  && \nonumber
\centerline{\hspace{.5cm}\(
\begin{array}{c}
\includegraphics[width=1.5cm]{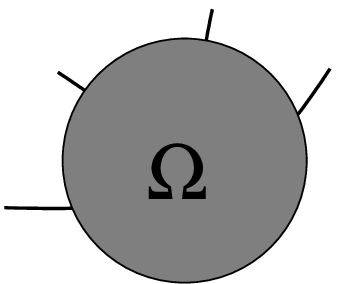}
\end{array}
= \)}\\ && \centerline{\hspace{-.5cm}\( = \alpha
\begin{array}{c}
\includegraphics[width=1.5cm]{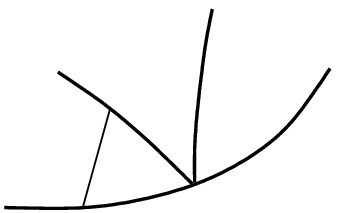}
\end{array}
 + \beta
\begin{array}{c}
\includegraphics[width=1.5cm]{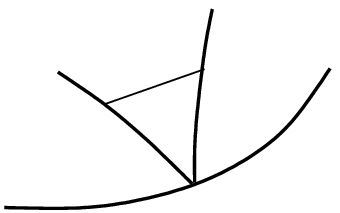}
\end{array}
+ \cdots + \gamma
\begin{array}{c}
\includegraphics[width=1.5cm]{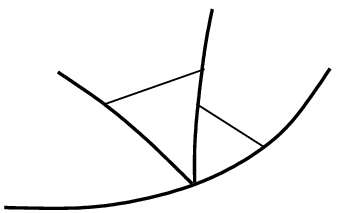}
\end{array}
+\cdots+\delta
\begin{array}{c}
\includegraphics[width=1.5cm]{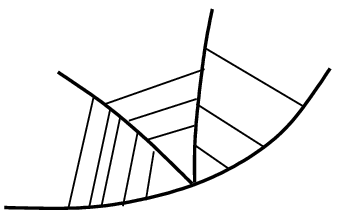}
\end{array}+ \cdots
\)} ,\ea where we have simplified the notation by dropping the
spin labels and the coefficients $\alpha, \beta \cdots \delta$
depend on the details of the definition of the scalar constraint
and the spin labelling of the corresponding spin networks. The
Greek letter $\Omega$ denotes the possible set of quantum numbers
labelling independent solutions at the dressed nodes.

Notice that the generic structure of these formal
solutions---which is shared by the exact solutions of explicit
quantizations---is on the one hand very appealing. The solutions
of all the constraints of LQG seem quite simple: they reduce to
simple algebraic relations to be satisfied by the coefficients of
(\ref{pirulo}). In some explicit cases\cite{ash10}, these
relations even reduced to finite dimensional matrix operations.

The structure of solutions also suggest the possibility of
defining a large variety of Dirac (i.e. physical) observables. For
instance, the fact that generic solutions can be characterized by
dressed spin-network states as in Figure \ref{dress} implies that
the spin labelling the links joining different dressed nodes are
indeed quantum numbers of Dirac observables (the operator
corresponding to these quantum numbers evidently commute with the
action of the constraint). There are infinitely many Dirac
(quasi-local) observables that one can construct for a given
quantization of the scalar constraint satisfying
(\ref{anom})\footnote{The intuitive idea here is presented in
terms of a given solution of the constraint based on a given
family of graphs: the dressed spin-networks.  One should keep in
mind that defining a quantum operators representing a Dirac
observable implies defining its action on the whole $\Hp$. A
simple example of Dirac operator whose action is defined far all
solutions is for instance $O_D=\sum_{e}j_e$, i.e., the sum of the
spins connecting dressed nodes\cite{lewa}.}. However, at present
it is not very  clear what the physical interpretation of these
observables would be.

On the other hand, the ultra-local character of the solutions has raised some
concerns of whether the quantum theories of the Thiemann type can
reproduce the physics of gravity in the classical
limit\cite{Smolin:1996fz}. In order to illustrate this point let
us go back to the classical theory. Since at the classical level
we can invert the transformations that lead to new variables in
Section \ref{hfnv} let us work with the constraints in ADM
variables. We simplify the discussion by considering the
York-Lichnerowicz conformal decompositions for initial data. For
the discussion here we can specialize to the time-symmetric case
$K_{ab}=0$ and we take the ansatz $q_{ab}=\psi^4 q^0_{ab}$ for
some given $q^0_{ab}$ defined up to a conformal factor. The vector
constraint is in this case identically zero and the only non
trivial constraint is the scalar constraint that can be shown to
be \be\label{elli} \Delta^0\psi-\frac{1}{8}R(q^0)=0, \end{equation} where $\Delta^0$ is
the covariant Laplacian defined with respect to $q^0_{ab}$. The
point is that the previous equation is manifestly {\em elliptic} which
is a general property of the scalar constraint written in this
form\cite{Cook:2000vr}. This means that if we give the value of
$\psi$ on a sphere, the scalar constraint (\ref{elli}) will
determine the value of $\psi$ inside.  The scalar constraint in
general relativity imposes a condition among unphysical degrees of
freedom (represented by $\psi$ in this case) that ``propagate''
along the initial value surface\footnote{With the time-symmetric
ansatz $K_{ab}=0$  for the five degrees of freedom in the
conformal metric $q^0_{ab}$ only one local physical degree of
freedom remains after factoring out the action of the constraints
as generating functions of gauge transformations. In this context
our truncation leads to only one free data per space-point.
Despite these restrictions there are many radiating spacetimes and
other interesting solutions in this sector.}. The only point in
writing the initial value problem in this way is to emphasize this
property of the scalar constraint.

Coming back to the quantum theory one can construct semi-classical
states by taking linear combinations of kinematical spin-network
states in order to approximate some classical geometry
$q_{ab}$\cite{Ashtekar:1992tm,Thiemann:2000bw}. In that context
one could define a sphere with some given semi-classical area. Now
because the scalar constraint acts only in the immediate vicinity
of nodes and does not change the value of the spins of the edges
that connect different {\em dressed} nodes it is not clear how the
elliptic character of the classical scalar constraint would be
recovered in this semi-classical context. In other words, how is it
that quantizations of the scalar constraint that are ultra-local in
the sense above can impose conditions restricting unphysical
degrees of freedom in the interior of the sphere once boundary
conditions defining the geometry of the sphere
are given? Because we still know very
little about the semi-classical limit these concerns should be taken
as open issues that  deserve a more precise analysis.

Motivated by these concerns different avenues of research have
been explored with the hope of finding alternatives and some
guiding principles that would lead to a clearer understanding of
the physics behind the quantum scalar constraint. For instance,
the previous concerns have lead to the exploration of the
formalism of consistent discretizations presented in these lectures
by Rodolfo Gambini\cite{rodolfo,pul0,pul00}.  The spin foam
approach has been motivated to a large extent by the hope of
solving the issue of ambiguities and ultra-locality from a
covariant perspective as well as by the search of a systematic
definition of the physical scalar product. An alternative strategy
to the quantization of the scalar constraint and the construction
of the physical inner product that in essence circumvents the
anomaly freeness condition (\ref{anom}) has been  recently proposed by
Thiemann\cite{Thiemann:2003zv}.

\subsection{An important application: computation of black hole entropy}

In the actual talks  we described the main ideas behind the
computation of black hole entropy in LQG. The philosophy was to
quantize a sector of the theory containing an isolated
horizon\cite{Ashtekar:2004cn} and then to count the number of
physical states $\cal N$ compatible with a given macroscopic area
$a_0$ of the horizon. The entropy $S$ of the black hole is defined
by $S={\rm ln}({\cal N})$. The counting can be made exactly when
$a_0>>\ell_p^2$. The result is \be
S=\frac{\gamma_0}{\gamma}\frac{a_0}{4\ell_p^2}+{\cal O}({\rm
ln}(\frac{a_0}{\ell_p^2})),\end{equation} where the real number
$\gamma_0=0.2375...$ follows from the counting
\cite{Domagala:2004jt,Meissner:2004ju}. The value of the Immirzi parameter comes
from the fact that $\gamma$ appears as a pre-factor in the spectrum
of the area operator. It is important to emphasize that  the
computation of $S$ is independent of the details of the
quantization of the scalar constraint. Semi-classical
considerations lead to $S=a_0/(4 \ell_p^2)$, the computation above
can be used to fix the value of the Immirzi parameter, namely \be
\gamma=\gamma_0. \end{equation} The above computation can be performed for
any black hole of the Kerr-Newman family and the result is
consistent with the chosen value of $\gamma$.

The reader interested in the details of this calculation is
encouraged to study the original
papers\cite{c5,Krasnov:1996wc,Ashtekar:1997yu,Ashtekar:2000eq,Ashtekar:1999wa},
the resent review by Ashtekar and Lewandowski, or the book of
Rovelli\cite{book}. Attention is drawn to the resent results of
Domagala and Lewandowski\cite{Domagala:2004jt} and
Meissner\cite{Meissner:2004ju}.

\section{Spin Foams: the path integral representation of the dynamics in loop
  quantum gravity}\label{SFM}

The spin foam approach was motivated by the need to shed new light
on the issue of the dynamics of loop quantum gravity by attempting
the construction of the path integral representation of the
theory. In this section we will introduce the main ideas behind
the approach by considering simpler systems in some detail. For a
broader overview see the review articles\cite{a18,ori2} and
references therein.

The solutions of the scalar constraint can be characterized by the
definition of the generalized projection operator $\PP$ from the
kinematical Hilbert space $\Hk$ into the kernel of the scalar
constraint $\Hp$. Formally one can write $P$ as
\begin{equation}\label{P}
P =``\prod_{x \subset \ \Sigma} \delta(\So (x))"=\int D[N] \ {\rm
exp}\left[ i\int \limits_{\Sigma}  N(x) \widehat {
S(x)}\right].\end{equation} A formal argument\cite{hart1,c6} shows
that $P$ can also be defined in a manifestly covariant manner as a
regularization of the formal path integral of general relativity
(in 2d gravity this is shown in\cite{a5}, here we will show that
this is the case in 3d gravity). In first order variables it
becomes
\begin{equation}\label{PI} P=\int \ D[e]\ D[A]\ \mu[A,e]\ {\rm
exp}\left[ i S_{\va GR}(e,A) \right]
\end{equation}
\vskip-.1cm \noindent where $e$ is the tetrad field, $A$ is the
spacetime connection, and $\mu[A,e]$ denotes the appropriate
measure.

In both cases, $P$ characterizes the space of solutions of {\em
quantum Einstein equations} as for any arbitrary state $|\phi>\in
\Hk$ then $P|\phi>$ is a (formal) solution of (\ref{QEE}).
Moreover, the matrix elements of $P$ define the physical inner
product ($<\ ,\ >_p$) providing the vector space of solutions of
(\ref{QEE}) with the Hilbert space structure that defines $\Hp$.
Explicitly
\[<s,s^{\prime}>_p:=<Ps,s^{\prime}>,\]
for $s,s^{\prime} \in \Hk$.

\begin{figure}[h]\!\!\!\!\!\!
\centerline{\hspace{0.5cm} \(
\begin{array}{c}
\includegraphics[height=6cm]{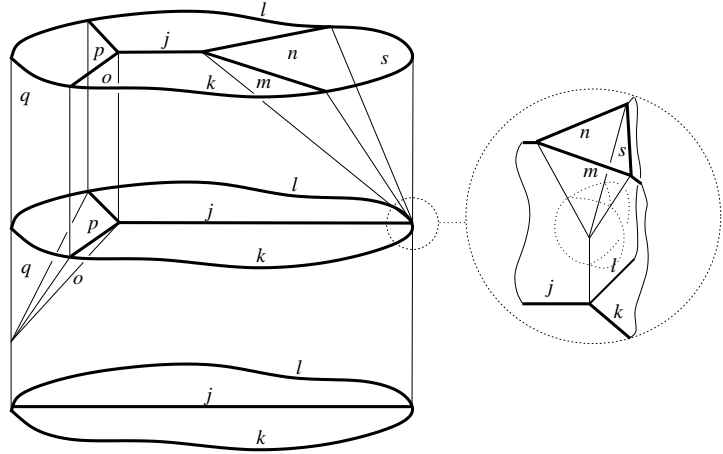}
\end{array}\) } \caption{A {\em spin foam} as the `colored' 2-complex
representing the transition between three different {\em spin
network} states. A transition vertex is magnified on the right.}
\label{spino}
\end{figure}

When these matrix element are computed in the {\em spin network}
basis (see Section \ref{GAUSS}), they can be expressed as a sum
over amplitudes of `{\em spin network} histories': {\em spin
foams} (Figure \ref{spino}). The latter are naturally given by
foam-like combinatorial structures whose basic elements carry
quantum numbers of geometry (see Section \ref{goo}). A {\em spin
foam} history\cite{baez7}, from the state $|s>$ to the state
$|s^{\prime}>$, is denoted by a pair $(F_{s\rightarrow
s^{\prime}}, \{j\})$ where $F_{s\rightarrow s^{\prime}}$ is the
2-complex with boundary given by the graphs of the {\em spin
network} states $|s^{\prime}>$ and $|s>$ respectively, and $\{j\}$
is the set of spin quantum numbers labelling its edges (denoted $e
\subset F_{s\rightarrow s^{\prime}}$) and faces (denoted $f
\subset F_{s\rightarrow s^{\prime}}$). Vertices are denoted $v
\subset F_{s\rightarrow s^{\prime}}$. The physical inner product
can be expressed as a sum over {\em spin foam} amplitudes
\begin{equation}\begin{array}{l}\label{SF}
<s^{\prime},s>_p = <Ps^{\prime}, s> = \\[3mm]
\qquad \sum \limits_{F_{s\rightarrow
s^{\prime}}}N(F_{s\rightarrow s^{\prime}})\sum\limits_{\{j\}} \
\prod_{f \subset F_{s\rightarrow s^{\prime}}} A_f(j_f) \prod_{e
\subset F_{s\rightarrow s^{\prime}}} A_e(j_e) \prod_{v\subset
F_{s\rightarrow s^{\prime}}} A_v(j_v),
\end{array}\end{equation}
where $N(F_{s\rightarrow s^{\prime}})$ is a (possible)
normalization factor, and $A_f(j_f)$, $A_e(j_e)$, and $A_v(j_v)$
are the 2-cell or face amplitude, the edge or 1-cell amplitude,
and the 0-cell of vertex amplitude respectively. These local
amplitudes depend on the spin quantum numbers labelling
neighboring cells in $F_{s\rightarrow s^{\prime}}$ (e.g. the
vertex amplitude of the vertex magnified in Figure \ref{spino} is
$A_v(j,k,l,m,n,s)$).

The underlying discreteness discovered in LQG is crucial: in the
{\em spin foam} representation, the functional integral for
gravity is replaced by a sum over amplitudes of combinatorial
objects given by foam-like configurations ({\em spin foams}) as in
(\ref{SF}). A {\em spin foam} represents a possible history of the
gravitational field and can be interpreted as a set of transitions
through different quantum states of space. Boundary data in the
path integral are given by the polymer-like excitations ({\em spin
network} states, Figure \ref{snb}) representing $3$-geometry
states in LQG.

\section{{\em Spin foams} in 3d quantum gravity}\label{pipo}

We introduce the concept of {\em spin foams} in a more explicit
way in the context of the quantization of three dimensional
Riemannian gravity. In Section \ref{SFH} and \ref{fcf} we will
present the definition of $P$ from the canonical and covariant
view point formally stated in the introduction by Equations
(\ref{P}) and (\ref{PI}) respectively\footnote{It is well known
that the physical inner product for 3d Riemannian gravity can
be defined using group averaging techniques\cite{Marolf:1997eb}, here
we review this  and use the approach to introduce the spin foam
representation\cite{a20,a21}.}. For other approaches to 3d quantum
gravity see the book of Carlip\cite{carlip}.

\subsection{The classical theory}

Riemannian gravity in $3$ dimensions is a theory with no local
degrees of freedom, i.e., a topological theory. Its action (in the
first order formalism) is given by
\begin{equation}\label{bfaction} S_{}(e,\omega)=\int
\limits_{ M}{\rm Tr}(e\wedge F(\omega)),
\end{equation}
where $M=\Sigma\times \R$ (for $\Sigma$ an arbitrary Riemann
surface), $\omega$ is an $SU(2)$-connection and the triad $e$ is
an $su(2)$-valued $1$-form. The gauge symmetries of the action are
the local $SU(2)$ gauge transformations
\begin{equation}\label{gauge1}
\delta e = \left[e,\alpha \right], \ \ \ \ \ \ \ \ \ \delta \omega
= d_{\omega} \alpha,
\end{equation}
where $\alpha$ is a ${{su(2)}}$-valued $0$-form, and the
`topological' gauge transformation
\begin{equation}\label{gauge2}
\delta e = d_{\va \omega} \eta, \ \ \ \ \ \ \ \ \ \delta \omega =
0,
\end{equation}
where $d_{\va \omega}$ denotes the covariant exterior derivative
and $\eta$ is a ${\tt su(2)}$-valued $0$-form. The first
invariance is manifest from the form of the action, while the
second is a consequence of the Bianchi identity, $d_{\va
\omega}F(\omega)=0$. The gauge symmetries are so large that all
the solutions to the equations of motion are locally pure gauge.
The theory has only global or topological degrees of freedom.

Upon the standard 2+1 decomposition, the phase space in these
variables is parametrized by the pull back to $\Sigma$ of $\omega$
and $e$. In local coordinates one can express them in terms of the
2-dimensional connection $A_a^{i}$ and the triad field
$E^b_j=\epsilon^{bc} e^k_c \eta_{jk}$ where $a=1,2$ are space
coordinate indices and $i,j=1,2,3$ are $su(2)$ indices. The
symplectic structure is defined by \be\{A_a^{i}(x),
E^b_j(y)\}=\delta_a^{\, b} \; \delta^{i}_{\, j} \;
\delta^{(2)}(x,y).\end{equation} Local symmetries of the theory are generated
by the first class constraints \be D_b E^b_j = 0, \ \ \
F_{ab}^i(A) = 0, \end{equation} which are referred to as the Gauss law and
the curvature constraint respectively---the quantization of these
is the analog of (\ref{QEE}) in 4d. This simple theory has been
quantized in various ways in the literature\cite{carlip}, here we
will use it to introduce the {\em spin foam} quantization.

\subsection{{\em Spin foams} from the Hamiltonian
formulation}\label{SFH}

The physical Hilbert space, $\Hp$, is defined by those `states' in
$\Hk$ that are annihilated by the constraints. As we discussed in
Section \ref{GAUSS}, spin network states solve the Gauss
constraint---$\widehat{ D_a E^a_i}|s>=0$---as they are manifestly
$SU(2)$ gauge invariant. To complete the quantization one needs to
characterize the space of solutions of the quantum curvature
constraints ($\widehat F^i_{ab}$), and to provide it with the
physical inner product. The existence of $\Hp$ is granted by the
following

\begin{theorem}
There exists a normalized positive linear form $P$ over $\rm Cyl$,
i.e. $P(\psi^{\star}\psi)\ge 0$ for $\psi\in {\rm Cyl}$ and
$P(1)=1$, yielding (through the GNS
construction\cite{Haag:1992hx}) the physical Hilbert space $\Hp$
and the physical representation $\pi_p$ of $\rm Cyl$.
\end{theorem}

The state $P$ contains a very large Gelfand ideal (set of zero
norm states) $J:=\{\alpha \in {\rm Cyl}\ \ {\rm s.t.} \ \
P(\alpha^{\star}\alpha)=0\}$. In fact the physical Hilbert space
$\Hp:={\overline{{\rm Cyl}/J}}$ corresponds to the quantization of
finitely many degrees of freedom. This is expected in 3d gravity
as the theory does not have local excitations (no `gravitons').
The representation $\pi_p$ of $\rm Cyl$ solves the curvature
constraint in the sense that for any functional $f_{\gamma}[A]\in
{\rm Cyl}$ defined on the sub-algebra of functionals defined on
contractible graphs $\gamma\subset \Sigma$, one has that
\be\pi_p[f_{\gamma}]\Psi=f_{\gamma}[0]\Psi.\end{equation} This equation
expresses the fact that `$\widehat F=0$' in $\Hp$ (for flat
connections parallel transport is trivial around a contractible
region). For $s,s^{\prime} \in \Hk$, the physical inner product is
given by \be\label{phyinn} <s,s^{\prime}>_p:=P(s^{\star}s), \end{equation}
where the $*$-operation and the product are defined in ${\rm
Cyl}$.
\begin{figure}[h]
\centerline{\hspace{0.5cm} \(
\begin{array}{c}
\includegraphics[width=10cm]{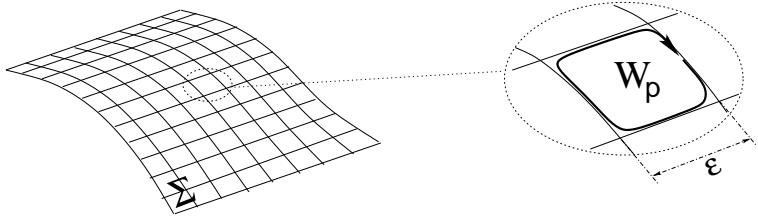}
\end{array}
\) }
\caption{Cellular decomposition of the space manifold $\Sigma$ (a
square lattice in this example), and the infinitesimal plaquette
holonomy $W_p[A]$.} \label{regu}
\end{figure}

The previous equation admits a `sum over histories'
representation\cite{a20}. We shall introduce the concept of the
{\em spin foam} representation as an explicit construction of the
positive linear form $P$ which, as in (\ref{P}), is formally given
by \be\PP=\int D[N] \ {\rm exp}(i\int
 \limits_{\Sigma} {\rm Tr}[ N \widehat{F}(A)])=\prod \limits_{x\subset \Sigma}
 \delta[\widehat {F(A)}],
\end{equation} where $N(x)\in {\rm su(2)}$. One can make the previous formal
expression a rigorous definition if one introduces a
regularization. Given a partition of $\Sigma$ in terms of
2-dimensional plaquettes of coordinate area $\epsilon^2$ one has
that \be \label{***} \int\limits_{\Sigma} {\rm Tr}[ N {F}(A)]=
\lim_{\epsilon\rightarrow 0}\ \sum_{p^i} \epsilon^2 {\rm
Tr}[N_{p^i} F_{p^i}],\end{equation} where $N_{p^i}$ and $F_{p^i}$ are values
of $N^i$ and $\epsilon^{ab}F_{ab}^i[A]$ at some interior point of
the plaquette $p^i$ and $\epsilon^{ab}$ is the Levi-Civita tensor.
Similarly the holonomy $W_{p^i}[A]$ around the boundary of the
plaquette $p^i$ (see Figure \ref{regu}) is given by \be
W_{p^i}[A]=\mathbbm{1}+ \epsilon^2 F_{p^i}(A)+{\cal
O}(\epsilon^2),\end{equation} where $F_{p^i}=\tau_j \epsilon^{ab}
F^j_{ab}(x_{p^i})$ ($\tau_j$ are the generators of $su(2)$ in the
fundamental representation). The previous two equations lead to
the following definition: given $s \in {\rm Cyl}$ (think of {\em
spin network}  state based on a graph $\gamma$) the linear form
$P(s)$ is defined as \be\label{new}
 \PP(s) := \lim_{\epsilon\rightarrow 0} \ \ <\Omega \prod_{p^i} \ \int \ dN_{p^i} \
 {\rm exp}(i {\rm Tr}[ N_{p^i} {W}_{p^i}]), s>.
\end{equation} where $<\ ,>$ is the inner product in the AL-representation
and $|\Omega>$ is the `vacuum' ($1\in {\rm Cyl}$) in the
AL-representation. The partition is chosen so that the links of
the underlying graph $\gamma$ border the plaquettes. One can
easily perform the integration over the $N_{p^i}$ using the
identity (Peter-Weyl theorem) \be\label{pw}\int \ dN \
 {\rm exp}(i {\rm Tr}[ N {W}])=\sum_{j} \ (2j + 1) \ {\rm Tr}[\stackrel{j}{\Pi}\!(W)].\end{equation}
Using the previous equation \be\label{final}
 \PP(s) := \lim_{\epsilon\rightarrow 0} \ \ \prod_{p^i} \sum_{j({\va p^i})}
 (2j({\vani p^i})+1)\ <\Omega\
{\rm Tr}[\stackrel{ j({\va p^i})}{\Pi}\!({W}_{p^i})]), s>, \end{equation}
where $j({\va p^i})$ is the spin labelling elements of the sum
(\ref{pw}) associated to the $ith$ plaquette. Since the ${\rm
Tr}[\stackrel{j}{\Pi}\!(W)]$ commute the ordering of
plaquette-operators in the previous product does not matter. It
can be shown that the limit $\epsilon \rightarrow 0$ exists and
one can give a closed expression of $P(s)$.

Some remarks are in order:

\vspace{.2cm} \noindent {\em Remark 1:} The argument
of the limit in (\ref{final}) satisfies the following
inequalities \ba && \nonumber \left|\sum_{j({\va p^i})}
 (2j({\vani p^i})+1)\
 \mu_{AL}\left( \prod_{p^i} \chi_{j({\va p^i})}({W}_{p^i}[A])\ \overline{ s[A]} s^{\prime}[A]\right)\right|\le \\ \nonumber &&\le C \left|\sum_{j_{\va
p^i}}
 (2j({\vani p^i})+1)\
 \mu_{AL}\left( \prod_{p^i} \chi_{j({\va p^i})}({W}_{p^i}[A])\right)\right|=\\ &&= C \sum_{j}
 (2j+1)^{2-2g},\  \ea
where we have used (\ref{kip}),  $C$ is a real positive
constant, and the last equation follows immediately from the
definition of the Ashtekar-Lewandowski measure $\mu_{AL}$
\cite{a5}. The convergence of the sum for  genus $g\ge 2$ follows
directly.

\vspace{.2cm} \noindent {\em Remark 2:} The case of the sphere
$g=0$ is easy to regularize. In this case (\ref{final}) diverges
simply because of a redundancy in the product of delta
distributions in the notation of (\ref{twenty}). This is a
consequence of the discrete analog of the Bianchi identity. It is
easy to check that eliminating a single arbitrary plaquette
holonomy from the product in  (\ref{final}) makes $P$ well defined
and produces the correct (one dimensional) $\Hp$.

The case of the torus $g=1$ is more subtle; in fact our prescription must be modified in that case\cite{carlip}.

 \vspace{.2cm}

\noindent {\em Remark 3:} It is immediate  to see that
(\ref{final}) satisfies hermitian condition \be\ssp=\sspb.\end{equation}

\vspace{.2cm}

\noindent {\em Remark 4:} The positivity condition also follows
from the definition $\sspe\ \ge 0$.

\begin{figure}
\centerline{\hspace{0.5cm} \( {\rm Tr}[\stackrel{k}{\Pi}\!(W_{p})]
\rhd \!\!\!\!\!\!\!\!\!\!\!\!\!\!\!\!\begin{array}{c}
\includegraphics[width=3.6cm]{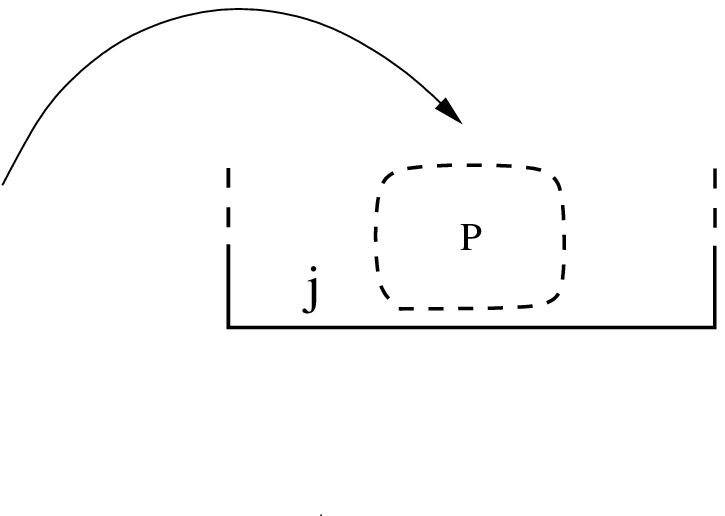}
\end{array}
=
\begin{array}{c}
\includegraphics[width=2.5cm]{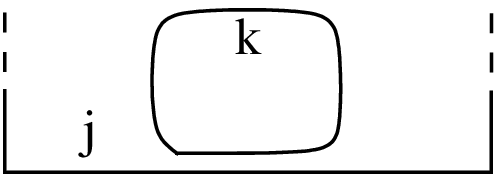}
\end{array}=\sum \limits_{m} N_{j,m,k}
\begin{array}{c}\includegraphics[width=2.5cm]{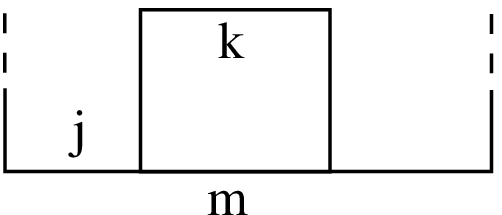}
\end{array}
\) }
\caption{Graphical notation representing the action of one
plaquette holonomy on a {\em spin network} state. On the right is
the result written in terms of the {\em spin network} basis. The
amplitude $N_{j,m,k}$ can be expressed in terms of Clebsch-Gordan
coefficients.} \label{pito}
\end{figure}

\begin{figure}
\centerline{\hspace{0.5cm} \( {\rm
Tr}[\stackrel{n}{\Pi}\!(W_{p})]\rhd
\!\!\!\!\!\!\!\!\!\!\!\!\!\!\!\!\begin{array}{c}
\includegraphics[width=2.5cm]{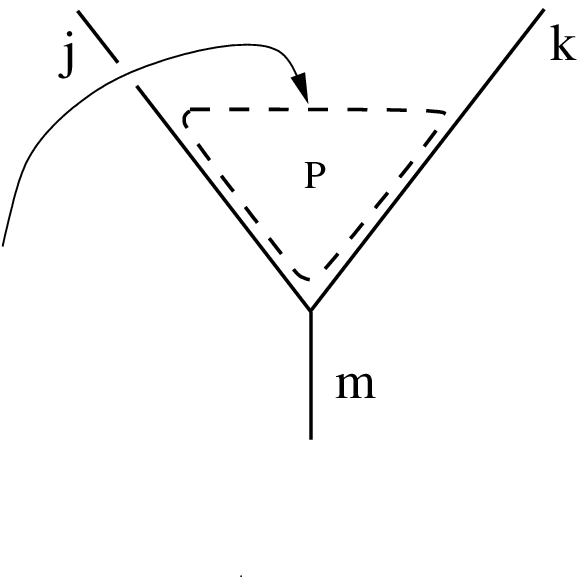}
\end{array}
=
\begin{array}{c}
\includegraphics[width=2.5cm]{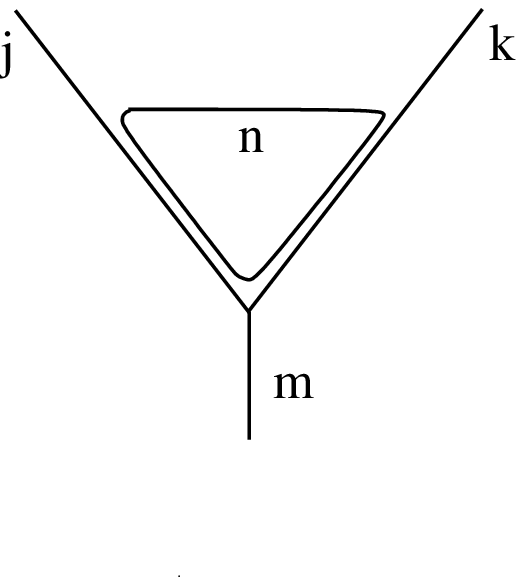}
\end{array}=\sum\limits_{o,p} \frac{1}{\Delta_n \Delta_j \Delta_k \Delta_m}
\left\{\begin{array}{ccc}j\ \ k \ \ m\\ n\ \  o\ \  p
\end{array}\right\}
\begin{array}{c}
\includegraphics[width=2.5cm]{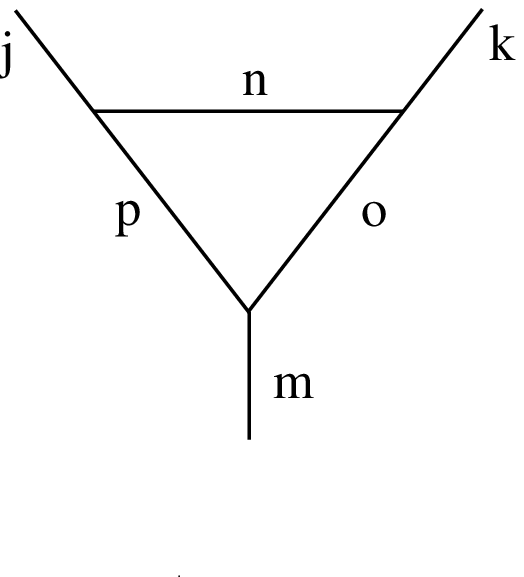}
\end{array}
\) }
\caption{Graphical notation representing the action of one
plaquette holonomy on a {\em spin network} vertex. The object in
brackets ($\{\}$) is a $6j$-symbol and $\Delta_j:=2j+1$.}
\label{pitolon}
\end{figure}

\begin{figure}[h!!!!!!!!!!!!!!!!!!!!!!!!!!!!!!!!!!!!!!!!!!!!!!!!!!!!!!!!!!!!!!!!!!!]
 \centerline{\hspace{0.5cm}\(
\begin{array}{ccc}
\includegraphics[height=2.5cm]{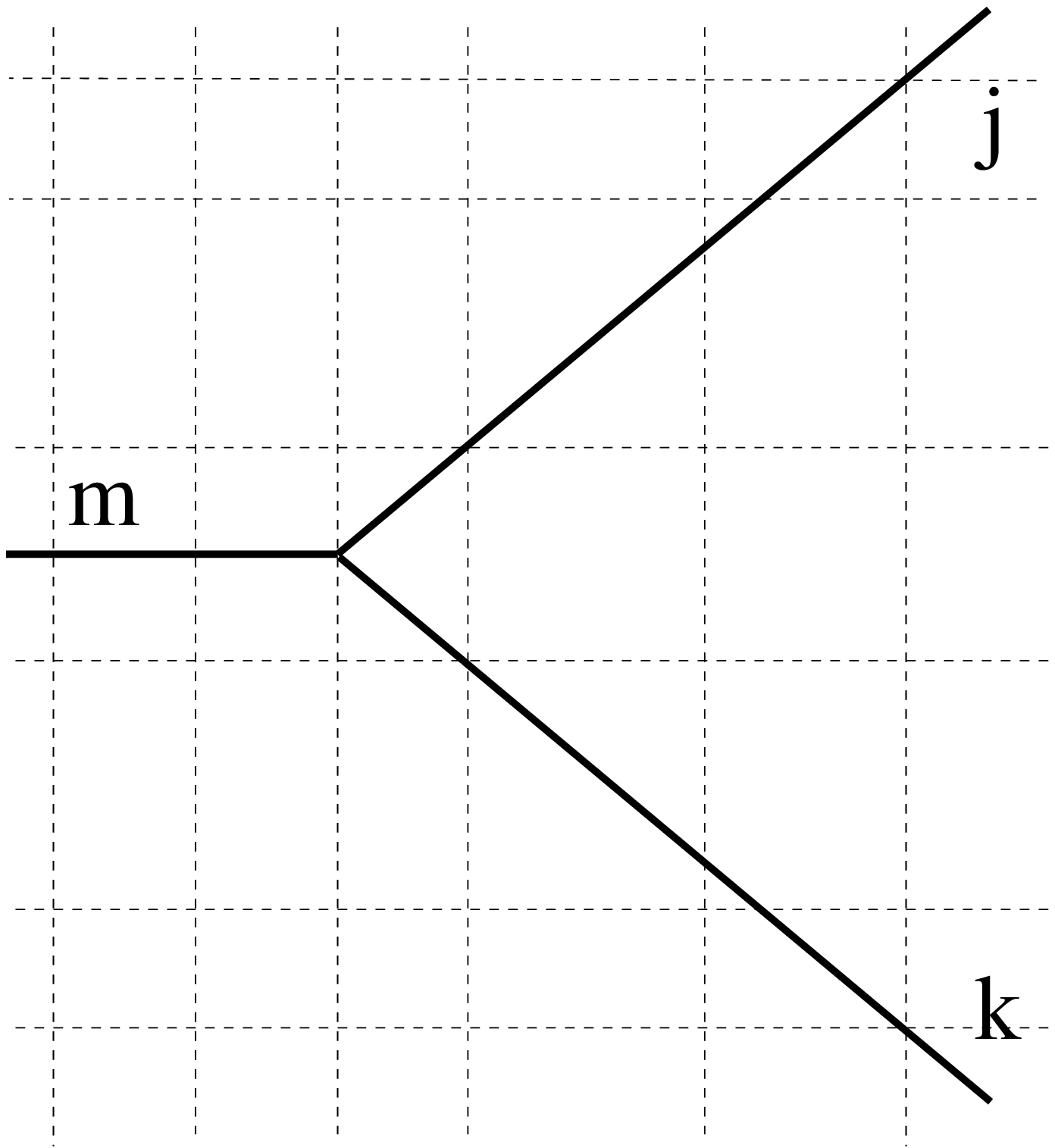} \\
\includegraphics[height=2.5cm]{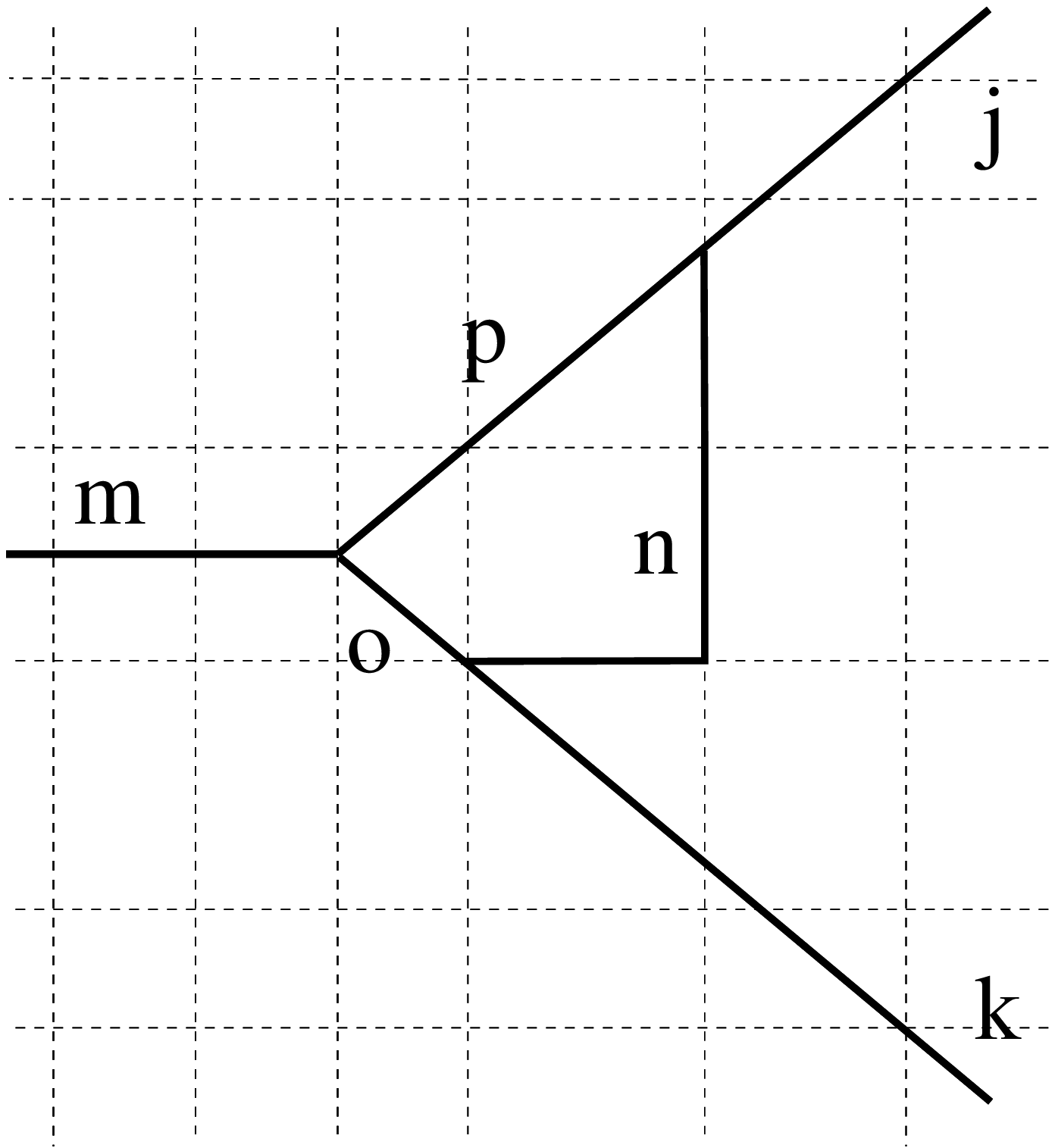} \\
\includegraphics[height=2.5cm]{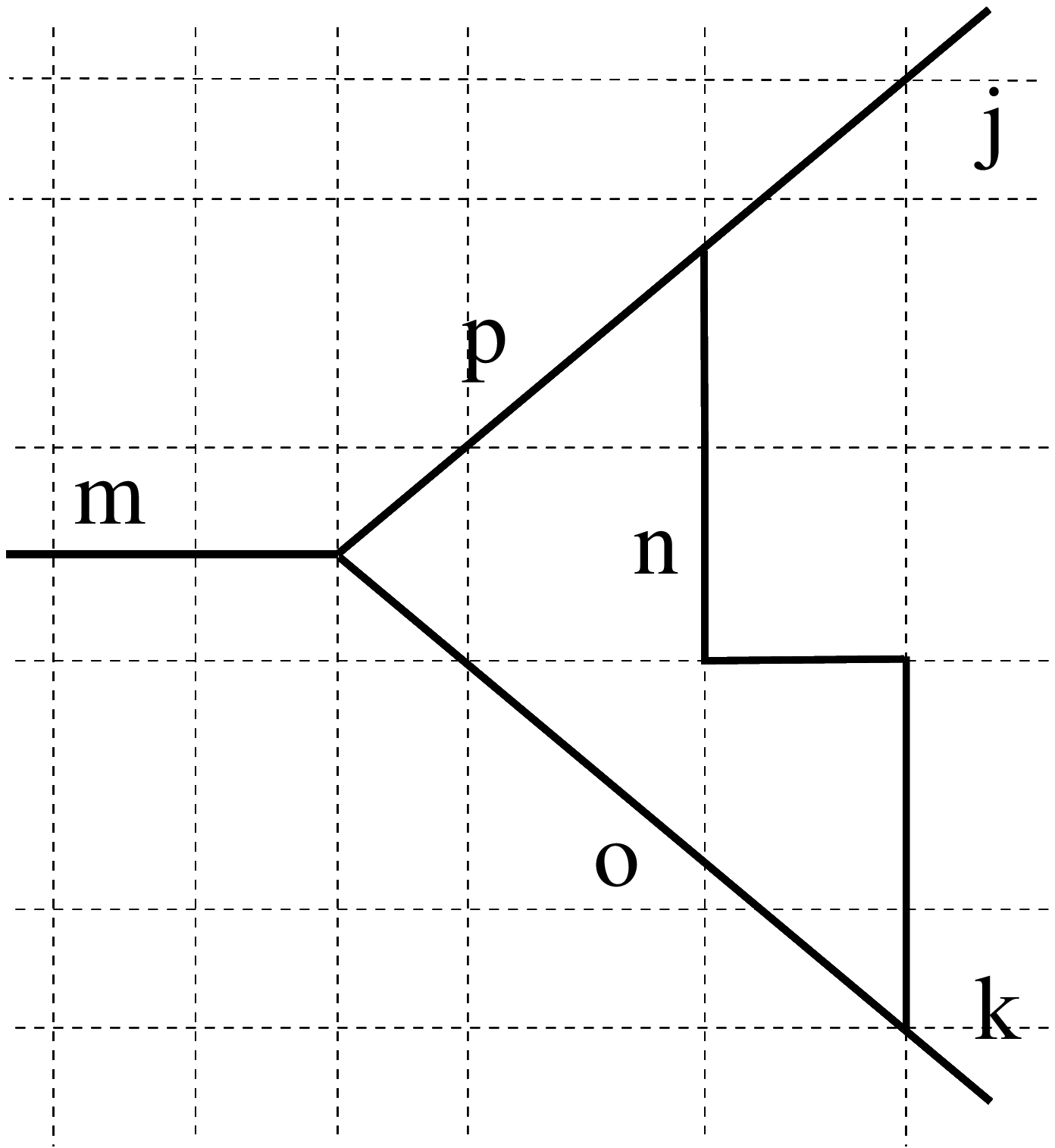}
\end{array}
\begin{array}{c}
\includegraphics[height=2.5cm]{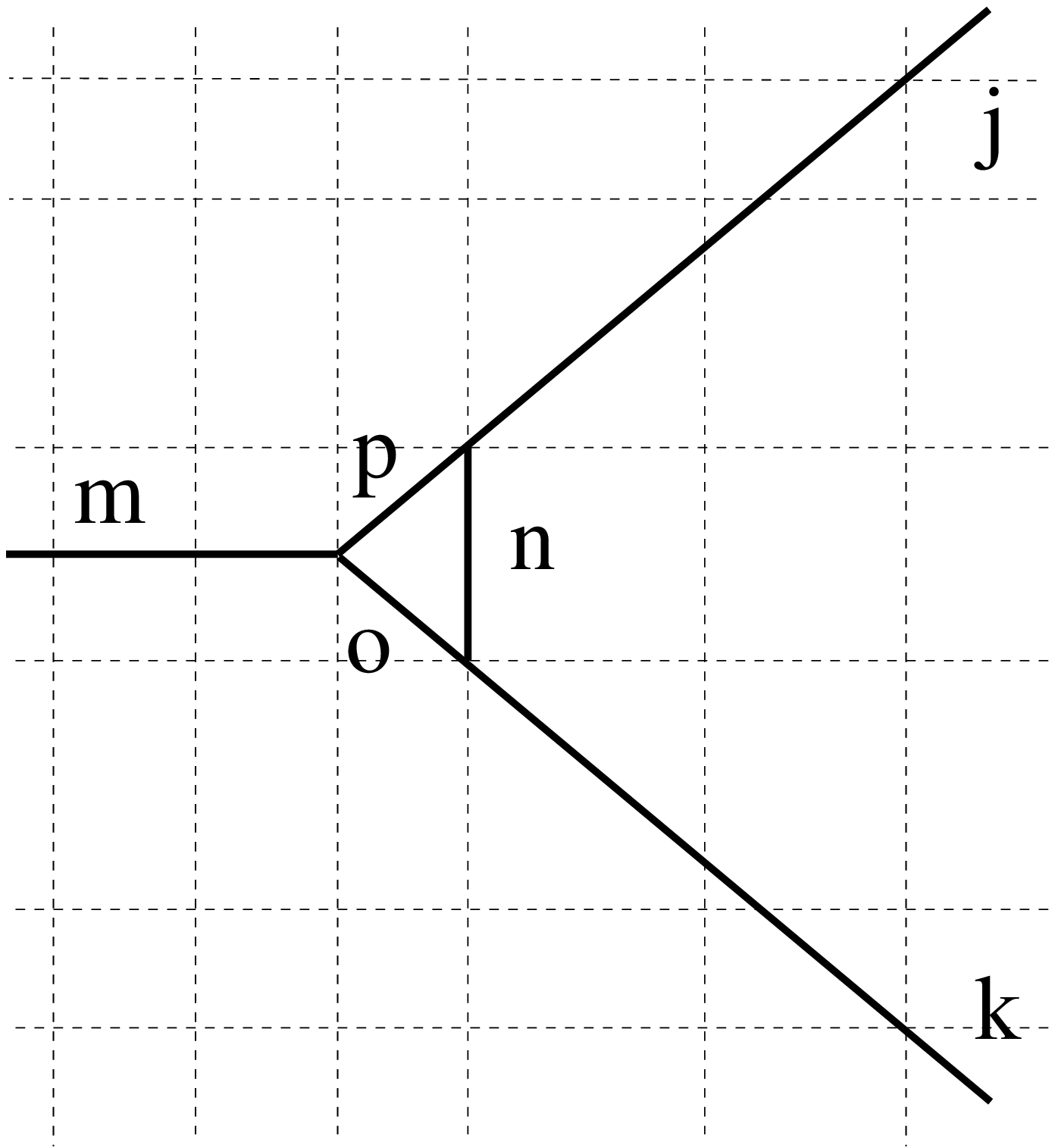}\\
\includegraphics[height=2.5cm]{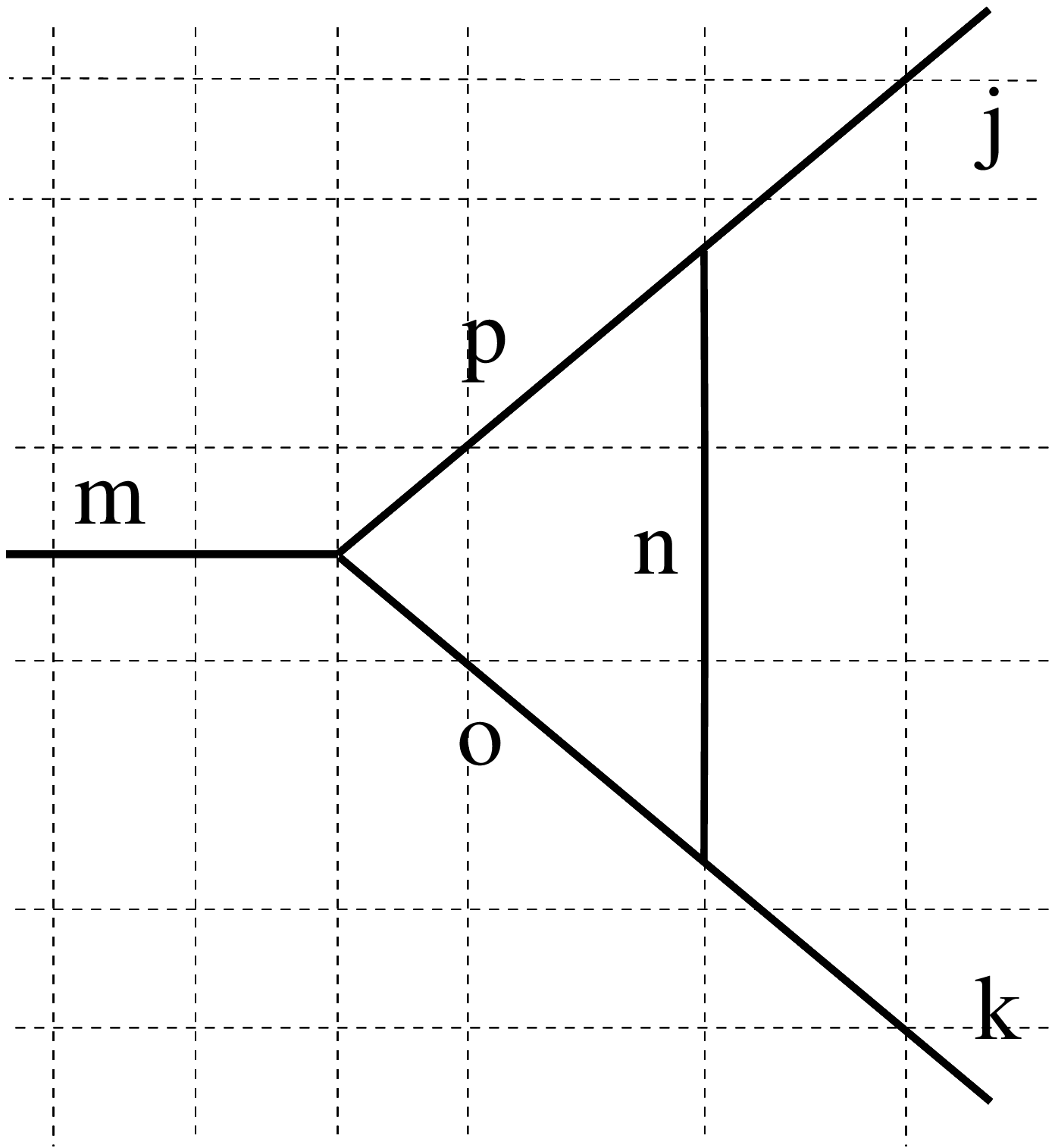}\\
\includegraphics[height=2.5cm]{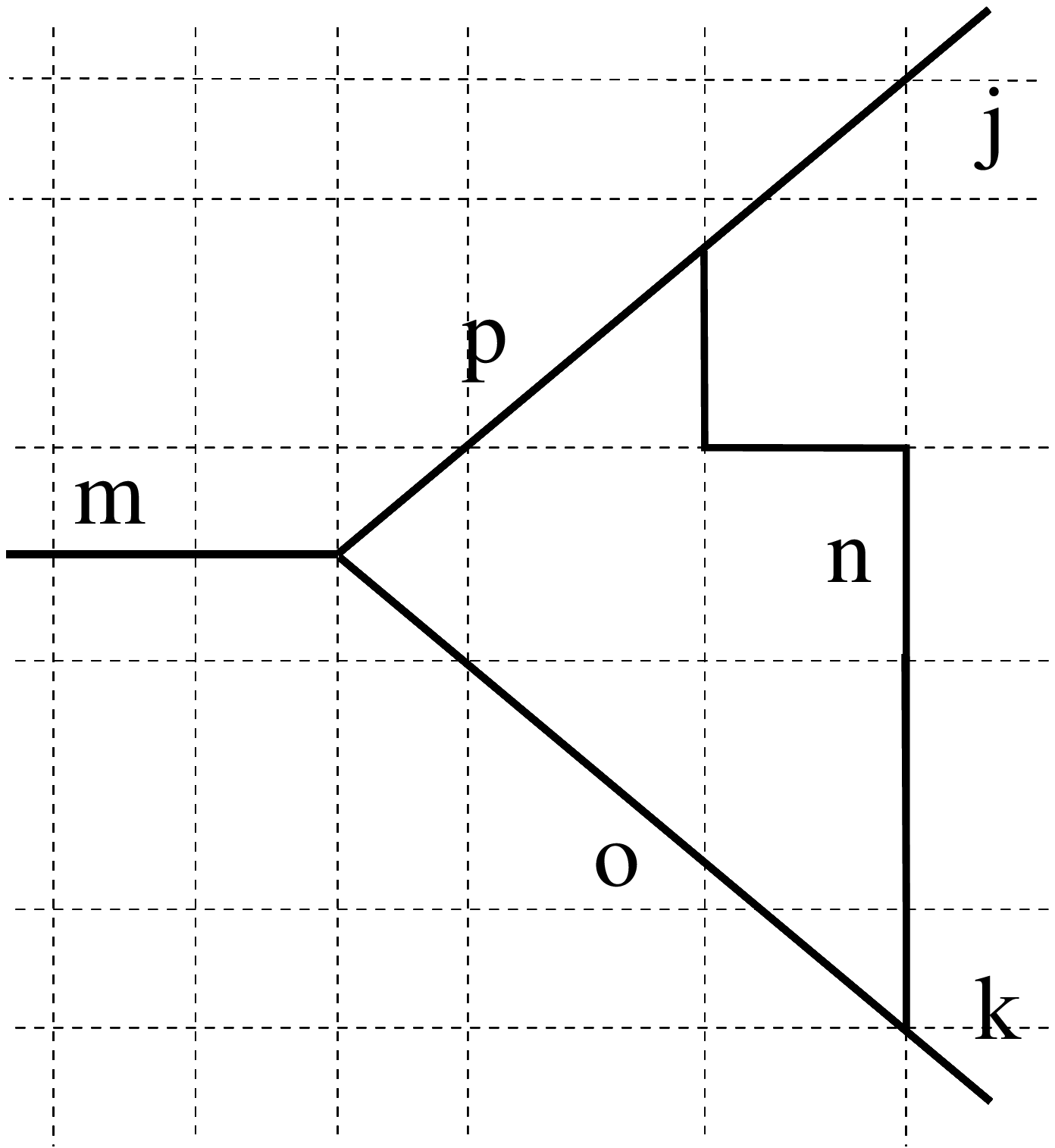}
\end{array}
\begin{array}{c}
\includegraphics[height=2.5cm]{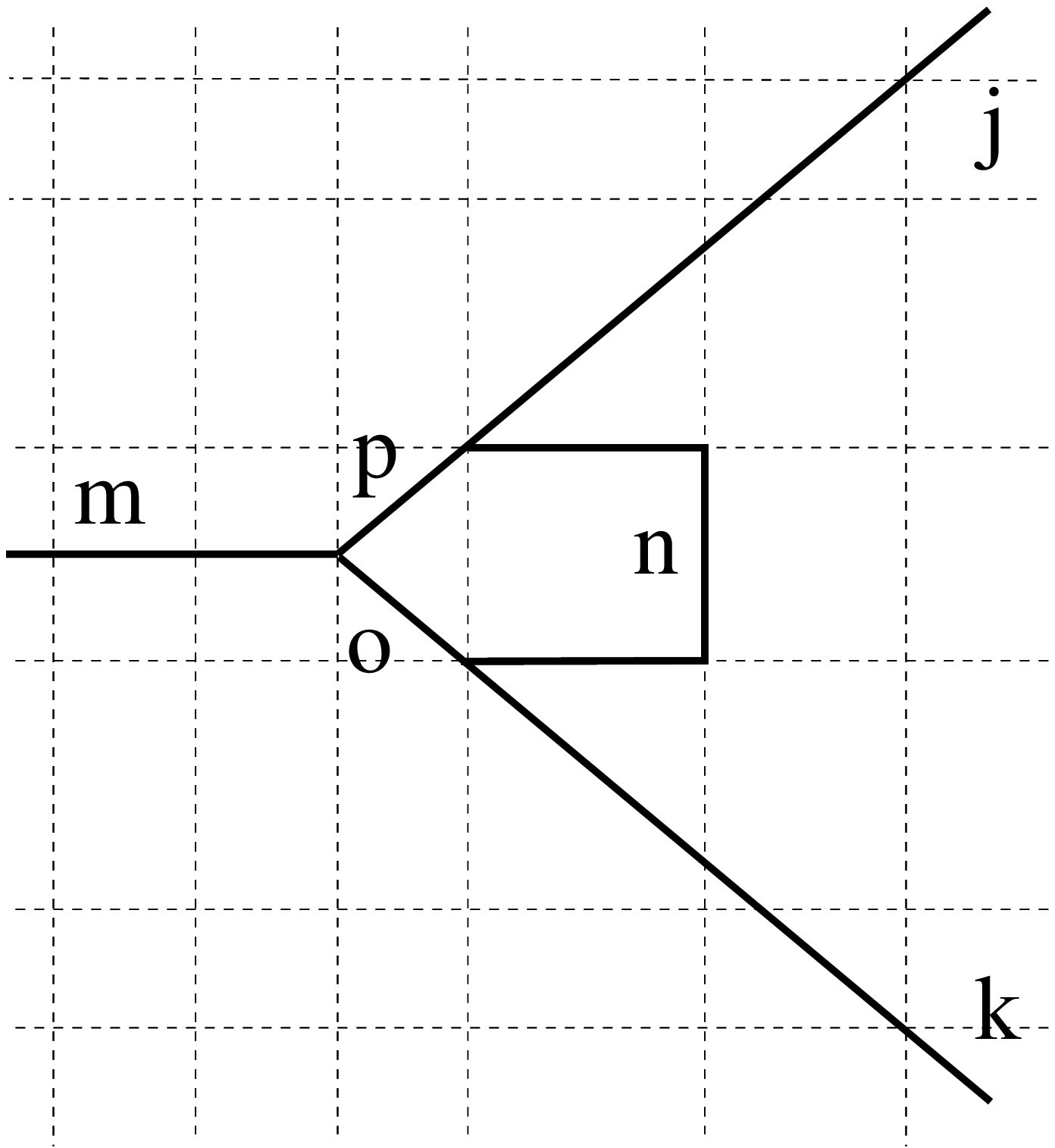}\\
\includegraphics[height=2.5cm]{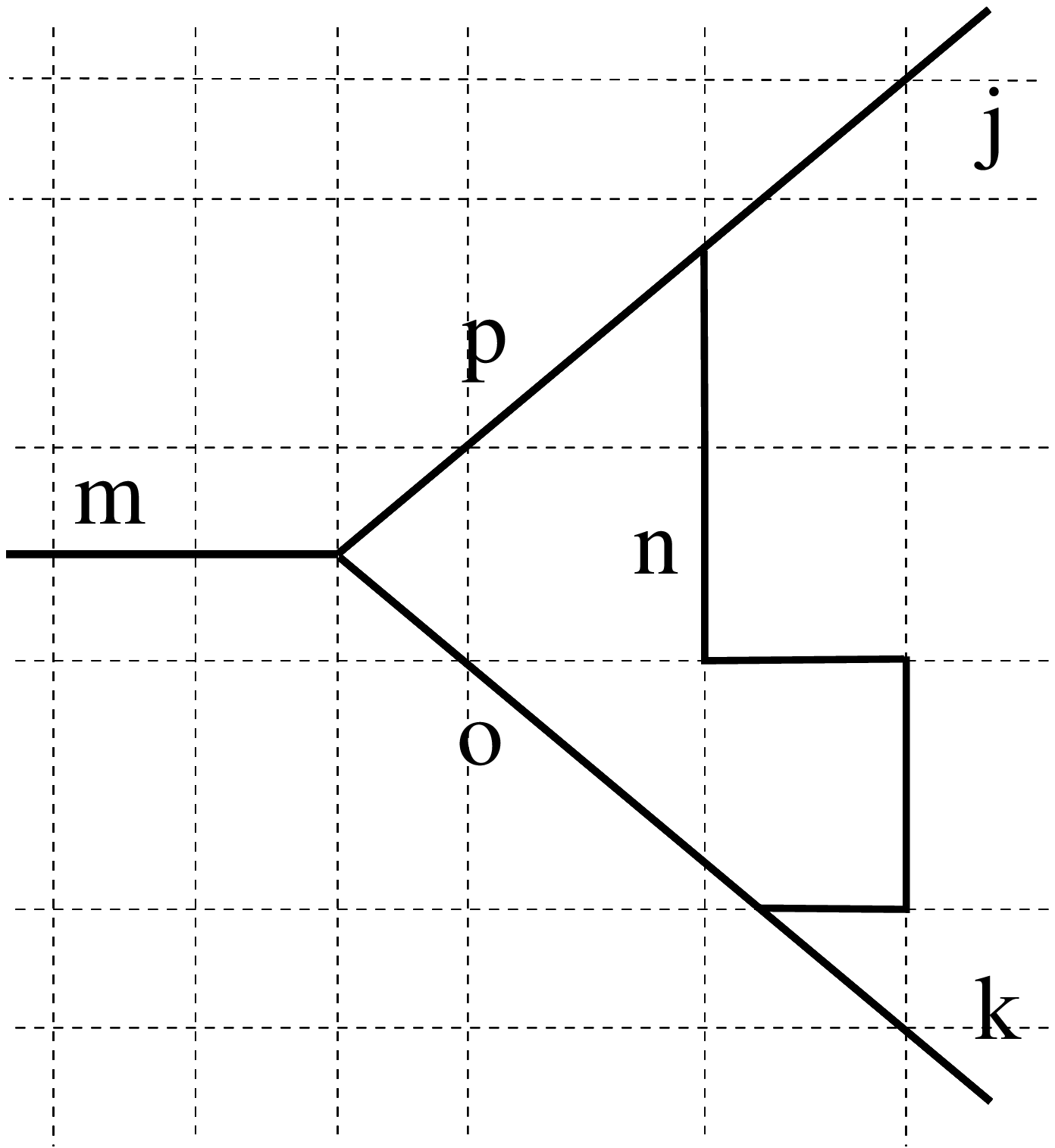}\\
\includegraphics[height=2.5cm]{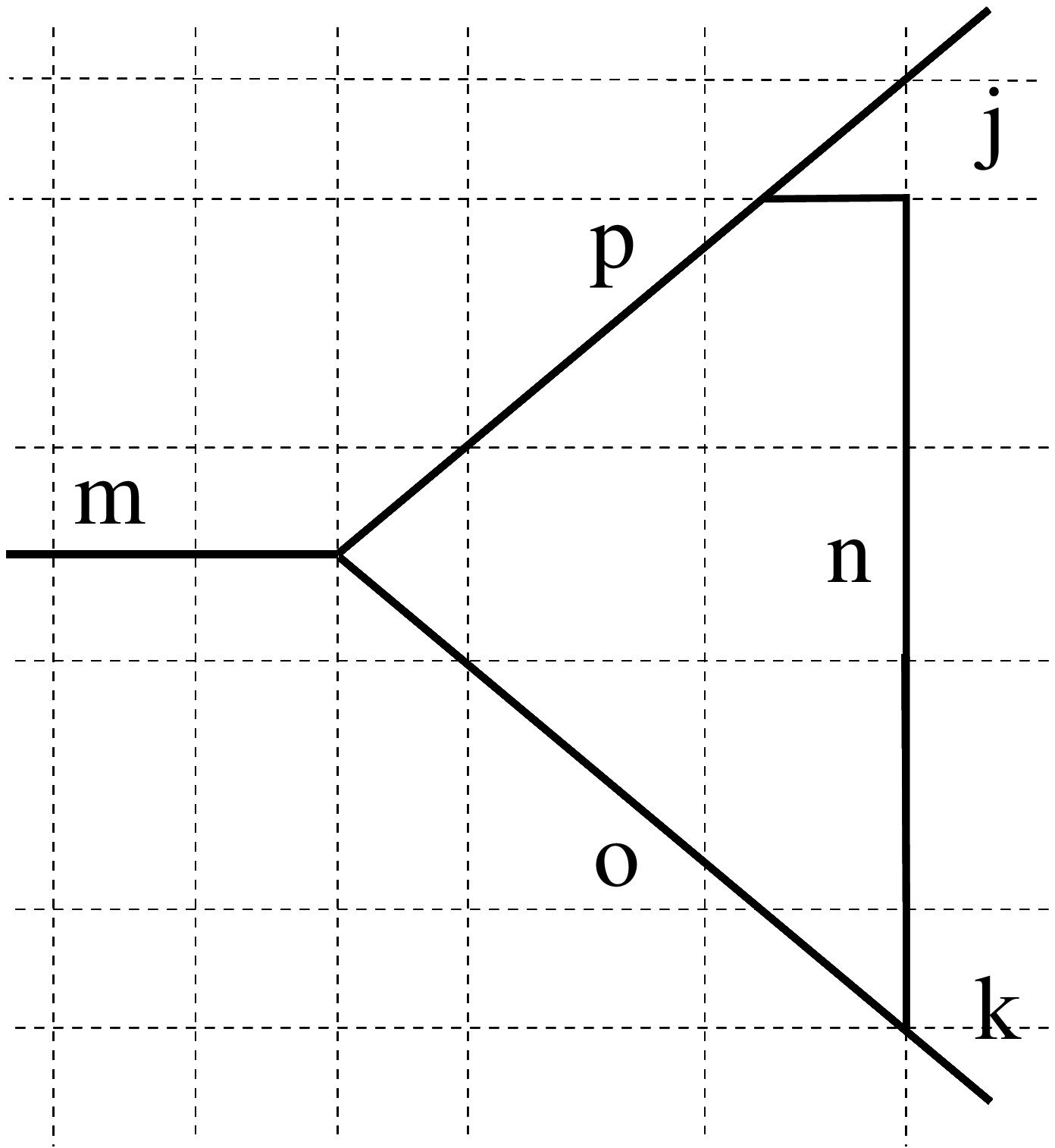}
\end{array}\begin{array}{c}
\includegraphics[height=.5cm]{flecha.eps}
\end{array}
\begin{array}{c}
\includegraphics[height=5cm]{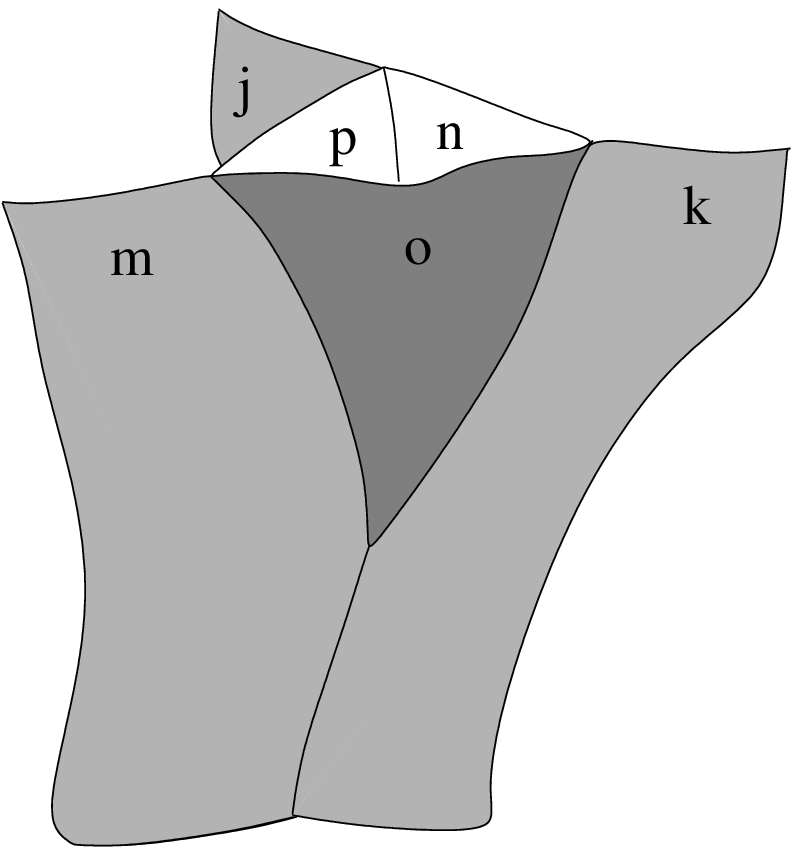}
\end{array}
\)} \caption{A set of discrete transitions representing one of the
contributing histories at a fixed value of the regulator. On the
right, the continuous {\em spin foam} representation when the
regulator is removed.} \label{vani}
\end{figure}

Now in the AL-representation, each ${\rm
Tr}[\stackrel{j({\va p^{i}})}{\Pi}\!(W_{p^{i}})]$ acts by creating
a closed loop in the $j_{p^{i}}$ representation at the boundary of
the corresponding plaquette (Figures \ref{pito} and
\ref{pitolon}).
One can introduce a (non-physical) time parameter that works
simply as a coordinate providing the means of organizing the
series of actions of plaquette loop-operators in (\ref{final});
i.e., one assumes that each of the loop actions occur at different
`times'.
\begin{figure}[h!!!!!]
 \centerline{\hspace{0.5cm}\(
\begin{array}{ccc}
\includegraphics[height=1.9cm]{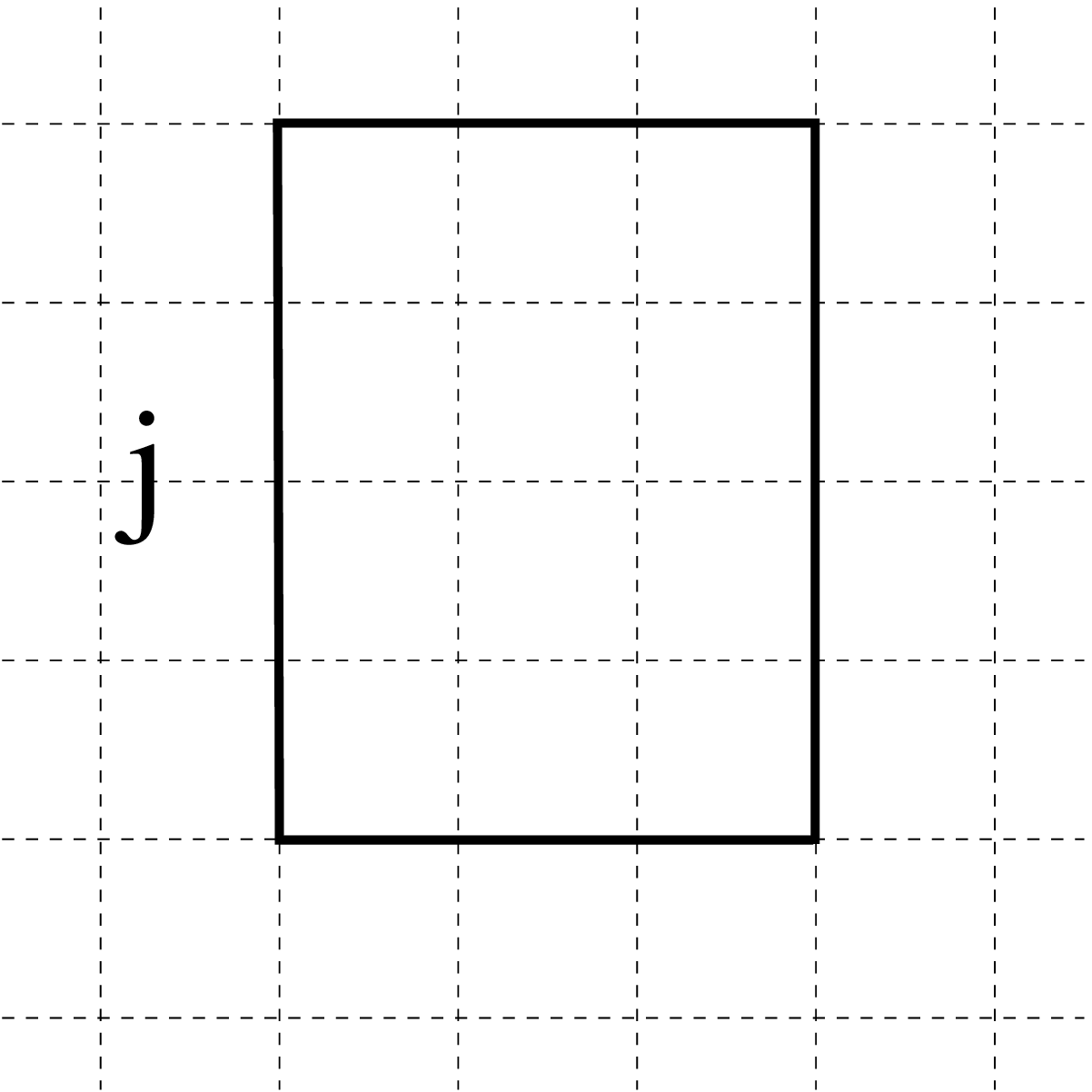} \\
\includegraphics[height=1.9cm]{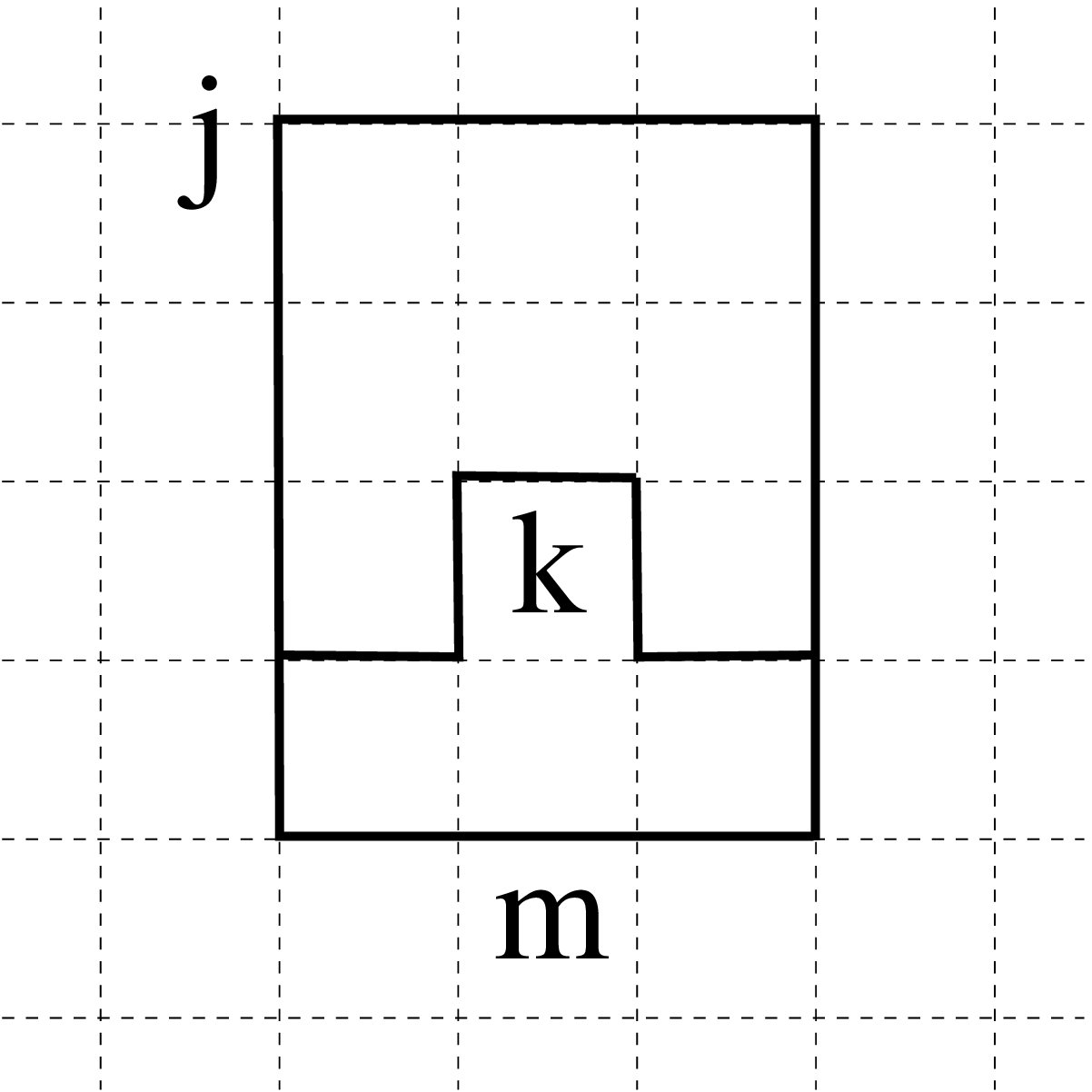}\\
\includegraphics[height=1.9cm]{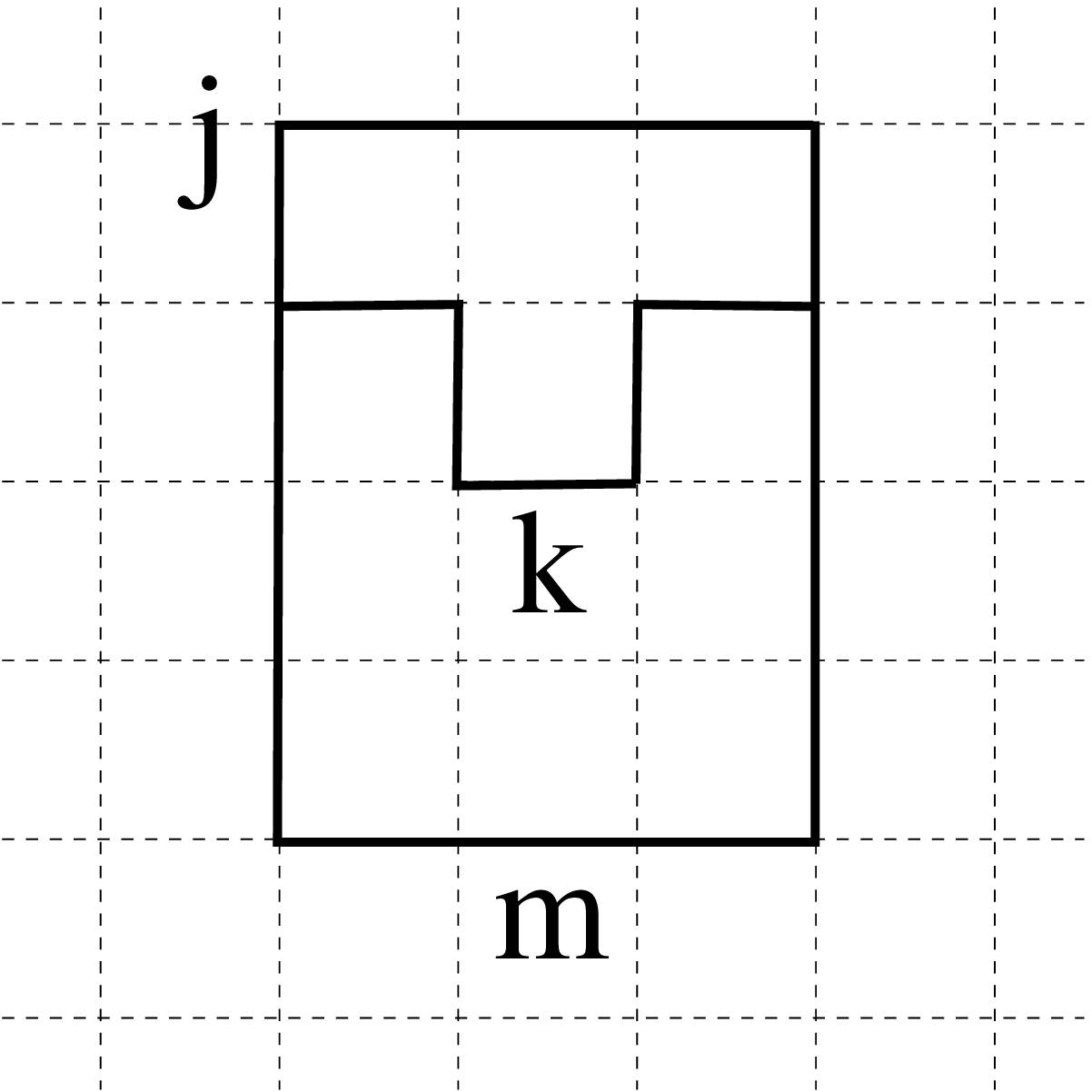}
\end{array}
\begin{array}{ccc}
\includegraphics[height=1.9cm]{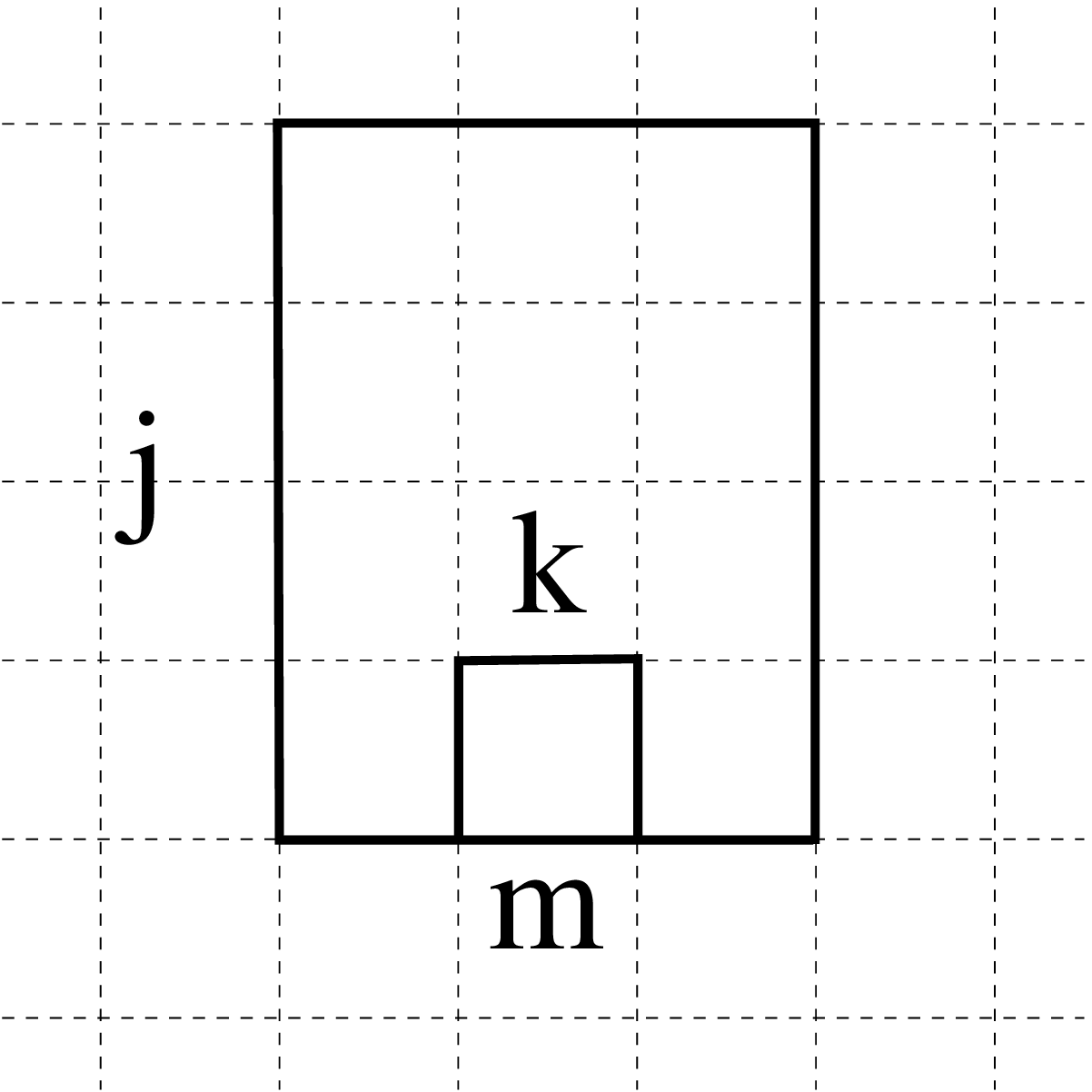} \\
\includegraphics[height=1.9cm]{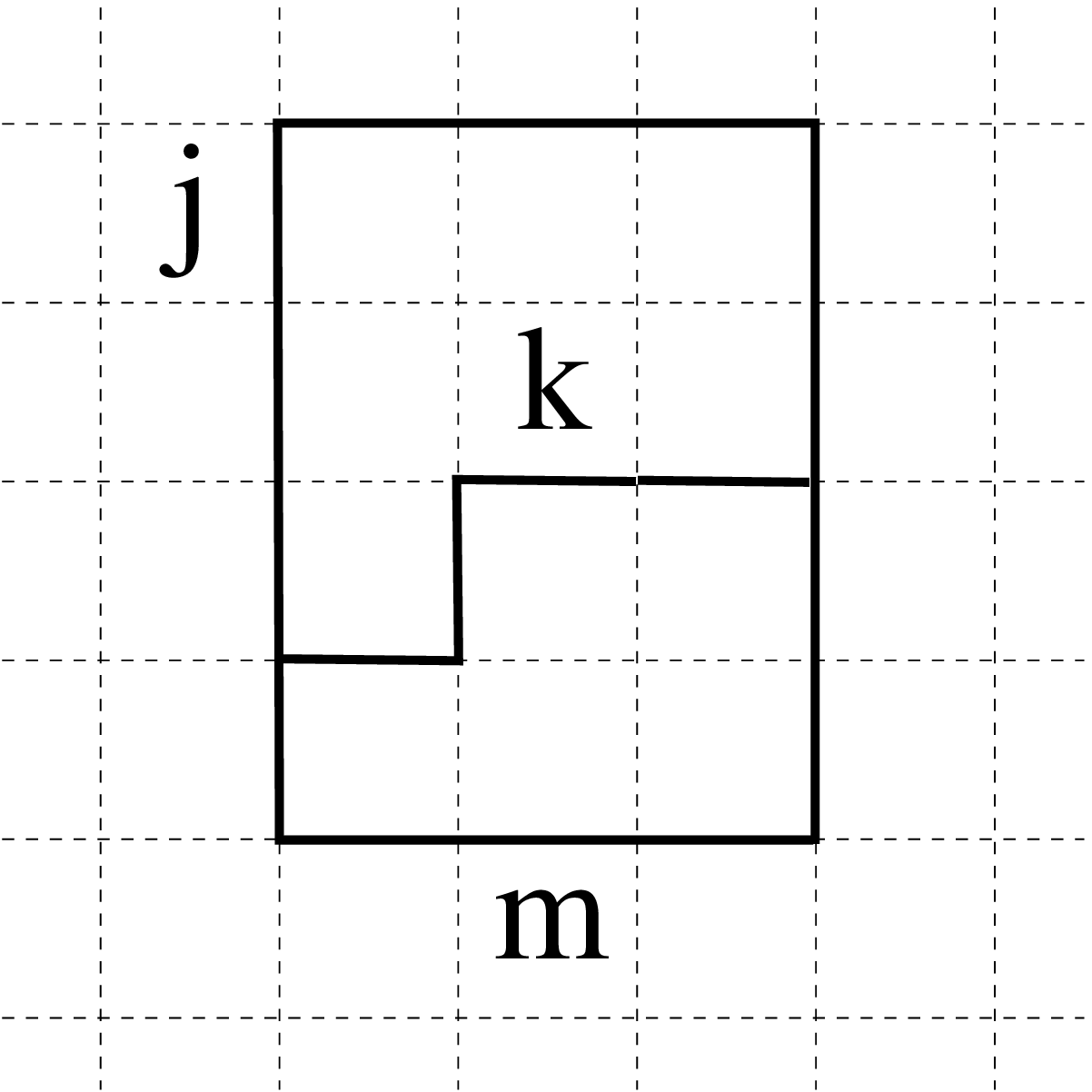}\\
\includegraphics[height=1.9cm]{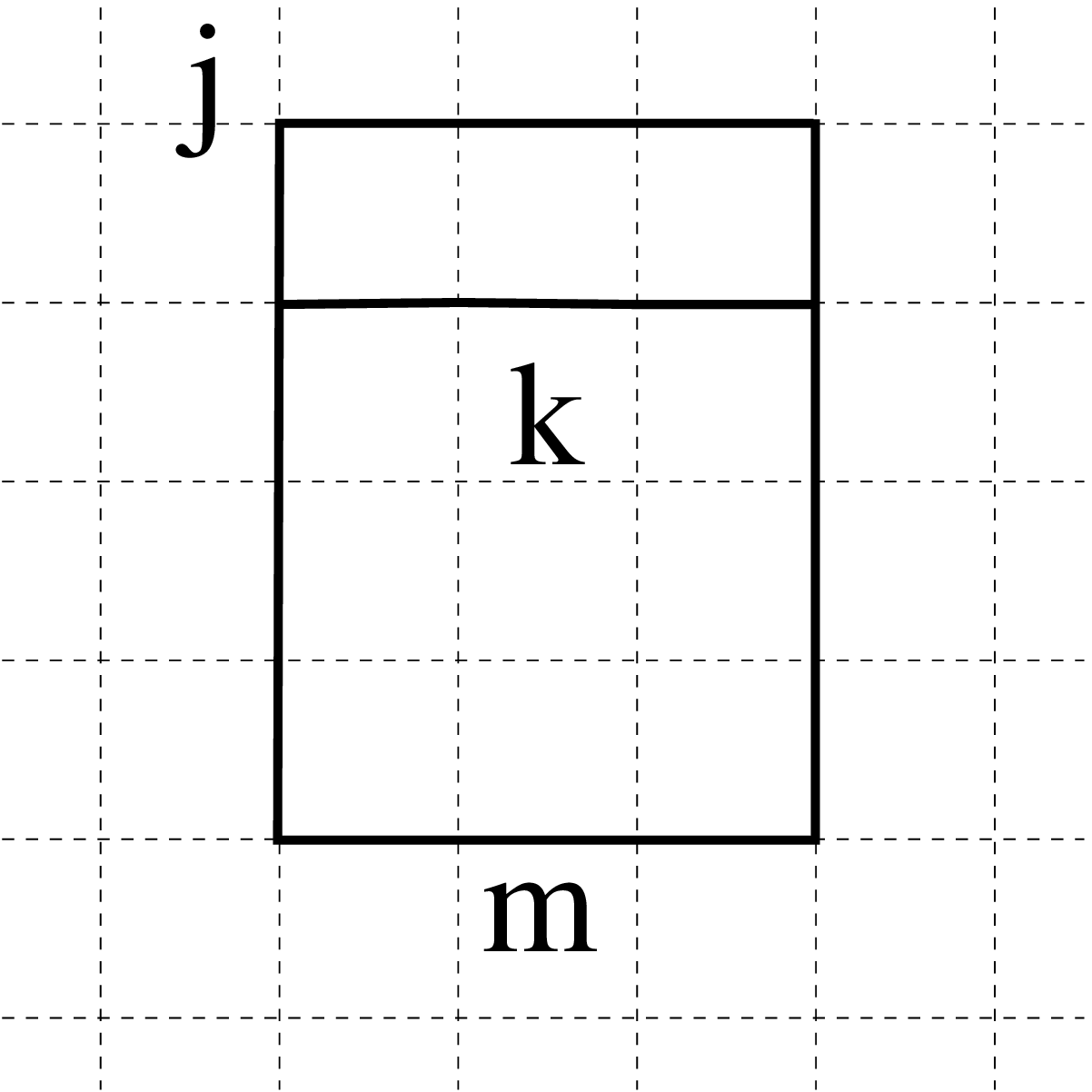}
\end{array}
\begin{array}{ccc}
\includegraphics[height=1.9cm]{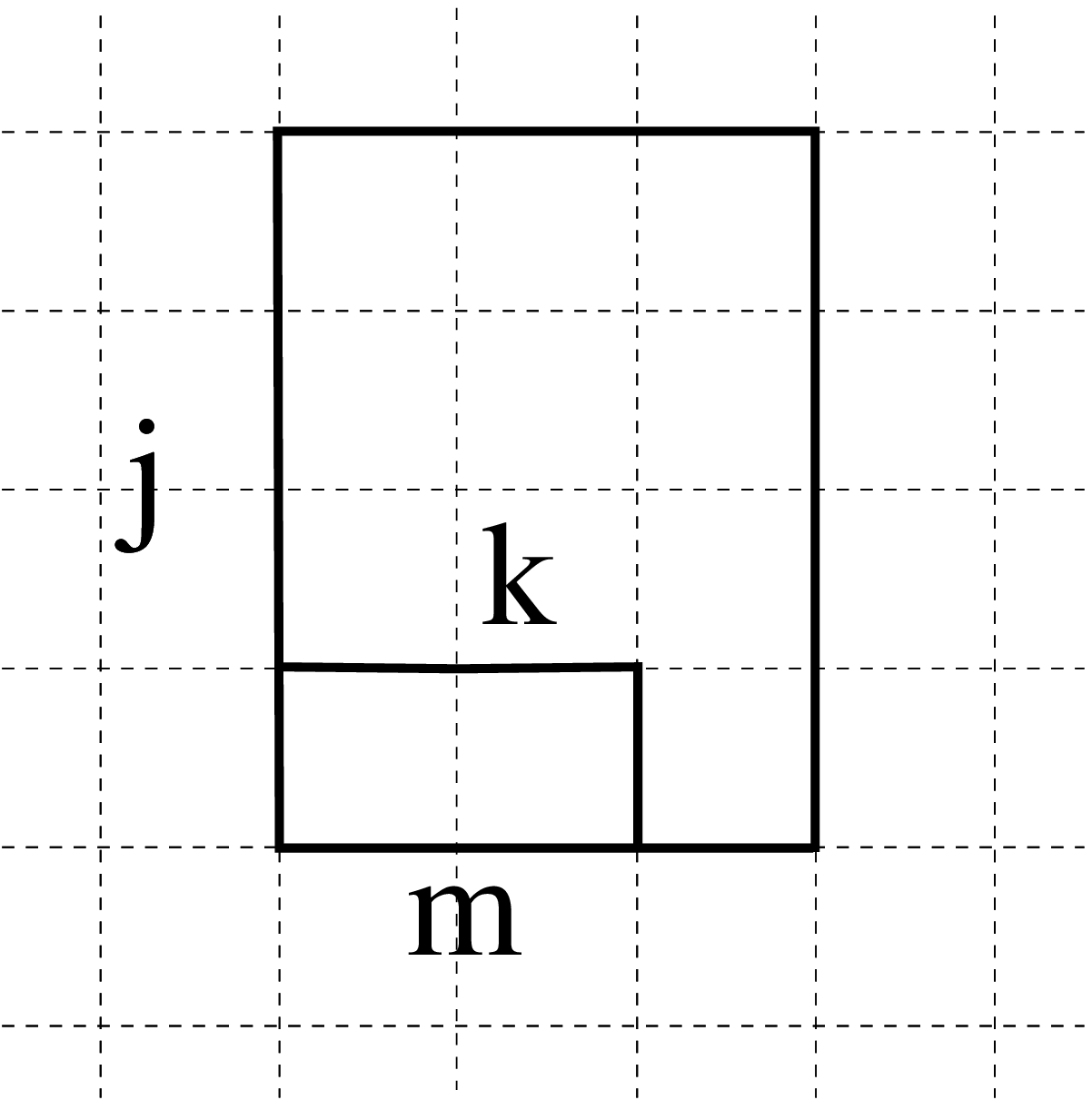} \\
\includegraphics[height=1.9cm]{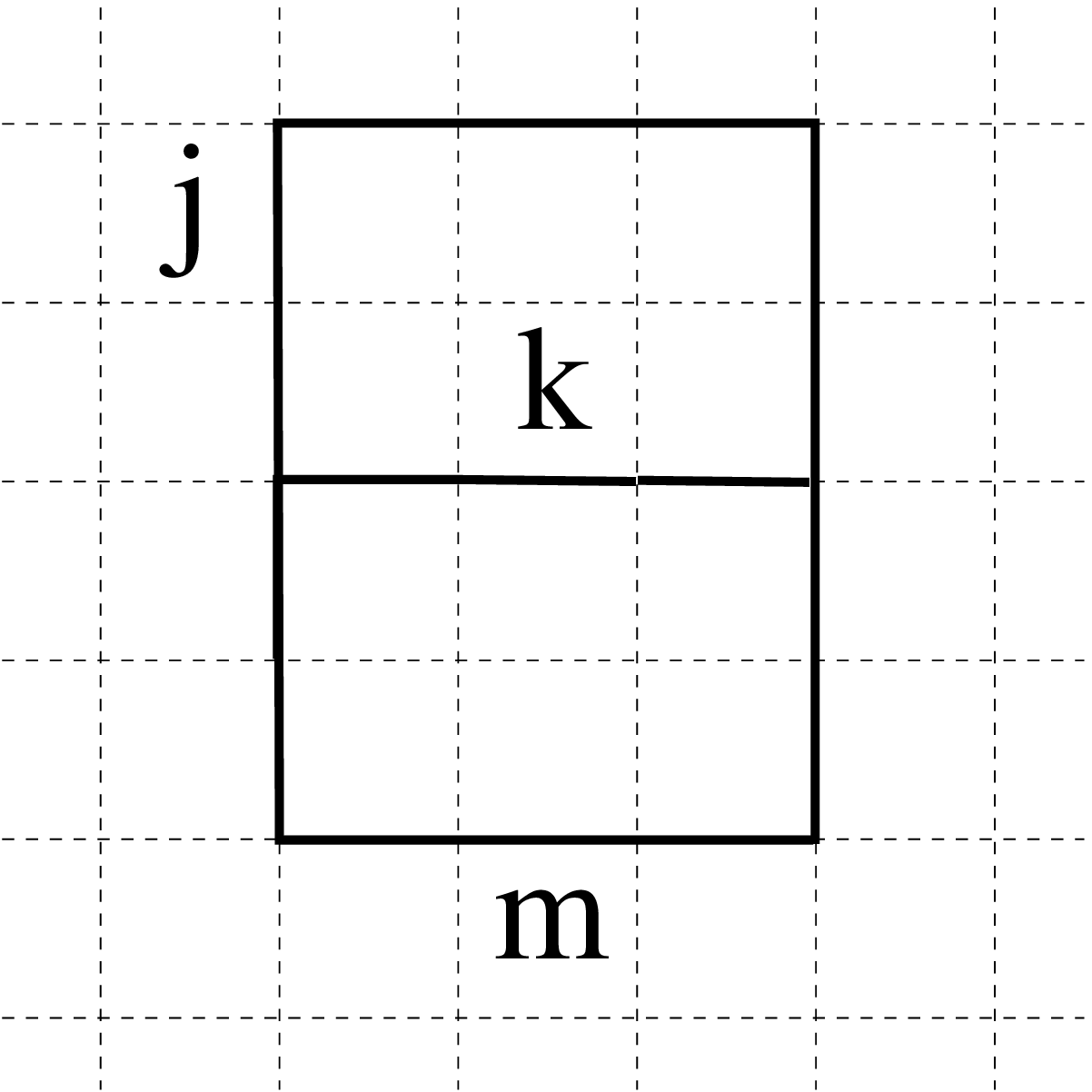}\\
\includegraphics[height=1.9cm]{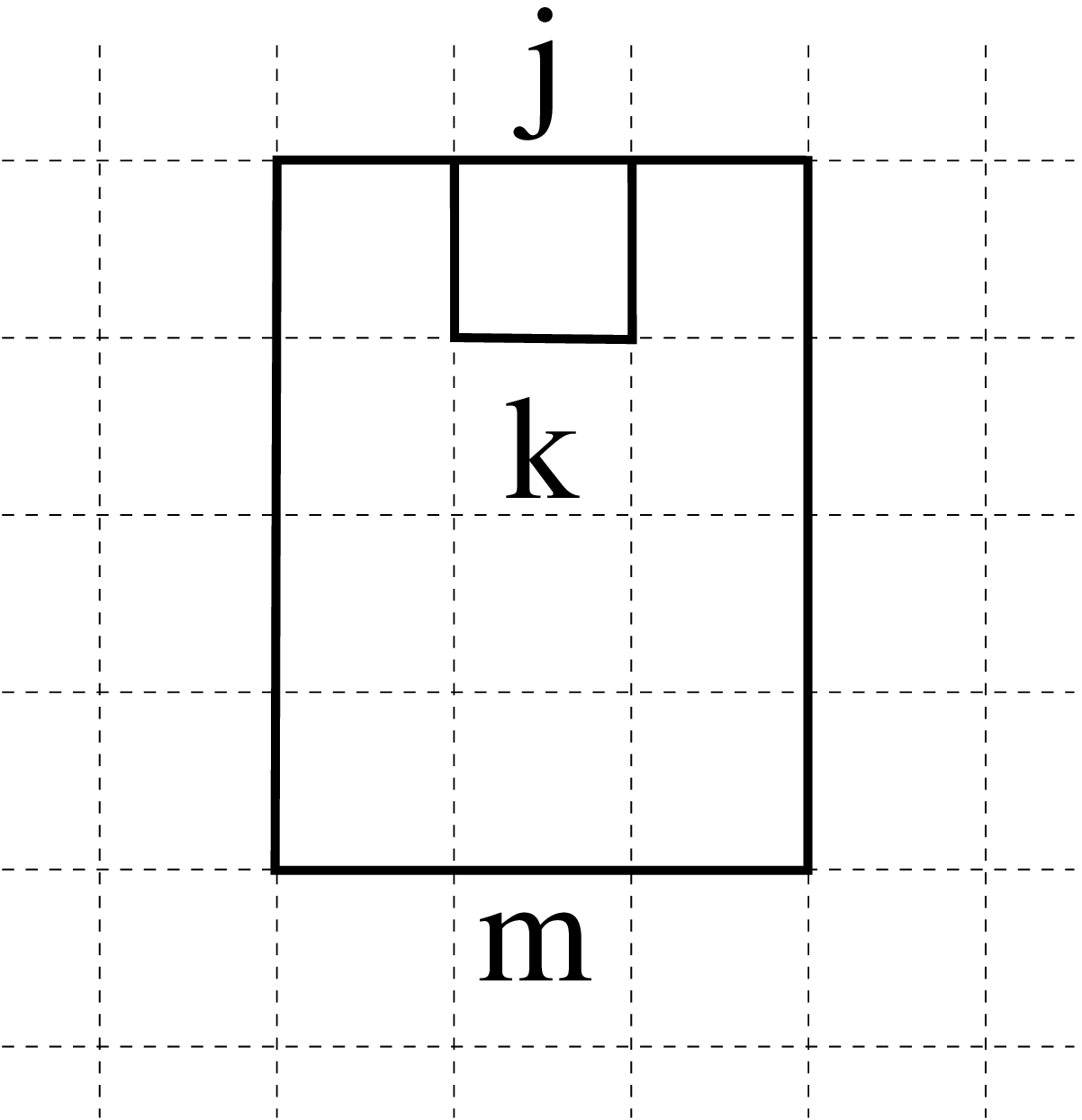}
\end{array}
\begin{array}{ccc}
\includegraphics[height=1.9cm]{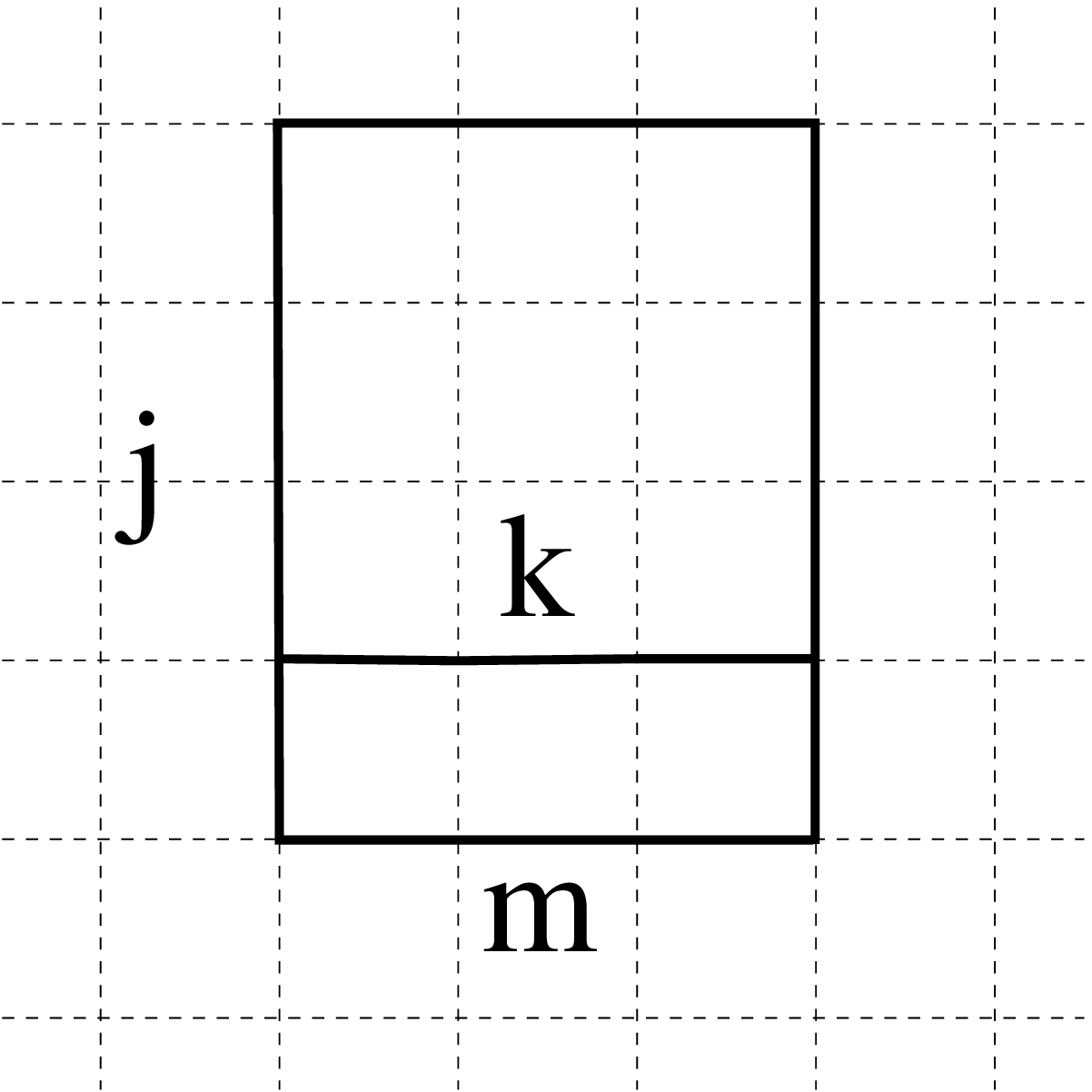} \\
\includegraphics[height=1.9cm]{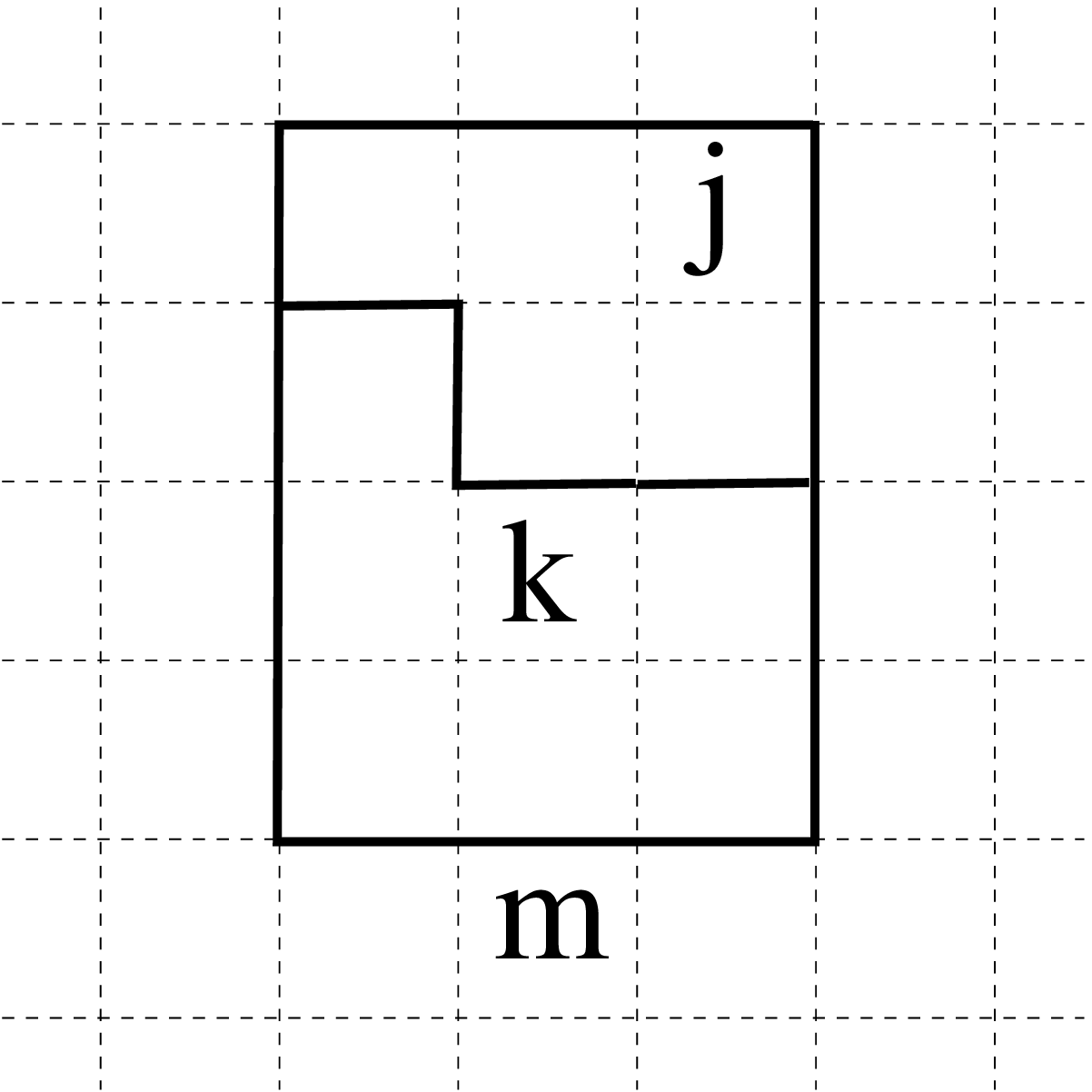}\\
\includegraphics[height=1.9cm]{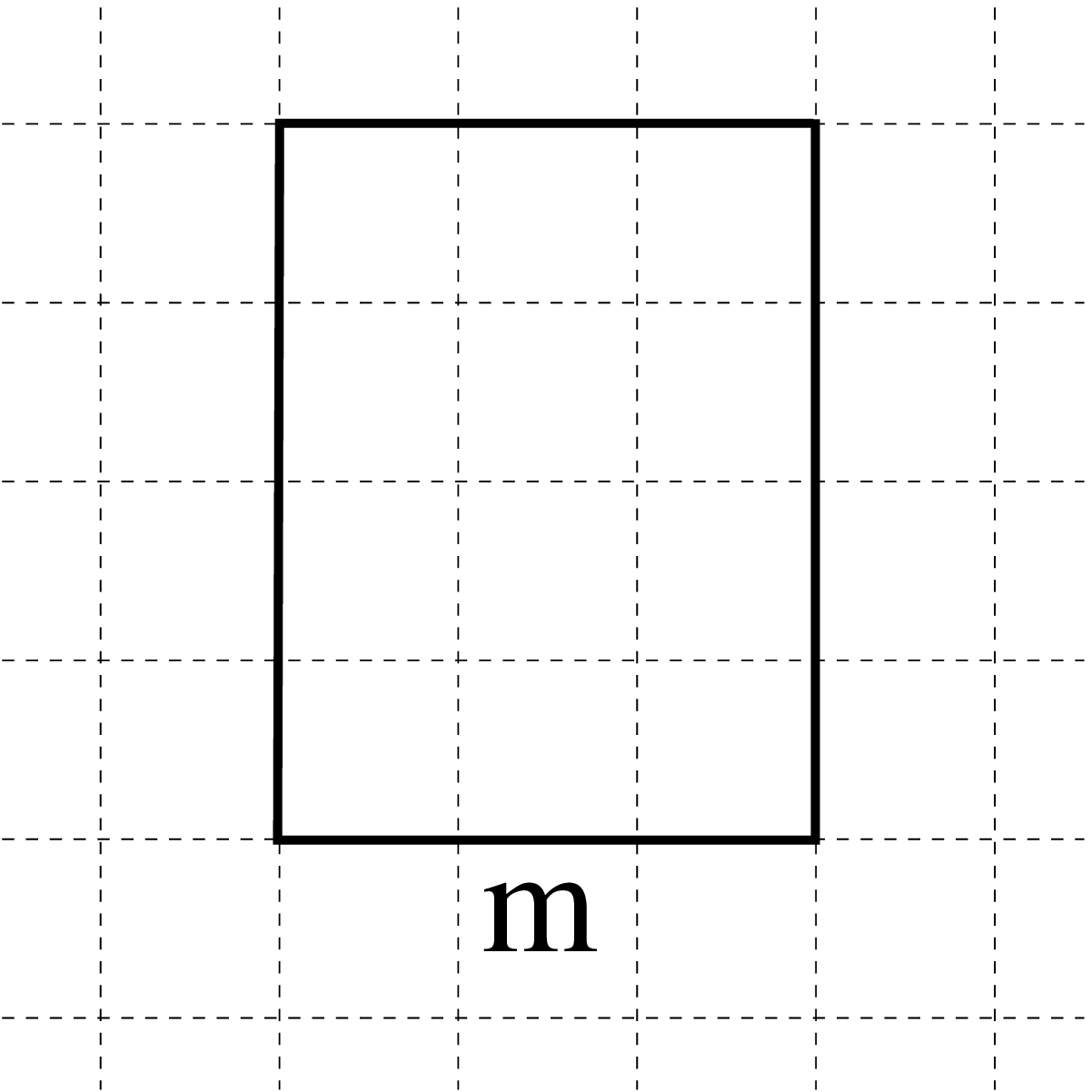}
\end{array}\begin{array}{c}
\includegraphics[height=.5cm]{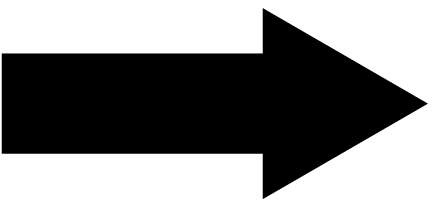}
\end{array}
\begin{array}{c}
\includegraphics[height=5cm]{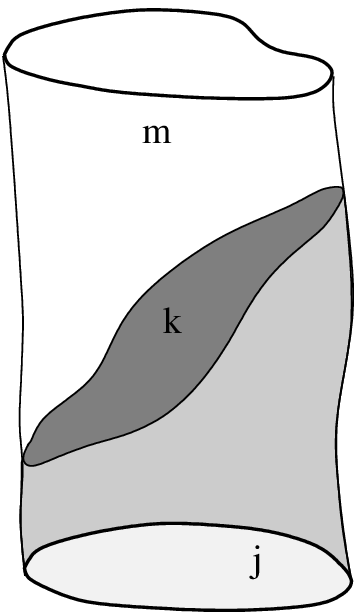}
\end{array}
\)} \caption{\small A set of discrete transitions in the
loop-to-loop physical inner product obtained by a series of
transitions as in Figure \ref{pito}. On the right, the continuous
{\em spin foam} representation in the limit $\epsilon\rightarrow
0$.} \label{lupy}
\end{figure}
We have introduced an auxiliary time slicing (arbitrary
parametrization). If one inserts the AL partition of unity \be
\mathbbm{1}=\ \sum \limits_{\gamma\subset \Sigma} \
\sum\limits_{\{j\}_{\gamma}} \ |\gamma,\{j\}><\gamma,\{j\}|,
\label{e} \end{equation} where the sum is over the complete basis of {\em
spin network}  states $\{|\gamma,\{j\}>\}$---based on all graphs
$\gamma \subset \Sigma$ and with all possible spin
labelling---between each time slice, one arrives at a sum over
spin-network histories representation of $P(s)$. More precisely,
$P(s)$ can be expressed  as a sum over amplitudes corresponding to a
series of transitions that can be viewed as the `time evolution'
between the `initial' {\em spin network}  $s$ and the `final'
`vacuum state' $\Omega$. The physical inner product between {\em
spin networks} $s$, and $s^{\prime}$ is defined as
\[<s,s^{\prime}>_p:=P(s^{\star}s^{\prime}),\]
and can be expressed as a sum over amplitudes corresponding to
transitions interpolating between the `initial' {\em spin network}
$s^{\prime}$ and the `final' {\em spin network}  $s$ (e.g. Figures
\ref{lupy} and \ref{vani}).

Spin network nodes evolve into edges while {\em spin network}
links evolve into 2-dimensional faces. Edges inherit the
intertwiners associated to the nodes and faces inherit the spins
associated to links. Therefore, the series of transitions can be
represented by a 2-complex whose 1-cells are labelled by
intertwiners and whose 2-cells are labelled by spins. The places
where the action of the plaquette loop operators create new links
(Figures \ref{pitolon} and \ref{vani}) define  0-cells or
vertices. These foam-like structures are the so-called {\em spin
foams}. The {\em spin foam} amplitudes are purely combinatorial
and can be explicitly computed from the simple action of the loop
operator in the AL-representation (Section \ref{alr}).
A particularly simple case arises when the {\em spin network}
states $s$ and $s^{\prime}$ have only 3-valent nodes. Explicitly
\be \label{3dc} <s,s^{\prime}>_p:=P(s^{\star}s^{\prime}) = \sum
\limits_{\{j\}} \ \prod_{f \subset F_{s\rightarrow s^{\prime}}} (2
j_f+1)^{\frac{\nu_f}{2}}
                \prod_{v\subset F_{s\rightarrow s^{\prime}}}
                \begin{array}{c}
                \includegraphics[width=3cm]{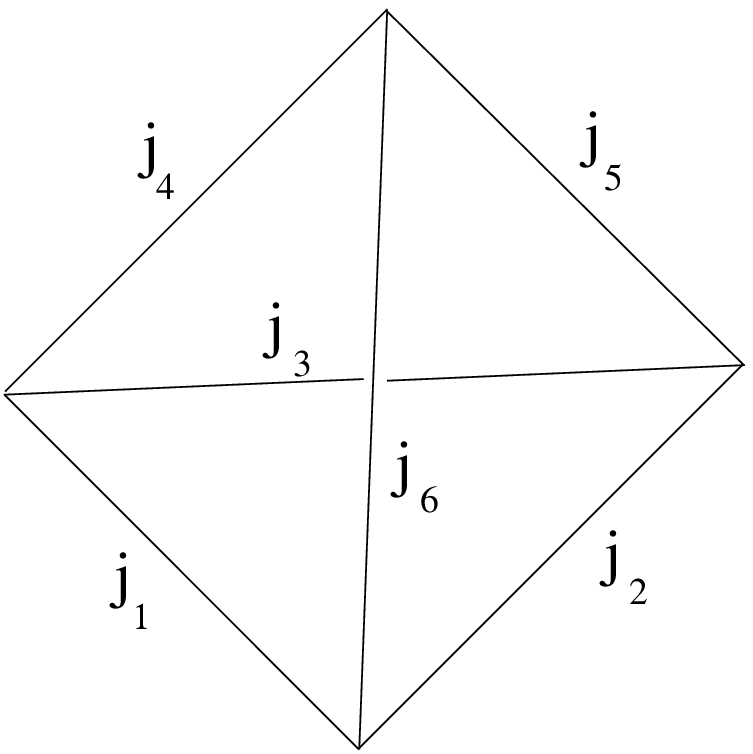}
                \end{array},
\end{equation} where the notation is that of (\ref{SF}), and $\nu_f=0$ if $f
\cap s \not= 0 \wedge f \cap s^{\prime}\not= 0$, $\nu_f=1$ if $f
\cap s \not= 0 \vee f \cap s^{\prime}  \not= 0$, and $\nu_f=2$ if
$f \cap s = 0 \wedge f \cap s^{\prime}= 0$. The tetrahedral
diagram denotes a $6j$-symbol: the amplitude obtained by means of
the natural contraction of the four intertwiners corresponding to
the 1-cells converging at a vertex.  More generally, for arbitrary
{\em spin networks}, the vertex amplitude corresponds to
$3nj$-symbols, and $<s,s^{\prime}>_p$ takes the general form
(\ref{SF}).

\subsection{{\em Spin foams} from the covariant path integral}\label{fcf}

In this section we re-derive the SF-representation of the physical
scalar product of 2+1 (Riemannian)\footnote{A generalization of
the construction presented here for Lorentzian 2+1 gravity has
been studied by Freidel\cite{fre1}.} quantum gravity directly as a
regularization of the covariant path integral\cite{iwa3,iwa4}. The
formal path integral for 3d gravity can be written as
\begin{equation}\label{zbf}
P = \int  {D}[e] {D}[A]\ {\rm exp}\left[{i \int_{\va M} {\rm
Tr}[e\wedge F(A)]}\right].
\end{equation}
Assume ${M}=\Sigma \times I$, where $I\subset\R$ is a closed
(time) interval (for simplicity we ignore boundary terms).
\begin{figure}[h]
\centerline{\hspace{0.5cm} \(
\begin{array}{c}
\includegraphics[width=12cm]{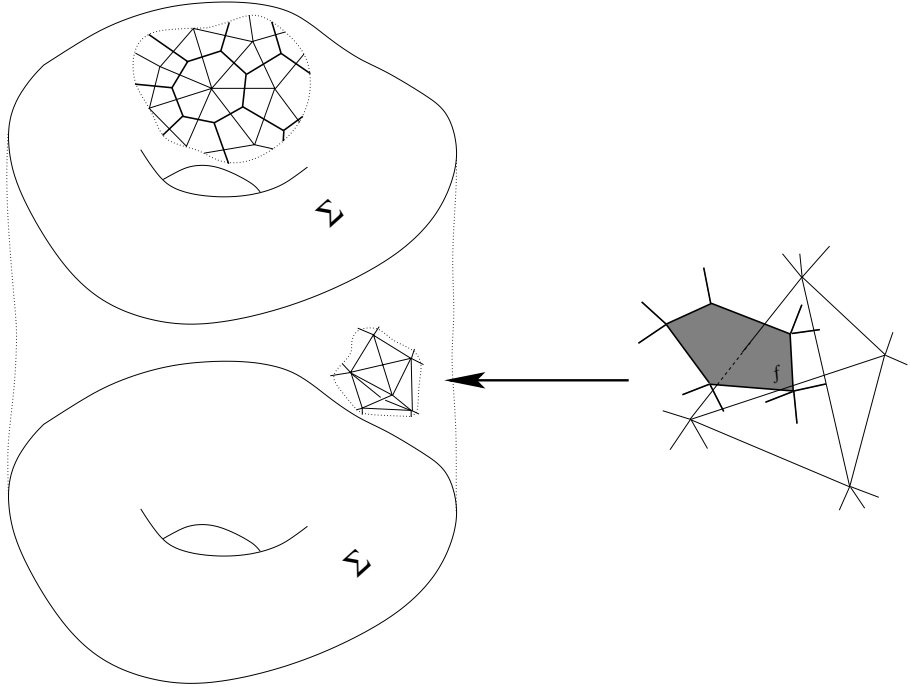}
\end{array}
\) }
\caption{The cellular decomposition of $M=\Sigma\times I$
($\Sigma=T^2$ in this example). The illustration
  shows part of the induced graph on the boundary and the detail of a
  tetrahedron in $\Delta$ and a face $f\subset\Delta^{\star}$ in the bulk.} \label{mani}
\end{figure}

In order to give a meaning to the formal expression above one
replaces the $3$-dimensional manifold (with boundary) ${M}$ with
an arbitrary cellular decomposition $\Delta$. One also needs the
notion of the associated dual 2-complex of $\Delta$ denoted by
$\Delta^{\star}$. The dual 2-complex ${\Delta}^{\star}$ is a
combinatorial object defined by a set of vertices $v\subset
{\Delta}^{\star}$ (dual to 3-cells in $\Delta$) edges $e\subset
{\Delta}^{\star}$ (dual to 2-cells in $\Delta$) and faces
$f\subset {\Delta}^{\star}$ (dual to $1$-cells in $\Delta$). The
intersection of the dual 2-complex $\Delta^{\star}$ with the
boundaries defines two graphs $\gamma_1, \gamma_2 \subset \Sigma$
(see Figure \ref{mani}). For simplicity we ignore the boundaries
until the end of this section.
\begin{figure}[h]
\centerline{\hspace{0.5cm} \(
\begin{array}{c}
\includegraphics[width=5cm]{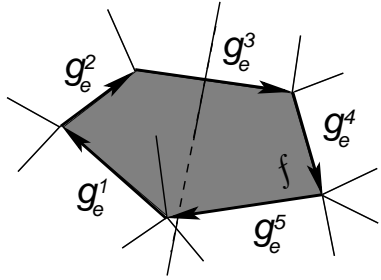}
\end{array}
\) }
\caption{A (2-cell) face $f\subset \Delta^{\star}$ in a cellular
decomposition of the spacetime manifold $M$.
Also the corresponding
dual 1-cell. The connection field is discretized by the assignment
of the parallel transport group-elements $g_e^i\in SU(2)$ to edges
$e\subset \Delta^{\star}$ ($i=1,\cdots 5$ in the face shown here).
} \label{caracara}
\end{figure}
The fields $e$ and $A$ are discretized as follows. The
$su(2)$-valued $1$-form field $e$ is represented by the assignment
of $e_f \in {su(2)}$ to each $1$-cell in $\Delta$. We use the fact
that faces in ${\Delta}^{\star}$ are in one-to-one correspondence
with $1$-cells in $\Delta$ and label $e_f$ with a face subindex
(Figure \ref{mani}). The connection field $A$ is represented by
the assignment of group elements $g_e \in SU(2)$ to each edge in
$e \subset {\Delta}^{\star}$ (see Figure \ref{caracara}).

With all this (\ref{zbf}) becomes the regularized version
$P_{\Delta}$ defined as
\begin{equation}\label{Zdiscrete}
{P}_{\Delta}=\int \prod_{f \subset {\Delta}^{\star}} de_f \
\prod_{e \subset {\Delta}^{\star}} dg_e  \ {\rm exp}\left[{i {\rm
Tr} \left[e_f W_f\right]}\right],
\end{equation}
where $de_f$ is the regular Lebesgue measure on $\R^3$, $dg_e$ is
the Haar measure on $SU(2)$, and $W_f$ denotes the holonomy around
(spacetime) faces, i.e., $W_f=g^1_e\dots g^{\va N}_e$ for $N$
being the number of edges bounding the corresponding face (see
Figure \ref{caracara}). The discretization procedure is
reminiscent of the one used in standard lattice gauge theory. The
previous definition can be motivated by an analysis equivalent to
the one presented in (\ref{***}).

Integrating over $e_f$, and using (\ref{pw}), one obtains
\begin{equation}\label{coloring}
P_{\Delta}= \sum \limits_{\{j\}} \int \ \prod_{e \subset
{\Delta}^{\star}} dg_e \ \prod_{f \subset {\Delta}^{\star}} (2
j_f+1) \ {\rm Tr}\left[\stackrel{j_{f}}{\Pi}(g^1_e\dots g^{\va
N}_e)\right].
\end{equation}

Now it remains to integrate over the lattice connection $\{g_e\}$.
If an edge $e\subset {\Delta}^{\star}$ bounds $n$ faces $f\subset
\Delta^{\star}$ there will be $n$ traces of the form ${\rm
Tr}[\stackrel{j_{f}}{\Pi}(\cdots g_e\cdots)]$ in (\ref{coloring})
containing $g_e$ in the argument. In order to integrate over $g_e$
we can use the following identity
\begin{equation}\label{3dp}
I^{n}_{inv}:= \int dg\ {\stackrel{j_{1}}{\Pi}(g)}\otimes
\stackrel{j_{2}}{\Pi}(g) \otimes \cdots \otimes
\stackrel{j_{n}}{\Pi}(g)= \sum_{\iota} {C^{\va \iota}_{\va j_1 j_2
\cdots j_n} \ C^{*{\va \iota}}_{\va j_1 j_2 \cdots j_n}},
\end{equation}
where $I^{n}_{inv}$ is the projector from the tensor product of
irreducible representations  ${\cal H}_{j_1\cdots j_n}=j_1\otimes
j_2 \otimes \cdots \otimes j_n$ onto the invariant component
${\cal H}^0_{j_1\cdots j_n}={\rm Inv}[j_1\otimes j_2 \otimes
\cdots \otimes j_n]$. On the r.h.s. we have chosen an orthonormal
basis of invariant vectors (intertwiners) in ${\cal H}_{j_1\cdots
j_n}$ to express the projector. Notice that the assignment of
intertwiners to edges is a consequence of the integration over the
connection. Using the (\ref{3dp}) one can write $P_{\Delta}$ in
the general SF-representation form (\ref{SF})
\begin{equation}\label{statesumx}
{P}_{\Delta}=\sum \limits_{\{ f\}}\ \prod_{f \subset
{\Delta}^{\star}} (2 j_f+1) \prod_{v\subset {\Delta}^{\star}}
A_v(j_v),
\end{equation}
where $A_v(\iota_v,j_v)$ is given by the appropriate trace of the
intertwiners corresponding to the edges bounded by the vertex. As
in the previous section this amplitude is given in general by an
$SU(2)$ $3Nj$-symbol. When $\Delta$ is a simplicial complex all
the edges in ${\Delta}^{\star}$ are $3$-valent and vertices are
$4$-valent. Consequently, the vertex amplitude is given by the
contraction of the corresponding four $3$-valent intertwiners,
i.e., a $6j$-symbol. In that case the path integral takes the
(Ponzano-Regge\cite{ponza}) form
\begin{equation}\label{statesum}
{P}_\Delta =\sum \limits_{ \{j\}} \ \prod_{f \subset
{\Delta}^{\star}} (2 j_f+1) \prod_{v\subset {\Delta}^{\star}}
\begin{array}{c}
\includegraphics[width=3cm]{tetras.eps}\end{array}.
\end{equation}
The labelling of faces that intersect the boundary naturally
induces a labelling of the edges of the graphs $\gamma_1$ and
$\gamma_2$ induced by the discretization. Thus, the boundary
states are given by {\em spin network}  states on $\gamma_1$ and
$\gamma_2$ respectively. A careful analysis of the boundary
contribution shows that only the face amplitude is modified to
$(2j_{f}+1)^{\nu_f/2}$, and that the {\em spin-foam} amplitudes
are as in  Equation (\ref{3dc}).

A crucial property of the path integral in 3d gravity (and of the
transition amplitudes in general) is that it does not depend on
the discretization $\Delta$---this is due to the absence of local
degrees of freedom in 3d gravity and not expected to hold in 4d.
Given two different cellular decompositions $\Delta$ and
$\Delta^{\prime}$ one has
\begin{equation}\label{rucu} \tau^{-n_0}
P_{\Delta}=\tau^{-n^{\prime}_0} P_{\Delta^{\prime}},
\end{equation}
where $n_0$ is the number of 0-simplexes in $\Delta$, and
$\tau=\sum_j (2j+1)^2$. This trivial scaling property of
transition amplitudes allows for a simple definition of transition
amplitudes that are independent of the discretization\cite{za1}.
However, notice that since $\tau$ is given by a divergent sum the
discretization independence statement is formal. Moreover, the sum
over spins in (\ref{statesum}) is typically divergent. Divergences
occur due to infinite gauge-volume factors in the path integral
corresponding to the topological gauge freedom (\ref{gauge2}).
Freidel and Louapre\cite{frei9} have shown how these divergences
can be avoided by gauge-fixing un-physical degrees of freedom in
(\ref{Zdiscrete}). In the case of 3d gravity with positive
cosmological constant the state-sum generalizes to the Turaev-Viro
invariant\cite{TV,tur,kau,a1} defined in terms of the quantum
group $SU_q(2)$ with $q^n=1$ where the representations are
finitely many and thus $\tau < \infty$. Equation (\ref{rucu}) is a
rigorous statement in that case. No such infrared divergences
appear in the canonical treatment of the previous section.

\section{{\em Spin foams} in four dimensions}

\subsection{SF from the canonical formulation}

There is no rigorous construction of the physical inner product of
LQG in four dimensions. The {\em spin foam} representation as a
device for its definition has been introduced formally by
Rovelli\cite{c2} and Rovelli and Reisenberger\cite{reis5}. In
4-dimensional LQG difficulties in understanding dynamics are
centered around the quantum scalar constraint $\widehat
S=\widehat{{\sqrt{{\rm det}E}}^{-1}{E_i^a E_j^b
F^{ij}_{ab}(A)}}+\cdots$ (see (\ref{QEE}))---the vector constraint
$\widehat V_a(A,E)$ is solved in a simple manner. The physical
inner product formally becomes
\begin{eqnarray} \label{vanin}&&\nonumber \left<Ps,
s^{\prime}\right>_{\vani diff}=\prod_{x} \delta[\widehat S(x)]=
\int {D}[N]<{\rm exp}\left[i \int
\limits_{\Sigma} N(x) \widehat {S}(x)\right] \ s, s^{\prime}>_{\vani diff}\\
&&\ \ \ \ \ \ \ \ \ \ \ \ \ \ \ \ \ \ \ \ \ \ \ \ \ \  =\int
{D}[N] \sum \limits^{\infty}_{n=0} \frac{i^{n}}{n!}<\left[\int
\limits_{\Sigma} N(x) \widehat {S}(x)\right]^n \ s,
s^{\prime}>_{\vani diff},
\end{eqnarray}
\vskip-.1cm \noindent where $<\ ,\ >_{\vani diff}$ denotes the
inner product in the Hilbert space of solutions of the vector
constraint, and  the exponential has been expanded in powers in
the second line.

\begin{figure}[h]\!\!\!\!\!\!
\centerline{\hspace{0.5cm} \( \!\!\!\!\!\! \!\!\!\!\!\! \int
\limits_{\Sigma} N(x) \widehat {S}(x) \ \rhd
\!\!\!\!\!\!\!\!\!\!\!\!
\begin{array}{c}
\includegraphics[height=3.5cm]{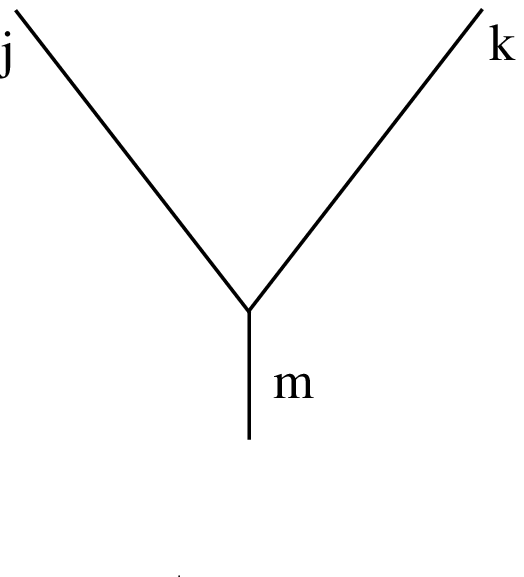}
\end{array} \!\!\!\!\!\! =\sum \limits_{nop} N(x_n) S_{nop} \!\!\!\!\!\!
\begin{array}{c}
\includegraphics[height=3.5cm]{tero3.eps}
\end{array}\ \ \ \
\begin{array}{c}
\includegraphics[height=.5cm]{flecha.eps}
\end{array}\ \ \
\begin{array}{c}
\includegraphics[height=4cm]{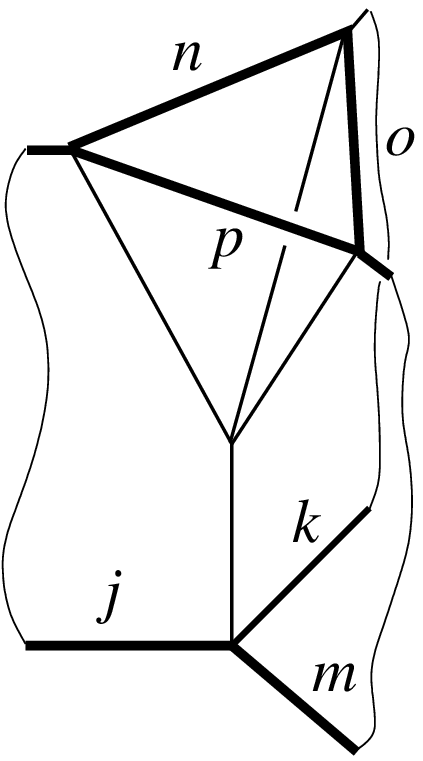}
\end{array}
\) } \caption{The action of the scalar constraint and its {\em
spin foam} representation. $N(x_n)$ is the value of $N$ at the
node and $S_{nop}$ are the matrix elements of $\widehat S$. }
\label{actionH}
\end{figure}

From early on, it was realized that smooth loop states are
naturally annihilated by $\widehat {S}$(independently of any
quantization ambiguity\cite{jac,c8}). Consequently, $\widehat S$
acts only on {\em spin network} nodes. Generically, it does so by
creating new links and nodes modifying the underlying graph of the
{\em spin network} states (Figure \ref{actionH}).

Therefore, each term in the sum (\ref{vanin}) represents a series
of transitions---given by the local action of $\widehat {S}$ at
{\em spin network} nodes---through different {\em spin network}
states interpolating the boundary states $s$ and $s^{\prime}$
respectively. The action of $\So$ can be visualized as an
`interaction vertex' in the `time' evolution of the node (Figure
\ref{actionH}). As in the explicit 3d case, equation (\ref{vanin})
can be expressed as sum over `histories' of {\em spin networks}
pictured as a system of branching surfaces described by a
2-complex whose elements inherit the representation labels on the
intermediate states. The value of the `transition' amplitudes is
controlled by the matrix elements of $\widehat S$. Therefore,
although the qualitative picture is independent of quantization
ambiguities transition amplitudes are sensitive to them.

Before even considering the issue of convergence of (\ref{vanin}),
the problem with this definition is evident: every single term in
the sum is a divergent integral! Therefore, this way of presenting
{\em spin foams} has to be considered as formal until a well
defined regularization of (\ref{P}) is provided. That is the goal
of the {\em spin foam} approach.

Instead of dealing with an infinite number of constraints Thiemann
recently proposed to impose one single {\em master} constraint
defined as
\begin{equation}
M= \int \limits_{\Sigma} dx^3 \ \frac{S^2(x)- q^{ab} V_{a}(x)
V_{b}(x)}{\sqrt{{\rm det}\ q(x)}}.
\end{equation}
Using techniques developed by Thiemann this constraint can indeed
be promoted to a quantum operator acting on $\Hk$. The physical
inner product is given by
\begin{equation}
<s,s^{\prime}>_p:=\lim_{T\rightarrow \infty} <s,\int
\limits_{-T}^{T}\ dt \ e^{i t \widehat M}s^{\prime}>.
\end{equation}
A SF-representation of the previous expression could now be
achieved by the standard skeletonization that leads to the path
integral representation in quantum mechanics. In this context one
splits the $t$-parameter in discrete steps and writes
\begin{equation} e^{it \widehat M}=\lim_{N\rightarrow \infty}\ [e^{it
\widehat M/N}]^N =\lim_{N\rightarrow \infty}\ [1+it \widehat
M/N]^N.
\end{equation}
The SF-representation follows from the fact that the action of the
basic operator $1+it \widehat M/N$ on a {\em spin network}  can be
written as a linear combination of new {\em spin networks}  whose
graphs and labels have been modified by the creation of new nodes
(in a way qualitatively analogous to the local action shown in
Figure \ref{actionH}). An explicit derivation of the physical
inner product of 4d LQG along these lines is under current
investigation.

\subsection{{\em Spin foams} from the covariant formulation}

In four dimensions the {\em spin foam} representation of the
dynamics of LQG has been motivated by lattice discretizations of
the path integral of gravity in the covariant
formulation\cite{reis8,reis6,reis4,iwa1}. This has lead to a
series of constructions which are refereed to as {\em spin foam
models\cite{a18}}. These treatments are closer related to the
construction of Section \ref{fcf}. Here we illustrate the
formulation which has captured much interest in the literature:
the {\em Barrett-Crane model} (BC model)\cite{BC2,BC1}.

\subsubsection{{\em Spin foam} models for gravity as constrained quantum BF theory}

The \BC model is one of the most extensively studied {\em spin
foam} models for quantum gravity. To introduce the main ideas
involved  we concentrate on the definition of the model in the
Riemannian sector. The \BC model can be formally viewed as a {\em
spin foam} quantization of $SO(4)$ Plebanski's formulation of
GR\cite{pleb}. Plebanski's Riemannian action depends on an $so(4)$
connection $A$, a Lie-algebra-valued 2-form $B$ and Lagrange
multiplier fields $\lambda$ and $\mu$. Writing explicitly the
Lie-algebra indices, the action is given by \vskip-.3cm
\begin{equation}\label{pleb}
I[B,A,\lambda,\mu]=\int \left[B^{IJ}\wedge F_{IJ}(A) +
\lambda_{IJKL} \ B^{IJ} \wedge B^{KL} +\mu
\epsilon^{IJKL}\lambda_{IJKL} \right],
\end{equation}
\vskip-.1cm \noindent where $\mu$ is a 4-form and
$\lambda_{IJKL}=-\lambda_{JIKL}=-\lambda_{IJLK}=\lambda_{KLIJ}$ is
a tensor in the internal space. Variation with respect to $\mu$
imposes the constraint $\epsilon^{IJKL}\lambda_{IJKL}=0$ on
$\lambda_{IJKL}$. The Lagrange multiplier tensor $\lambda_{IJKL}$
has then $20$ independent components. Variation with respect to
$\lambda$ imposes $20$ algebraic equations on the $36$ components
of $B$. The (non-degenerate) solutions to the equations obtained
by varying the multipliers $\lambda$ and $\mu$ are \vskip-.3cm
\begin{equation}\label{ambi}
B^{IJ}=\pm \epsilon^{IJKL} e_K \wedge e_L, \ \ \ {\rm and}\ \ \
B^{IJ}=\pm e^I\wedge e^J,
\end{equation}
\vskip-.1cm \noindent in terms of the $16$ remaining degrees of
freedom of the tetrad field $e^I_a$. If one substitutes the first
solution into the original action one obtains Palatini's
formulation of general relativity; therefore on shell (and on the
right sector) the action is that of classical gravity.

The key idea in the definition of the model is that the path
integral for the theory corresponding to the action $I[B,A,0,0]$,
namely \vskip-.3cm \be P_{topo}=\int \ D[B] D[A] \ {\rm
exp}\left[i \int \left[B^{IJ}\wedge F_{IJ}(A)
\right]\right]\label{Topo}\end{equation} \vskip-.1cm \noindent can be given a
meaning as a {\em spin foam} sum, (\ref{SF}), in terms of a simple
generalization of the construction of Section \ref{pipo}. In fact
$I[B,A,0,0]$ corresponds to a simple theory known as BF theory
that is formally very similar to 3d gravity\cite{baez5}. The
result is independent of the chosen discretization because BF
theory does not have local degrees of freedom (just as 3d
gravity).

The \BC model aims at providing a definition of the path integral
of gravity pursuing a well-posed definition of the formal
expression \vskip-.3cm
\begin{equation}\label{heu}
P_{GR}=\int \ D[B] D[A] \ \delta\left[B\rightarrow \epsilon^{IJKL}
e_K \wedge e_L \right] \  {\rm exp}\left[i \int \left[B^{IJ}\wedge
F_{IJ}(A) \right]\right],\end{equation} \vskip-.1cm \noindent
 where $D[B] D[A] \delta(B\rightarrow \epsilon^{IJKL} e_K \wedge e_L)$ means that one must restrict the sum in
(\ref{Topo}) to those configurations of the topological theory
satisfying the constraints $B=*( e \wedge e)$ for some tetrad $e$.
The remarkable fact is that this restriction can be implemented in
a systematic way directly on the {\em spin foam} configurations
that define $P_{topo}$\cite{crane0,crane00}.

In $P_{topo}$ {\em spin foams} are labelled with spins
corresponding to the unitary irreducible representations of
$SO(4)$ (given by two spin quantum numbers ($j_R,j_L$)).
Essentially, the factor `$\delta(B\rightarrow \epsilon^{IJKL} e_K
\wedge e_L)$' restricts the set of {\em spin foam} quantum numbers
to the so-called simple representations (for which
$j_R=j_L=j$\cite{baez7,baez6}). This is the `quantum' version of
the solution to the constraints (\ref{ambi}). There are various
versions of this model, some versions satisfy intriguing
finiteness\cite{a2,a22,a7,a10} properties\footnote{For a
discussion of the freedom involved see\cite{myo}.}. The simplest
definition of the transition amplitudes in the \BC-model is given
by \be \label{4dc} P(s^{\star}s) = \sum \limits_{\{j\}} \ \prod_{f
\subset F_{s\rightarrow s^{\prime}}} (2 j_f+1)^{\nu_f}
                \prod_{v\subset F_{s\rightarrow s^{\prime}}}
                \sum \limits_{\iota_1\cdots \iota_5}
\begin{array}{c}
\includegraphics[width=7cm]{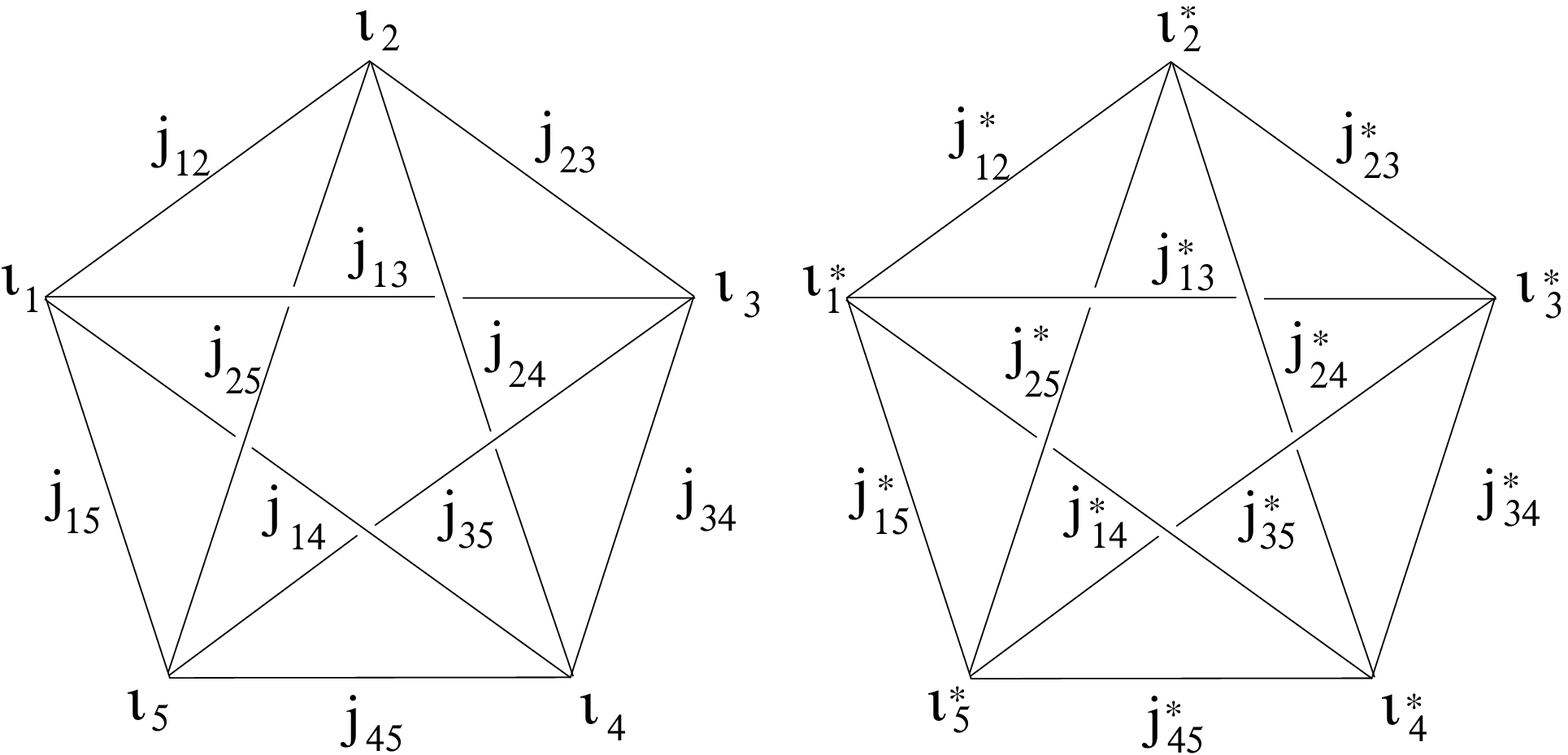}\end{array},
\end{equation} where we use the notation of (\ref{3dc}), the graphs denote
$15j$ symbols, and $\iota_i$ are half integers labelling $SU(2)$
normalized $4$-intertwiners\footnote{Reisenberger\cite{reis3}
proved that the 4-simplex BC amplitude is unique up to
normalization.}. No rigorous connection with the Hilbert space
picture of LQG has yet been established. The self-dual version of
Plebanski's action leads, through a similar construction, to
Reisenberger's model\cite{reis4}. A general prescription for the
definition of constrained BF theories\footnote{Gambini and  Pullin
studied an alternative modification of BF theory leading to a
simple model with intriguing properties\cite{pul2}.} on the
lattice has been studied by Freidel and Krasnov\cite{fre5}.
Lorentzian generalizations of the Barrett-Crane model have been
defined\cite{a8,a9}. A generalization using quantum groups was
studied by Noui and Roche\cite{Noui:2002ag}.
\begin{figure}[h]\!\!\!\!\!\!
\centerline{\hspace{0.5cm} \(
\begin{array}{c}
\includegraphics[height=3.5cm]{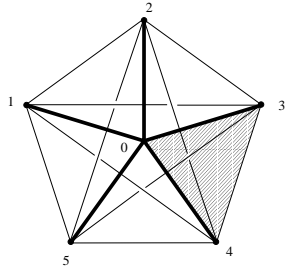}
\end{array}\) }
\caption{The dual of a 4-simplex. } \label{spi}
\end{figure}

The simplest amplitude in the \BC model corresponds to a single
4-simplex. A 4-simplex can be viewed as the simplest triangulation
of the 4-dimensional space time given by the interior of a
3-sphere (the corresponding 2-complex is shown in Figure
\ref{spi}). States of the 4-simplex are labelled by 10 spins $j$
(labelling the 10 edges of the boundary {\em spin network}, see
Figure \ref{spi}) which can be shown to be related to the area in
Planck units of the 10 triangular faces that form the 4-simplex. A
first indication of the connection of the model with gravity was
that the large $j$-asymptotics appeared to be dominated by the
exponential of the Regge action\cite{crane,bawi} (the action
derived by Regge as a discretization of GR). This estimate was
done using the stationary phase approximation to the
integral\cite{baba} that gives the amplitude of a 4-simplex in the
\BC model. However, more detailed calculations showed that the
amplitude is dominated by configurations corresponding to
degenerate 4-simplexes\cite{baez8,frei9,Barrett:2002ur}. This seem to
invalidate a simple connection to GR and is one of the main
puzzles in the model.

\subsection{{\em Spin foams} as Feynman diagrams}

The main problem with the models of the previous section is that
they are defined on a discretization $\Delta$ of $M$ and
that---contrary to what happens with a topological theory, e.g. 3d
gravity (Equation \ref{rucu})---the amplitudes depend on the
discretization $\Delta$. Various possibilities to eliminate this
regulator have been discussed in the literature but no explicit
results are yet known in 4d. An interesting proposal is a
discretization-independent definition of {\em spin foam} models
achieved by the introduction of an auxiliary field theory living
on an abstract group manifold---$Spin(4)^4$ and $SL(2,C)^4$ for
Riemannian and Lorentzian gravity
respectively\cite{fre2,reis1,reis2,pietri1,mik1}. The action of
the auxiliary group field theory (GFT) takes the form \vskip-.3cm
\be I[\phi]=\int_{G^4} \phi^2+\frac{\lambda}{5!}\int_{G^{10}}
M^{(5)}[\phi],\end{equation} \vskip-.1cm \noindent where $M^{(5)}[\phi]$ is a
fifth order monomial, and $G$ is the corresponding group. In the
simplest model
$M^{(5)}[\phi]=\\
\phi(g_1,g_2,g_3,g_4)\phi(g_4,g_5,g_6,g_7)
\phi(g_7,g_3,g_8,g_9)\phi(g_9,g_6,g_2,g_{10})\phi(g_{10},g_8,g_5,g_1)$.
The field $\phi$ is required to be invariant under the
(simultaneous) right action of the group on its four arguments in
addition to other symmetries (not described here for simplicity).
The perturbative expansion in $\lambda$ of the GFT {\em Euclidean}
path integral is given by \vskip-.3cm \be P=\int D[\phi]
e^{-I[\phi]}=\sum \limits_{{ F}_N} \frac{\lambda^N}{sym[{ F}_N]}
A[{F}_N],\end{equation} \vskip-.1cm \noindent where $A[{F}_N]$ corresponds to
a sum of Feynman-diagram amplitudes for diagrams with $N$
interaction vertices, and $sym[{F}_N]$ denotes the standard
symmetry factor. A remarkable property of this expansion is that
$A[{ F}_N]$ can be expressed as a sum over {\em {\em spin foam}}
amplitudes, i.e., 2-complexes labelled by unitary irreducible
representations of $G$. Moreover, for very simple interaction
$M^{(5)}[\phi]$, the {\em spin foam} amplitudes are in one-to-one
correspondence to those found in the models of the previous
section (e.g. the BC model). This {\em duality} is regarded as a
way of providing a fully combinatorial definition of quantum
gravity where no reference to any discretization or even a
manifold-structure is made. Transition amplitudes between {\em
spin network} states correspond to $n$-point functions of the
field theory\cite{a3}. These models have been inspired by
generalizations of matrix models applied to BF theory\cite{bu,oo}.

Divergent transition amplitudes can arise by the contribution of
`loop' diagrams as in standard QFT. In {\em spin foams}, diagrams
corresponding to 2-dimensional bubbles are potentially divergent
because spin labels can be arbitrarily high leading to unbounded
sums in (\ref{SF}). Such divergences do not occur in certain field
theories dual (in the sense above) to the Barrett-Crane (BC)
model. However, little is known about the convergence of the
series in $\lambda$ and the physical meaning of this constant.
Nevertheless, Freidel and Louapre\cite{frei10} have shown that the
series can be re-summed in certain models dual to lower
dimensional theories. Techniques for studying the continuum limit
of these kind of theories have been
proposed\cite{fot4,fot5,Oeckl:2002ia,Oeckl:2004yf}. There are
models defined in this context admitting matter degree of
freedom\cite{mik2,mik3}.

\subsection{Causal {\em spin foams}}\label{fotin}

Let us finish by presenting a fundamentally different construction
leading to {\em spin foams}. Using the kinematical setting of LQG
with the assumption of the existence of a micro-local (in the
sense of Planck scale) causal structure Markopoulou and
Smolin\cite{fot1,fot2,fot3} define a general class of (causal)
{\em spin foam} models for gravity. The elementary transition
amplitude $A_{s_I\rightarrow s_{I+1}}$ from an initial {\em spin
network} $s_{I}$ to another {\em spin network} $s_{I+1}$ is
defined by a set of simple combinatorial rules based on a
definition of causal propagation of the information at nodes. The
rules and amplitudes have to satisfy certain causal restrictions
(motivated by the standard concepts in classical Lorentzian
physics). These rules generate surface-like excitations of the
same kind one encounters in the previous formulations. {\em Spin
foams} ${F}^{N}_{s_i\rightarrow s_{f}}$ are labelled by the number
of times, $N$, these elementary transitions take place. Transition
amplitudes are defined as
\begin{equation}
\left<s_i,s_f\right>=\sum_{N} A({F}^{N}_{s_i\rightarrow s_{f}})
\end{equation}
which is of the generic form (\ref{SF}). The models are not
related to any continuum action. The only guiding principles in
the construction are the restrictions imposed by causality, and
the requirement of the existence of a non-trivial critical
behavior that reproduces general relativity at large scales. Some
indirect evidence of a possible non trivial continuum limit has
been obtained in certain versions of these models in $1+1$
dimensions.

\section{Some final bibliographic remarks}

We did not have time to discuss the applications of loop quantum
gravity to cosmology. The interested reader is referred to review
article\cite{marhugo} and the  references therein.

We did not have the chance to mention in this lectures the
important area of research in LQG devoted to the study of the low
energy limit of the theory. We refer the reader to the general
sources\cite{ash10,book,bookt} for bibliography and an overview of
results and outlook.

In Section  \ref{sect4} we introduced a representation of the basic kinematical
observables in the kinematical Hilbert space $\Hk$. 
That was the starting point for the definition of the theory.
The reader might wonder why one emphasizes so much the Hilbert space as a
fundamental object when in the context of standard
quantum field theory it is rather the algebra of observables what
plays the fundamental role. Hilbert spaces correspond to
representations of the algebra of observables which are chosen according to the physical situation at
hand. However, when extra symmetry is present it can happen that
there is no freedom and that a single representation is selected by
the additional symmetry. This is in fact the case in LQG if one 
imposes the condition  of diffeomorphism invariance on the state that defines
the representation of the holonomy and the flux operators
\cite{Sahlmann:2003qa,Sahlmann:2003in,Sahlmann:2002xv,Sahlmann:2002xu,Okolow:2003pk,Okolow:2004th,lost}.

\section*{Acknowledgments}

I would like to thank the organizer of the {\em Second International
Conference on Fundamental Interactions} for their support and for a
wonderful conference. Special thanks to  them also 
for their great hospitality and the wonderful time we had in Pedra Azul. 
I thank Abhay Ashtekar, Bernd Bruegmann, Rodolfo Gambini, Jurek
Lewandowski, Marcelo Maneschy, Karim Noui and Carlo Rovelli for
discussions and to Mikhail Kagan and Kevin Vandersloot for the careful
reading of the manuscript. I thank Hoi Lai Yu for listening to an early version 
to these lectures and for insightful questions. Many thanks to  Olivier Piguet and Clisthenis Constantinidis. 
This work has been supported by NSF grants
PHY-0354932 and INT-0307569 and the Eberly Research Funds of Penn
State University.


\end{document}